\documentclass[%
 reprint,
superscriptaddress,
nofootinbib,
 amsmath,amssymb,
aps,
pre,
]{revtex4-2}

\usepackage{graphicx}
\usepackage{subcaption}
\usepackage{dcolumn}
\usepackage{empheq}
\usepackage{bm}
\usepackage{hyperref}
\usepackage{color}

\usepackage{blkarray}
\usepackage{mathtools}
\usepackage{xcolor}

\newcommand{\aref}[1]{\hyperref[#1]{Appendix~\ref*{#1}}}

\definecolor{darkblue1}{RGB}{9,34,74}
\definecolor{yellow0}{RGB}{249,232,94}

\begin{document}


\title{Generative models for two-ground-truth partitions in networks}

\author{Lena Mangold}
 \email{lena.mangold@cmb.hu-berlin.de, she/her/hers}
\author{Camille Roth}%
 \affiliation{%
 Computational Social Science team, Centre Marc Bloch, \hbox{Friedrichstr.} 191, 10117 Berlin, Germany
}%
 \affiliation{%
 Centre national de la recherche scientifique (CNRS), 3 rue Michel-Ange, 75 016 Paris, France
}%
 \affiliation{Centre d'Analyse et de Math\'ematique Sociales (CAMS), \'Ecole des hautes \'etudes en sciences sociales (EHESS), 54 Bd Raspail, 75006 Paris, France
 \vspace{0.3cm}
 }
\date{\today}

\newcommand{\tb}[1]{\textcolor{blue}{#1}}

\begin{abstract}
A myriad of approaches have been proposed to characterise the mesoscale structure of networks – most often as a partition based on patterns variously called communities, blocks, or clusters. Clearly, distinct methods designed to detect different types of patterns may provide a variety of answers to the network’s mesoscale structure. Yet, even multiple runs of a given method can sometimes yield diverse and conflicting results, producing entire landscapes of partitions which potentially include multiple (locally optimal) mesoscale explanations of the network. Such ambiguity motivates a closer look at the ability of these methods to find multiple qualitatively different `ground truth' partitions in a network. Here, we propose the stochastic cross-block model (SCBM), a generative model which allows for two distinct partitions to be built into the mesoscale structure of a single benchmark network. We demonstrate a use case of the benchmark model by appraising the power of stochastic block models (SBMs) to detect implicitly planted coexisting bi-community and core-periphery structures of different strengths. Given our model design and experimental set-up, we find that the ability to detect the two partitions individually varies by SBM variant and that coexistence of both partitions is recovered only in a very limited number of cases. Our findings suggest that in most instances only one -- in some way dominating -- structure can be detected, even in the presence of other partitions. They underline the need for considering entire landscapes of partitions when different competing explanations exist and motivate future research to advance partition coexistence detection methods. Our model also contributes to the field of benchmark networks more generally by enabling further exploration of the ability of new and existing methods to detect ambiguity in the mesoscale structure of networks.
\end{abstract}

\maketitle


\section{\label{sec:level1}Introduction}
Network structure is frequently characterized at the mesoscale level by the configuration of what is broadly denoted as `communities' --- groupings of nodes that display some sort of similarity in terms of their connectivity in the network. Networks may exhibit a wide variety of mesoscale structures, such as densely connected or cohesive clusters, assortative or disassortative communities, core-periphery structures, equivalence classes, or combinations thereof \citep{leskovec_community_2009,rosvall_different_2019-1}. In turn, there is often more than one scientifically plausible way to divide the nodes of a real-world network, as demonstrated for instance by the coexistence of both cohesive clusters and core-periphery structures in multiple cases \cite{rombach_core-periphery_2017, zhang_identification_2015}.

Clearly, methods designed to identify distinct types of mesoscale structures yield different partitions for a given network. Perhaps more interestingly, results produced by different algorithms aimed at identifying one specific type of mesoscale structure may still vary considerably for the same network. A commonly studied empirical example is the Karate Club (KC) network, a friendship network of 34 members of a sports club which split into two new clubs after a fall-out between its members \citep{zachary_information_1977}. While the existing literature has repeatedly produced a partition of two cohesive groupings similar in terms of node membership to the division caused by the split of the club \citep{newman_finding_2004, newman_modularity_2006}, variability in what is detected as the ‘optimal’ partition of this network has been demonstrated not only in terms of community membership of nodes \citep{newman_finding_2004} but also of the total number of communities recovered \citep{duch_community_2005, blondel_fast_2008, evans_clique_2010}. Additionally, other types of mesoscale structures can be detected as plausible explanations for the KC network \cite{peixoto_revealing_2021}, including a core-periphery-type structure of leaders and followers.

Competing explanations of mesoscale structure in real networks, such as the KC example, motivate a further exploration of ambiguity on this scale; perhaps the reason for conflicting results is that multiple qualitatively different `ground truths' and partitions were responsible for the generative process of a network and its mesoscale configuration \cite{peel_ground_2017}. In fact, recent work on stochastic block models (SBMs), which have become increasingly popular for mesoscale network description, has emphasised the importance of exploring the variability of the entire partition landscapes that they return, instead of forcing a global consensus from a distribution of partitions (\hbox{i.e.} choosing one among many by maximising some objective) \cite{peixoto_revealing_2021}. A further important phenomenon in the context of mesoscale variability is that of detectability limits, where known structures are no longer detected due to lacking signal strength and which have been shown to exist due to the presence of phase transitions in networks generated by SBMs \cite{decelle_asymptotic_2011, decelle_inference_2011}.

On the whole, mesoscale variability may thus stem both from the intrinsic ambiguity in the generative processes of the network and from the stochastic ambiguity of the generative blockmodel.
In this paper we are generally interested in appraising the accuracy of mesoscale structure detection under ambiguity constraints. We choose (1) to start from the KC example as one of the simplest configurations possessing jointly a \textit{core-periphery} and a \textit{bi-community}  (two equally-sized communities) structure, and (2) to rely on SBMs as a theoretical framework accommodating many types of mesoscale structures, beyond clusters and including the two above ones, and that may further be used not only for generating but also for detecting mesoscale structures.
More specifically, we are motivated to explore the ability of SBMs to detect certain structures when we implicitly introduce some level of measurable ambiguity into the mesoscale. We propose a framework for a generative benchmark model, the stochastic cross-block model (SCBM), which can have such ambiguity built into its mesoscale structure by allowing for two qualitatively distinct partitions (\hbox{i.e.} two `ground truths') to be planted into the same network. Using this framework, the two partitions are defined respectively through block matrices that specify the connectivity within and between blocks which -- similar to the planted partition model \cite{condon_algorithms_2001} -- facilitates the analysis of SBMs (or any community detection algorithm) for varying strengths of the planted mesoscale structures. We use our framework to plant two qualitatively different partitions into a synthetic network and try to recover them using two SBM variants. In this way, we analyse the ability of SBMs to detect two competing structures present in a network, appraising both the extent to which each partition is recovered individually as well as the successful detection of the \textit{coexistence} of both partitions: the appearance of two implicitly planted structures within the posterior distribution of inferred partitions of a given graph.

\bigskip
This paper is structured as follows. In \autoref{sec:background}, we provide some background on conflicting explanations of mesoscale structure in networks (\ref{sec:dissensus}), the interplay between ambiguity and detectability of block structures (\ref{sec:ambiguity}), as well as a more general overview of the existing literature on SBMs (\ref{sec:sbms}) and generative benchmark models of mesoscale structures (\ref{sec:generative}). We then introduce our model framework in \autoref{sec:framework}, covering the derivation of the model in the two-partition case and two variants of generative processes of edge placement in the network. In \autoref{sec:simulations}, we illustrate a use case for our model in form of a set of simulations, the results of which we discuss in \autoref{sec:results}. We summarise our main results in \autoref{sec:conclusion} and briefly touch on possible future work that could extend on our simulations and on the model itself.

\section{\label{sec:background} Background}
\subsection{\label{sec:dissensus}Community detection \& partition landscapes}
Community detection often adopts a clustering perspective and focuses on cohesive communities, which denote groups of nodes more densely connected to other nodes of the same group than to nodes in other groups \citep{alba_graph-theoretic_1973}, as opposed to other types of meso-level structures more generally \citep{rosvall_different_2019-1}. Corresponding methods to identify community structure are designed to perform best with specific types of data and networks \citep{peel_ground_2017} and many authors have chosen distinct routes to optimise for the most plausible partition (see \cite{fortunato_community_2010} for a review of different methods). Existing algorithms therefore have at their basis a multitude of measures, such as modularity \citep{newman_modularity_2006}, spectral properties \citep{newman_spectral_2013}, generative models \citep{karrer_stochastic_2011}, betweenness centrality \citep{newman_finding_2004}, or information-theoretic methods \citep{rosvall_maps_2008}, which is only one of the causes for the diversity in results from algorithms that use different approaches. 

Similarly, the detection of core-periphery structure -- the division of a network into a well-connected, cohesive core group and a sparsely connected peripheral group, first rigorously formulated by \citet{borgatti_models_2000} -- has been approached in a number of different ways \cite{holme_core-periphery_2005}, including methods based on edge density \cite{borgatti_models_2000, rombach_core-periphery_2017, lee_density-based_2014}, path length \cite{cucuringu_detection_2016, lee_density-based_2014} and generative network models \cite{zhang_identification_2015, gallagher_clarified_2021}. Some works also explore the coexistence of community and core-periphery structures in a nested way, in particular in the form of communities that exhibit core-periphery structures internally \cite{rombach_core-periphery_2017, yan_multicores-periphery_2019, tunc_unifying_2015, kojaku_finding_2017}.

The diversity in methods for similar node aggregation tasks unsurprisingly results in a diversity in partitions returned by such methods. While most community detection algorithms of the 20th century (from cliques to k-cores through CONCOR \cite{luce_method_1949, seidman_network_1983, breiger_algorithm_1975}) as well as the Girvan-Newman algorithm \cite{girvan_community_2002} popular in the 2000s were deterministic, this challenge has been further exacerbated by the advent of approaches whose results are non-deterministic by nature, such as Louvain \cite{blondel_fast_2008} and SBM \cite{holland_stochastic_1983, karrer_stochastic_2011, peixoto_nonparametric_2017}. In that case, even multiple results yielded by one given community detection algorithm may not clearly indicate a consensus partition, leading to a partition selection problem on top of the above-mentioned issue around model selection.
In existing work, this has been addressed by finding some kind of consensus in a distribution of partitions to identify an ‘optimal’ partition, for example by averaging over results from multiple runs of the same algorithm \citep{lancichinetti_consensus_2012, tandon_fast_2019}. 

Such consensus-seeking methods run into problems, however, when multiple partitions that are close to the optimum are qualitatively different from each other, revealing the need for considering multiple local consensus partitions that may provide different, similarly likely explanations to the network structure at hand \citep{peixoto_revealing_2021}. The issue of multiple locally optimal partitions was also addressed by \citet{peel_ground_2017}, who demonstrated that many real-world networks have multiple plausible (high likelihood) partitions and that different sets of node meta-data may correlate with different aspects of the structure of the network. These recent results suggest that by accommodating for a diversity of ground truths in the generative process, the stochastic nature of some community detection methods is, in fact, not only an issue to deal with but a feature of these methods. In this work, we propose a model that accepts any combination of two partitions, including structures that somewhat resemble nested core-periphery pairs (as in \autoref{sec:simulations}), yet without being limited to these specific structures. We also aim to extend the literature by proposing a model that ensures the consistency with two distinct planted structures and thus allows for the exploration of networks in which connectivity patterns induced by multiple block structures are satisfied (\hbox{i.e.} the coexistence of multiple structures) and the extent to which both are detected and detectable.

\subsection{\label{sec:ambiguity}Community detectability \& ambiguity}
The recent focus on the importance of analysing the variability of partition distributions calls for an exploration of what we will call mesoscale ambiguity. In an effort to identify prototypical network partitions representative of various regions of an entire partition landscape \citet{kirkley_representative_2022} aimed at introducing (and detecting) what they call `ambiguity' on the mesoscale of synthetic networks. By specifying such ambiguity through certain edge probabilities between blocks in an SBM (see \autoref{sec:sbms} for a review of SBMs) they demonstrated the ability of their model to detect a set of representative partitions which identify different aspects of the introduced ambiguity.

This specific example of ambiguity calls for an investigation of the distinction between truly ambiguous block structures on the one hand, and weak or noisy signals which prevent algorithms from correctly detecting mesoscale structures. The stochastically generated SBM ensemble may exhibit some variability in their block structure, but if only one `ground truth' partition is planted, are qualitatively different recovered partitions the result of real ambiguity or merely of detectability issues? And should the detection of partitions which could not have been generated from the planted model (\hbox{i.e.} that lie outside of the distribution of possible networks with the specified parameters) be viewed as a failure of the partitioning algorithm rather than successfully recovered ambiguity?

To distinguish between the correct recovery of some type of ambiguity on the mesoscale and the inability of a community detection algorithm to identify the true partition due to noise that is `blurring' the signal of the block structure, we need to provide some understanding of the detectability phase transitions that have been demonstrated to exist in community detection. Overall, it has been shown that the detectability of block structure in networks depends on the overall density of the network, the difference between the connectivity of the blocks, as well as the number of blocks (see \citep{moore_computer_2017, abbe_community_2017} for extensive reviews). When the structural signal in a network exists but is too weak or too noisy, it becomes impossible for community detection algorithms to identify such structures. At a certain phase transition, algorithms will mistake a network for a random graph when the structural traces of underlying communities are not sufficiently tangible in the actual network. Prior work on the detectability of modules in network has shown - both analytically as well as heuristically - the existence \citep{reichardt__2008} and positions of such phase transitions, notably for spectral community detection methods \citep{decelle_inference_2011} and for methods using Bayesian maximum-likelihood \citep{nadakuditi_graph_2012}. Much of this early work on phase transitions focused on the symmetric case of the traditional (Poisson degree-distributed) SBM. Since then, others have worked on networks with heterogeneous node degree distributions and have argued for the existence of phase transitions in such cases \citep{zhang_spectra_2014} and demonstrated that heterogeneity in networks facilitates the detection of communities in the case of modularity maximisation \citep{radicchi_detectability_2013}. 
While many efforts have gone into the appraisal of phase transitions for community detection for decisive (albeit sometimes noisy) structures, and others have demonstrated that such detectability thresholds do not exist in core-periphery structures \cite{zhang_identification_2015}, little work has focused on the limits of SBMs cases where some level of ambiguity is introduced specifically. Exploring this further seems particularly important, since the application of community detection methods is primarily aimed at real-world networks, which arguably exhibit more `ambiguity' and for which the possible existence of multiple locally optimal solutions has been demonstrated repeatedly (see above). As mentioned previously, empirical networks have also been shown to have different types of mesoscale structures all at once - complicating the issue even further. 

Overall, the question around a clear differentiation between the issues of detectability of certain block structures and `true' ambiguity in the sense of multiple different ground truths appears challenging and is -- to the best of our knowledge -- an open research question, that carries with it the question of how such ambiguity can be described. Owing to the lack of a clear definition, we from now on characterise \emph{ambiguity} as the simultaneous existence of multiple planted partitions in one single network. We denote this simultaneous existence of implicit structures by \emph{coexistence} of structures, where the network's connectivity aligns consistently and concurrently with the connectivity of the said structures.

\subsection{\label{sec:sbms}Stochastic block models}
In this work, we exploit the features of SBMs two-fold. On the one hand, we use SBMs for \textit{generating} synthetic networks with planted mesoscale structure. On the other hand, we explore the issue of ambiguity in mesoscale structure in networks by fitting SBMs to synthetic graphs and thus using SBMs as a way of \textit{detecting} mesoscale structures. Bayesian inference methods, such as SBMs, are especially suited for the exploration of partition distributions due to their stochastic nature.

SBMs originate in mathematical sociology, where they built on the concept of node similarity expressed through equivalent connectivity patterns of \textit{blocks} of nodes, coining the term \textit{blockmodeling} for the grouping of such nodes \citep{lorrain_structural_1971, white_social_1976, everett_regular_1994}. Early strict notions of this type of equivalence were later relaxed in form of \textit{stochastic equivalence}, which holds for nodes that connect to other node sets with the same probability \citep{holland_stochastic_1983}. The latter work also manifested the first appearance of \emph{stochastic block models}, generative models that create networks (or entire distributions over networks) by first dividing nodes into blocks and then placing edges between node pairs with a probability depending solely on the block membership of each node.

The SBM is therefore an extension of the simple random graph model, where constant edge probabilities are specific to block pairs rather than being the same for the entire network. An SBM takes as parameters (a) a block membership vector, with entries indicating the block membership of each node, and (b) a square connectivity matrix of size equal to the number of blocks, whose elements indicate the probability of a connection between the respective blocks (or within a block for the elements on the diagonal). 
Since their first appearance \citep{lorrain_structural_1971, white_social_1976, everett_regular_1994, holland_stochastic_1983}, SBMs have been repurposed repeatedly to function as a baseline model for addressing the community detection issue as an inference problem \cite{hastings_community_2006, karrer_stochastic_2011, peixoto_nonparametric_2017}. It is increasingly popular in theoretical and applied network science research, partly due to its flexibility grounded in a relatively general definition of what it means for nodes to be similar i.e., to belong to a block or a community. The idea is that one can `reverse' the generative process of an SBM for the purpose of block structure detection: statistical inference methods can be used to fit SBMs to network data, to recover the parameters of the model (essentially block memberships) that offer the most likely explanation of the generative processes of a network. 

Famously, edge placement between two nodes in the `traditional' SBM only depends on the nodes' block assignment. It therefore does not resemble the structure of many real networks, which tend to have heterogeneous node degree distributions. One way to model degree heterogeneity is by adding `degree correction' into the SBM, through which edge placements also depends on the respective degree of each node. Using the degree-corrected version as the generative model assumed in the process of detecting block structure considerably improved the ability of SBMs to pick up community structures in real networks \cite{karrer_stochastic_2011}. More SBM extensions have since been developed, including hierarchical \cite{peixoto_hierarchical_2014}, overlapping \cite{airoldi_mixed_2008, peixoto_model_2015} and multilayer \cite{peixoto_inferring_2015} variants, many of which were demonstrated to be an improvement in the goodness of fit for certain types of networks, compared even to the degree-corrected version. Others have exploited the flexibility of SBMs (in terms of the types of mesoscale structures that can be detected) to demonstrate the diversity of core-periphery structures in real networks \cite{gallagher_clarified_2021}. In the existing SBM literature, most work focuses on the recovery of a single partition that is identified by optimising some model selection criterion; however, some recent work has gone beyond the single-partition approach and has emphasised the need to consider the entire partition landscape returned by SBMs \cite{peixoto_revealing_2021, kirkley_representative_2022, peel_ground_2017}. In this work, we intend to contribute to this particular subfield of the SBM literature, by drawing attention to the existence and detectability of more than one planted structure in a single network.

One existing strand of work within the SBM literature that is particularly relevant in the context of planting multiple ground truth structures is the mixed membership SBM (MMSBM) \cite{airoldi_mixed_2008}. This model allows nodes to belong to multiple blocks to varying degrees, expressed by mixed membership vectors assigned to each node, the elements of which denote the probabilities of the node belonging to the different blocks. In an MMSBM, nodes may therefore embody connectivity patterns from more than one block at a time, making it relevant to our case of planting multiple partitions. There is a certain conceptual similarity between the MMSBM and our method, and the MMSBM can be shown to be equivalent to our method in some cases; however, we will see later that an MMSBM-based approach to the issue has significant limitations that can be overcome with our method.

\subsection{\label{sec:generative}Generative benchmark models}
As opposed to real-world networks whose exact generative processes are not known, synthetic networks are a natural choice to plant a specific structure and serve as a benchmark, in particular for appraising the success of a certain method in recovering various mesoscale structures, including communities. Such benchmark frameworks allow for certain mesoscale structures to be `built into' a synthetic network, on which the performance of algorithms can be tested by measuring the extent to which the predefined structure is successfully recovered. In general, one or several parameters may be adjusted to explore the potential limits of an algorithm and to imitate the features of certain types of real networks. One of the earliest such models is the Girvan-Newman (GN) benchmark \cite{girvan_community_2002}, a network of 128 nodes divided into four equally-sized groups and relying on one parameter controlling inter-group connectivity strength through the external (out-community) degree of nodes.

To overcome some of its shortcomings, such as its general inflexibility, small size and unrealistic features, \citet{lancichinetti_community_2009} proposed a benchmark accounting for heterogeneous degree and community size distributions. Other existing benchmarks allow for the specification of the within- and between-group connectivity through the use of SBMs, such as the planted partition model \cite{condon_algorithms_2001} which has been extended to other special cases including multilayer networks \cite{bazzi_framework_2020}. In general, the aim of generative benchmark models is to resemble features of empirical networks and while many of the existing benchmarks account for one or several such features, the ambiguity in mesoscale structures has as yet been neglected. Our contribution is to focus on this particular aspect and to complement single ground truth benchmarks by a framework that generates networks with multiple built-in ground truths. 

\section{\label{sec:framework} Model framework}
We propose a generative network model whose parameters aim to simultaneously respect two partitions: edges are placed between node pairs in a way such that the resulting network exhibits a block structure that takes into account each of these two partitions at the same time. We discuss later how this framework can be extended to more complex structures with more than two partitions but we focus on the simple two-partition case in the majority of this work.

The difficulty in generating a network that exhibits two coexisting structures primarily lies in connecting node pairs in a way that is consistent with the connectivity patterns of \emph{both} planted structures. The probability of placing an edge between each node pair must depend on the block memberships of each node in each of the planted structures. We thus aim to design a generative process that specifies the appropriate probabilities for the desired connectivity patterns. An obvious choice would be a constrained MMSBM, in which nodes are members of two blocks with equal probability. It turns out that finding the appropriate normalisation constants to make the additive edge probabilities of the MMSBM consistent with the two planted structures requires extra calculations that we do not need if we implement a simpler single-membership SBM approach that considers the overlaps of the blocks as simple blocks. In the following, we first lay out the outline of the method, explain the limitations of the MMSBM-based approach, and finally introduce the \textit{stochastic cross-block model} (SCBM) as an alternative approach.

\begin{figure}
\centering
\begin{subfigure}{0.23\textwidth}
    \centering
    \includegraphics[width=1\textwidth]{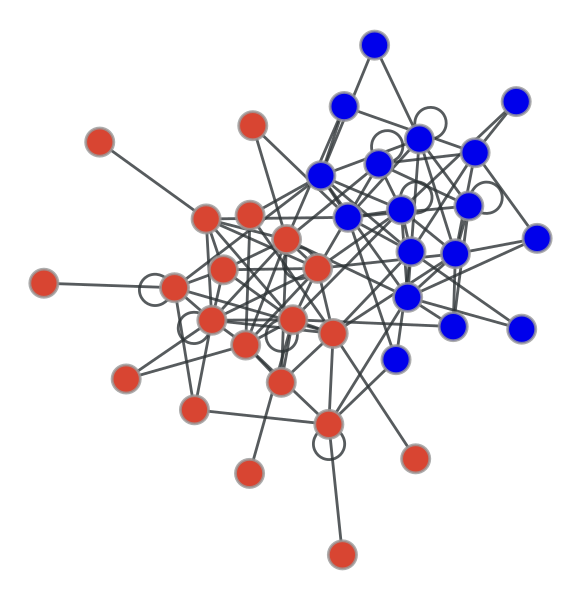}
    \caption{Partition 1 (bi-communities)}
    \label{fig:schem_p1}
\end{subfigure}
\begin{subfigure}{0.23\textwidth}
    \centering
    \includegraphics[width=1\textwidth]{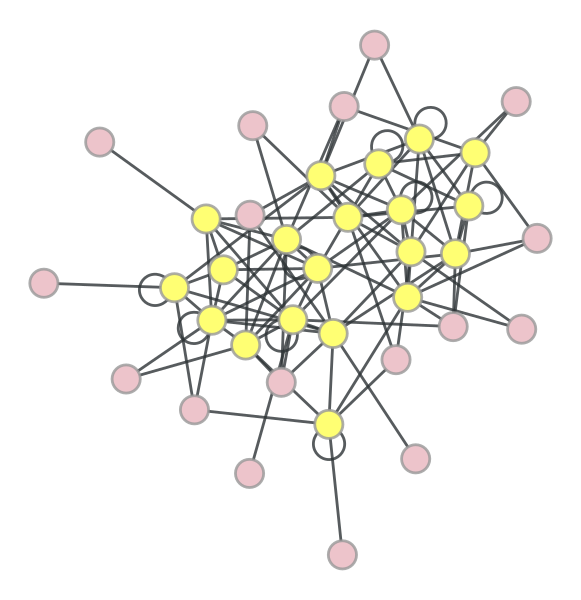}
    \caption{Partition 2 (core-periphery)}
    \label{fig:schem_p2}
\end{subfigure}
\\
\begin{subfigure}{0.23\textwidth}
    \centering
    \includegraphics[width=1\textwidth]{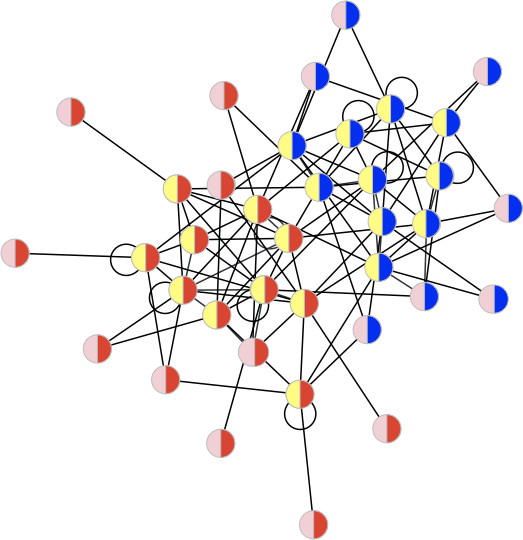}
    \caption{MMSBM}
    \label{fig:schem_mmsbm}
\end{subfigure}
\begin{subfigure}{0.23\textwidth}
    \centering
    \includegraphics[width=1\textwidth]{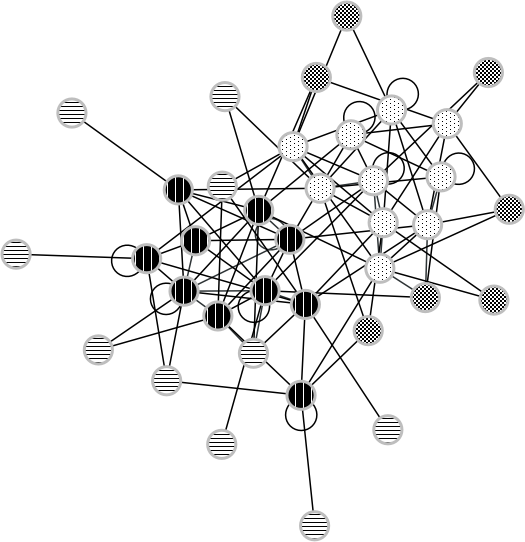}
    \caption{SCBM}
    \label{fig:schem_crosspartition}
\end{subfigure}
\caption{An example graph with nodes coloured according to their block memberships following different ways of formulating the partitions: The planted bi-community partition (\protect\subref{fig:schem_p1}), the planted core-periphery partition (\protect\subref{fig:schem_p2}), a visualisation of the MMSBM formulation of the two-partition-problem, where nodes belong to two blocks with equal probability (\protect\subref{fig:schem_mmsbm}) and the SCBM approach (\protect\subref{fig:schem_crosspartition}).}
\label{fig:schem}
\end{figure}

As stated above, we are generally guided by the KC example where both a bi-community (a partition of the network into two assortative communities) and a core-periphery structure may be found and where it is likely that the two structures jointly explain the network formation. \autoref{fig:schem_p1} and \autoref{fig:schem_p2} show the same example graph with nodes coloured according to a bi-community and core-periphery partition respectively. 

\subsection{Formal outline} We generate a network with $N$ nodes and $E$ edges, into which we plant a set of $P$ partitions with $K_p$ blocks in partition $p$. Edges between nodes are described by the adjacency matrix $\mathbf{Y}$ of size $N \times N$. We focus on the undirected, unweighted case, whereby $Y_{ij} = Y_{ji} = 1$ if $i$ and $j$ are connected by an edge and $0$ otherwise. The block structure in each partition $p$ is defined as an SBM with two parameters: (a) the set of block membership vectors $\mathbf{b}^p_{i}$ of length $K_p$ assigned to each node $i$ and (b) the block matrix $\boldsymbol{\theta}_p$ of size $K_p \times K_p$, where diagonal (\hbox{resp.} off-diagonal) elements indicate the probability of an edge within (\hbox{resp.} between) blocks. Note that each $\mathbf{b}^p_{i}$ is a one-hot vector, which is a binary vector with only one element set to $1$ (indicating the block membership of node $i$ in partition $p$) and all others set to $0$. We denote by $\mathbf{B}_p$ the $K_p \times K_p$ matrix of expected edge counts, which -- for graphs with self-loops -- has elements $\theta_{p_{rs}}n_{p_r}n_{p_s}$, where $n_{p_r}$ is the number of nodes in block $r$ in partition $p$. Note that block matrices are symmetric since the graphs we are generating are undirected and that for $r=s$ the elements of $\mathbf{B}_p$ denote twice the number of expected edge counts, for convenience of calculation and notation.

For illustrative purposes, we focus on the simple case of $P=2$ partitions and $K_1 = K_2 = 2$ blocks in each partition and we define the block matrices for the two implicitly planted partitions by $\boldsymbol{\theta}_1 = \{\theta_{1_{rs}}\}$ and $\boldsymbol{\theta}_2 = \{\theta_{2_{rs}}\}$. In \autoref{sec:morepartitions}, we outline how our framework can be extended to more complex partition combinations.

\subsection{MMSBM formulation}
Given the parameters needed to plant two different partitions, we need to define the generative process which yields a network that satisfies the connectivity patterns of both. One possible approach is to formulate the two-partition scenario as a special case of an MMSBM with appropriate normalisation. In the original MMSBM formulation \cite{airoldi_mixed_2008}, each node belongs to all latent groups with certain probability expressed through a mixed membership vector specific to each node. The existence of an edge between two nodes depends on the block memberships of the two nodes, which is repeatedly drawn from the mixed membership vector for each node-pairing; nodes thus inherit connectivity patterns from multiple blocks and the expected density at the `overlap' of multiple blocks is a weighted average of their individual densities \cite{peixoto_model_2015}. In order to frame our two-ground-truth partition problem as an MMSBM, we consider the blocks in the two planted partitions as four latent blocks but we constrain the generative process in a way that forces certain overlaps to be empty. In particular, we need to (a) constrain the mixed membership vectors so that the probability of nodes being members of certain combinations of blocks is zero, and (b) specify the within- and between-block edge probabilities in a way such that the connectivity of the generated network is consistent with that of the two planted partitions given by $\boldsymbol{\theta}_1$ and $\boldsymbol{\theta}_2$. The schematic in \autoref{fig:schem_mmsbm} visualises these constraints by showing nodes coloured according to their block memberships in two planted partitions.

It turns out that the required normalisation of the additive block probabilities induced by the MMSBM is not straightforward. In particular, one needs to determine a normalisation constant for each combination of block pairs, for which one needs to obtain the solution to an underdetermined system of equations. As we detail in \aref{app:mmsbmformulation} the minimum norm solution to this system (that can be obtained with a least squares solver) is negative for certain combinations of planted structures. This means different methods for approximate solutions are needed for different planted structures, which is likely to have unexpected side effects to an exploration of planted structures.

In order to be able to have enough flexibility for an exploration of a sufficiently large range of structure combinations, we thus propose an alternative to the MMSBM-based approach: it turns out that we can circumvent the extra step involved in finding the normalisation constants by formulating the two-partition problem as a single membership SBM and by replacing the additive probabilities imposed by the MMSBM by multiplicative ones. Instead of generating the network through the mixed membership of nodes in two blocks, our proposed model is a simplification of the problem which considers the overlaps between the blocks of the individual planted partitions as the blocks of an SBM.

\subsection{Stochastic cross-block model}
In the stochastic cross-block model (SCBM), a node is assigned a combination of the blocks it inhibits in multiple partitions, called a \textit{cross-block}. We denote the set of cross-blocks by $\boldsymbol{\Gamma} = \{(r, r')\}$, where $r$ and $r'$ are the blocks in partition $1$ and $2$ respectively. We can then rephrase our problem as follows: To plant two partitions in one single network, we generate a network in which we explicitly plant one single \textit{cross-partition} with $K_{\text{SCBM}}=K_{1}K_{2}$ cross-blocks in a way that is consistent with the expected densities from the block matrices $\boldsymbol{\theta}_1$ and $\boldsymbol{\theta}_2$. \autoref{fig:schem_crosspartition} illustrates the four cross-blocks in the cross-partition resulting from the partitions visualised in \autoref{fig:schem_p1} and \autoref{fig:schem_p2}. Note that the cross-partition -- the explicit division of the network into $K_{\text{SCBM}}$ cross-blocks -- is different from the \textit{partition coexistence} which refers to the property of the network to be consistent with the connectivity of the two implicitly planted structures with $K_{1}$ and $K_{2}$ blocks respectively.

Our generative network then simply becomes a standard SBM where the probability of an edge between two nodes $i$ and $j$ is determined entirely by the probability of a node between the cross-block $u$ of $i$ and the cross-block $v$ of node $j$. To generate the final network, we straightforwardly create the one-hot cross-block membership vectors $\mathbf{b}^{\text{SCBM}}_i$ for each node $i$ from vectors $\mathbf{b}^{p}_i$. For the placement of edges, we need to determine the connectivity between and within the cross-blocks by defining $\boldsymbol{\theta}_{\text{SCBM}}$ and $\mathbf{B}_{\text{SCBM}}$. As above, element $\theta_{uv}$ denotes the probability of an edge between cross-blocks $u$ and $v$ and $B_{uv}$ denotes the average expected number of such edges, where we have dropped the subscript. The block matrix is created in a way in which the edge probabilities are consistent with the elements of the block matrices $\boldsymbol{\theta}_p$ for each planted partition $p$. The deciding difference between this cross-partition approach and an MMSBM formulation of our problem is that we replace the additive probabilities that result from the MMSBM by normalised multiplicative probabilities. In other words, for each cross-block, the probability of an edge is calculated by multiplying the edge probabilities in the original blocks that cause the overlap and by normalising appropriately.

\begin{figure}
\centering
\begin{subfigure}{0.20\textwidth}
    \centering
    \includegraphics[width=1\textwidth]{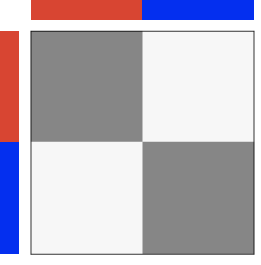}
    \caption{$\boldsymbol{\theta}_1$}
    \label{fig:schem_matrix_p1}
\end{subfigure}
\hspace{1.5em}%
\begin{subfigure}{0.20\textwidth}
    \centering
    \includegraphics[width=1\textwidth]{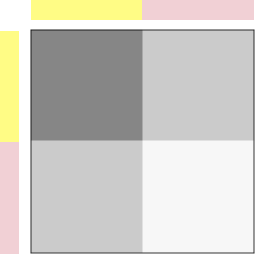}
    \caption{$\boldsymbol{\theta}_2$}
    \label{fig:schem_matrix_p2}
\end{subfigure}

\vspace{1.1em}%
\begin{subfigure}{0.20\textwidth}
    \centering
    \includegraphics[width=1\textwidth]{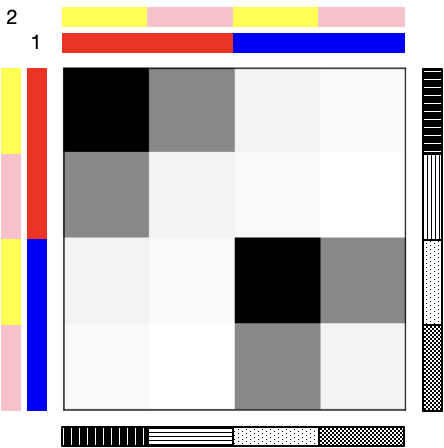}
    \caption{$\boldsymbol{\theta}_{\text{SCBM}}$}
    \label{fig:schem_matrix_crosspartition}
\end{subfigure}
\hspace{1.5em}%
\begin{subfigure}{0.20\textwidth}
    \centering
    \includegraphics[width=1\textwidth]{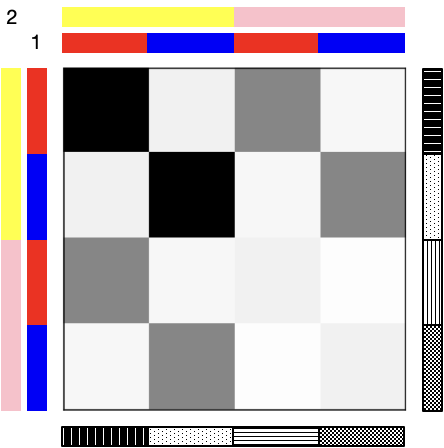}
    \caption{$\boldsymbol{\theta}_{\text{SCBM}}$}
    \label{fig:schem_matrix_crosspartition_rev}
\end{subfigure}
\caption{Block matrix visualisations for two planted partitions (\protect\subref{fig:schem_matrix_p1} and \protect\subref{fig:schem_matrix_p2}) and resulting cross-partition (\protect\subref{fig:schem_matrix_crosspartition} and \protect\subref{fig:schem_matrix_crosspartition_rev}). Note that \protect\subref{fig:schem_matrix_crosspartition} and \protect\subref{fig:schem_matrix_crosspartition_rev} are equivalent but ordered differently, as visualised by the coloured bars (blocks in partitions $1$ and $2$) and patterned bars (cross-blocks). Matrix elements are coloured according to the expected edge densities in the block matrices, with black elements representing the maximum edge probability and white elements representing an edge probability of $0$; note that the colours are consistent so that they are comparable across the three graphs.}
\label{fig:schem_matrices}
\end{figure}

In \autoref{fig:schem_matrices}, we visualise two example block matrices $\boldsymbol{\theta}_1$ and $\boldsymbol{\theta}_2$ and the resulting cross-block matrix $\boldsymbol{\theta}_{\text{SCBM}}$; note that we include two differently ordered visualisations of the same cross-block matrix in \autoref{fig:schem_matrix_crosspartition} and \autoref{fig:schem_matrix_crosspartition_rev}, to emphasise that the cross-partition is consistent with both $\boldsymbol{\theta}_1$ and $\boldsymbol{\theta}_2$. The particular choice of $\boldsymbol{\theta}_1$ and $\boldsymbol{\theta}_2$ visualised here is also responsible for the partitions visualised in \autoref{fig:schem_p1} and \autoref{fig:schem_p2} and the cross-block matrices thus correspond with the graph visualised in \autoref{fig:schem_crosspartition}.

\paragraph*{\label{sec:connectivitymat} Equal block sizes}
In the first instance, we focus on the case of equal block sizes, both in the planted partitions as well as the cross-partition, so that $n_r=n$ for all blocks $r$ and $2n=N$. We allow for self-loops, so we have $B_{p_{rs}}=n_rn_s\theta_{p_{rs}}=n^2\theta_{p_{rs}}$ for $p \in \{1,2\}$. We denote the vector of cross-block sizes by $\boldsymbol{\nu}=\{\nu_u\}$ and constrain cross-blocks to be equally-sized: $\nu_u = \nu$ for $u \in \boldsymbol{\Gamma}$ and thus $2\nu=n$. We construct the edge probabilities within and between cross-blocks through the product of the probabilities of the blocks in the two partitions, with an appropriate normalisation. In particular, we define the block matrix of the SCBM as $\boldsymbol{\theta}_{\text{SCBM}} = \{\theta_{uv}\} = x\{\theta_{1_{rs}}\theta_{2_{r's'}}\}$ with cross-blocks $u=(r, r') \in \boldsymbol{\Gamma}$ and $v=(s, s') \in \boldsymbol{\Gamma}$, where the constant $x$ needs to be chosen in a way such that the expected number of edges within and between certain cross-block pairs in $\mathbf{B}_{\text{SCBM}}$ adds up to the respective elements of $\mathbf{B}_1$ and $\mathbf{B}_2$, so that the connectivity in the generated network is consistent with that of partitions 1 and 2. In the case of equal block sizes, this is satisfied for $x=1/\rho$ where $\rho=2E/N^2$ is the overall expected edge density of a network with self-loops (see \aref{app:connectivity} for the derivation). The elements of the expected edge count matrix are denoted by the matrix elements $B_{uv}=\nu^2\theta_{uv}$, where we have dropped the SCBM subscripts in $B$ and $\theta$.

\paragraph*{\label{sec:generalcase}Varying block sizes}
In any other more general case in terms of varying block sizes, a constant $x$ satisfying both planted partitions only exists if we introduce other constraints, such as equal edge probabilities across all blocks of the two planted partitions, \hbox{i.e.} planting a random graph (see \aref{app:varyingsizes}). To determine $\boldsymbol{\theta}_{\text{SCBM}}$ for non-trivial partitions with varying block sizes, we need to find a set of normalisation constants $\{x_{uv}\}$, one for each cross-block pairs, so that $\boldsymbol{\theta}_{\text{SCBM}} = \{\theta_{uv}\} = \{x_{uv}\theta_{1_{rs}}\theta_{2_{r's'}}\}$. To find $\{x_{uv}\}$ in a way in which $\boldsymbol{\theta}_{\text{SCBM}}$ generates a network consistent with $\boldsymbol{\theta}_1$ and $\boldsymbol{\theta}_2$ we must, again, solve an underdetermined system of equations for which we can compute a minimum norm solution, which is unique and always exists if any solution to the system exists (see \aref{app:solving}). In this case, the minimum norm solution returns a set of entirely positive $\{x_{uv}\}$ across our entire parameter space in \autoref{sec:simulations}; while we explore the equal-block case for our simulations, we have shown here that a more general set-up is also possible using the SCBM approach.

\paragraph*{\label{sec:morepartitions}Multiple blocks and partitions}

In our simulations in \autoref{sec:simulations}, we focus on the simple two-partition case described above to demonstrate detectability issues of two `ground truth' partitions. However, it may in some cases be interesting to explore more complex structures with more coexisting partitions and/or more blocks. While a thorough exploration of such cases is beyond the scope of this work, we briefly outline an extension of the described simple case. In fact, keeping the number of partitions at $P=2$ but planting more than two blocks in one or both partitions can be done straightforwardly with the same normalisation constant in the case of equal block sizes and with a larger system of equations having to be solved in the case of varying block sizes. Since planting more than two partitions can also be re-framed as recursively planting sets of two partitions until a final cross-partition is reached, this is also possible. In the case of equal block sizes, the SCBM block matrix can then be generalised as $\boldsymbol{\theta}_{\text{SCBM}} = 1/\rho^{p-1}\{\prod_p \theta_{p_{rs}}\}$. In the case of varying block sizes, one needs to consider the possible limitation of scalability that arises for large numbers of partitions and/or blocks, since the number of cross-blocks is $K = \prod_p K_p$ and we need to find $K(K+1)/2$ normalisation constants. The least square method for finding the minimum norm solution involves computing the pseudo-inverse of the coefficient matrix, which can be computationally expensive for large $n \times m$ matrices (with a computational complexity of approximately $O(n^2m)$). Limitations of computational time and memory thus need to be considered for very complex combinations of planted partitions.

\subsection{\label{sec:modelvariants}Generative SBM}
We generate the final network from the connectivity matrices according to the `traditional' SBM \citep{holland_stochastic_1983}, which uses matrix $\boldsymbol{\theta}_{\text{SCBM}}$ alongside cross-block membership vectors $\mathbf{b}$ to determine whether or not an edge exists between two nodes. More specifically, we will place an edge between each pair of nodes $(i,j)$ independently at random, with probability $\theta_{uv}$, where $u, v \in \boldsymbol{\Gamma}$ are the cross-blocks of $i$ and $j$ respectively. We thus sample the value of the interaction between $i$ and $j$ with ${Y_{ij} \sim \textrm{Bernoulli}(\mathbf{\mathbf{b}_i}^T \boldsymbol{\theta}_{\text{SCBM}} \mathbf{b_j})}$. In this version of the SBM, the expected edge counts in $\mathbf{B}_{\text{SCBM}}$ are satisfied on average. 

We also consider the microcanonical SBM \citep{peixoto_nonparametric_2017}, in which the (rounded) elements of the given matrix $\mathbf{B}_{\text{SCBM}}$ are satisfied exactly (rather than on average) and which is based on the configuration model \citep{fosdick_configuring_2018}. Specifically, we consider the degree-corrected extension of this microcanonical SBM, in which the probability of an edge being placed between two nodes does not depend solely on the elements of a connectivity matrix but also on a given degree sequence or distribution. This SBM variant has been demonstrated to have characteristics that more closely resemble empirical networks, by producing synthetic networks with the type of within-block degree variability that is more likely to occur in real networks \citep{karrer_stochastic_2011}. Given a degree sequence $\{k_i\}$ in which $k_i$ denotes the degree of node $i$, this works by assigning $k_i$ half-edges to node $i$ and then choosing two half-edges in the network at random (allowing for self-edges) and connecting them under the condition that the expected given within- and between-block edge counts are satisfied. However, the elements of $\mathbf{B}_{\text{SCBM}}$ can be real numbers and must therefore be rounded in the network generation process. This introduces small differences between the (implicitly planted) expected edge count matrices $\mathbf{B}_p$ and the generated networks in terms of the total number of edges as well as the within- and between- cross-block edge counts. It also means that a given degree sequence can only be satisfied exactly if $\mathbf{B}_{\text{SCBM}} \in \mathbb{Z}^{N \times N}$; in our simulations we sample node degrees from a power law distribution rather than satisfying the exact degree sequence. Note that in the graphs we generate in this way in our simulations below, any pair of nodes is connected by a maximum of one unweighted edge; removing this constraint and producing multigraphs instead is straightforward. 

The version of our model which generates networks according to the traditional SBM will from now on be called the \textit{canonical model} to distinguish it from the latter version, which we will call the \textit{microcanonical model}. This is to avoid confusion in the notation between the SBM variants we use to \textit{generate} our networks from those we use to infer partitions.

\section{\label{sec:simulations} Simulations}
We now explore the extent to which two built-in ground truths are recovered by SBMs, by generating a set of networks in which we implicitly plant two partitions. Clearly, there are many interesting two-partition structures one may explore; as indicated before, we are interested in the type of structure present in our motivating example, the KC network. For this network, samples of the posterior distribution of inferred partitions yield a number of plausible explanations of the mesoscale structure \cite{peixoto_revealing_2021}; one local consensus partition is the famous two-faction division of the network into two assortative communities, another is a leader-follower partition that resembles a core-periphery structure. We are interested in the recovery limits of these two types of structures in networks and we therefore plant similar structures into an ensemble of synthetic networks according to our generative framework.
In our simulations, we build both a bi-community as well as core-periphery structure into a set of graphs and we fit two different SBM variants to our networks to infer the posterior distribution of partitions for each of them. We finally calculate the similarity between the recovered partitions and the planted partitions and present the results for the partitions planted by each model variant and recovered by each SBM variant.

\subsection{\label{sec:parameters}Parameters}
We focus on the case of equal block sizes here, for which the multiplicative probabilities in $\boldsymbol{\theta}_{\textbf{SCBM}}$ can be normalised simply by the constant $x=1/\rho$ and we do not have to rely on the minimum norm solution (see \autoref{sec:connectivitymat}). We plant networks with $N=400$ and we run three sets of simulations with varying average degree $c = 5, 10, 20$. The total expected edge density is therefore different between the three sets of simulations but held constant within each set. This is to ensure that any differences we are seeing in the detectability of the two planted partitions are due to a different distribution of the edges within the network (which we will induce by varying the block connectivities), rather than differences in total edge density. Nodes are chosen uniformly at random and assigned to each block to create block membership vectors $\mathbf{b}_1$ and $\mathbf{b}_2$. 
To plant the two partitions, we define the symmetric block matrices $\boldsymbol{\theta}_1$ and $\boldsymbol{\theta}_2$ in \autoref{eq:ac1} and \autoref{eq:cp1} where $\beta=2E/n^2$ with $E$ being the total number of edges in the network. Partition 1 ($\boldsymbol{\theta}_1$) configures the bi-community structure whereby a parameter $\mu$ controls the expected intra- \hbox{vs.} inter-block connectivity strength for two planted equal-sized communities. Partition 2 ($\boldsymbol{\theta}_2$) configures the core-periphery (CP) partition, in which the expected edge density within the core and the expected edge density among peripheral nodes is controlled by parameter $\lambda$. Note that the edge probability between blocks in partition $2$ is fixed in this way so that for $\lambda \in [0, 0.5)$, we always have $\theta_{2_{11}} > \theta_{2_{12}} > \theta_{2_{22}}$, in line with common definitions of core-periphery structure, and to have equal probabilities within- and between blocks for both $\mu=0.5$ and $\lambda=0.5$.

\begin{subequations}
\begin{empheq}[left=\empheqlbrace]{align}
\label{eq:ac1}
& \boldsymbol{\theta}_1= \beta \begin{pmatrix} 1 - \mu & \mu \\ \mu & 1 - \mu\end{pmatrix} \\
\label{eq:cp1}
& \boldsymbol{\theta}_2= \beta \begin{pmatrix} 1 - \lambda & \frac{1}{2} \\ \frac{1}{2} & \lambda \end{pmatrix}
\end{empheq}
\end{subequations}

We use our model to generate multiple sets of networks, sweeping parameters $\mu$ and $\lambda$ from $0.01$ to $0.5$ at increments of $0.01$ in each case. Note that we exclude $\mu=0$ as it would yield a disconnected graph of two components and exclude $\lambda=0$ for symmetry in the two dimensions. 

\begin{figure}
\centering
\begin{subfigure}{0.08\textwidth}
    \centering
    \includegraphics[width=1\textwidth]{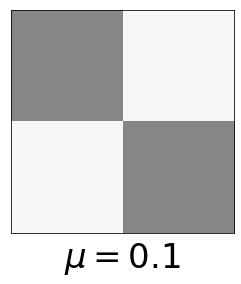}
    \caption{$\boldsymbol{\theta}_1$}
    \label{fig:schem_sweep_fixed_mu}
\end{subfigure}
\begin{subfigure}{0.39\textwidth}
    \centering
    \includegraphics[width=1\textwidth]{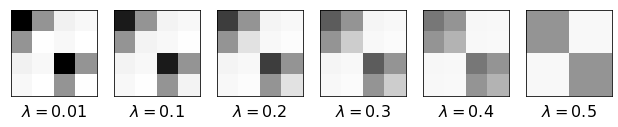}
    \caption{$\boldsymbol{\theta}_{\text{SCBM}}$, varying $\lambda$ in $\boldsymbol{\theta}_2$}
    \label{fig:schem_sweep_varying_lambda}
\end{subfigure}
\\
\begin{subfigure}{0.08\textwidth}
    \centering
    \includegraphics[width=1\textwidth]{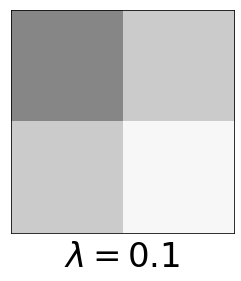}
    \caption{$\boldsymbol{\theta}_2$}
    \label{fig:schem_sweep_fixed_lambda}
\end{subfigure}
\begin{subfigure}{0.39\textwidth}
    \centering
    \includegraphics[width=1\textwidth]{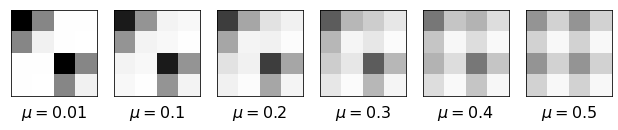}
    \caption{$\boldsymbol{\theta}_{\text{SCBM}}$, varying $\mu$ in $\boldsymbol{\theta}_1$}
    \label{fig:schem_sweep_varying_mu}
\end{subfigure}
\caption{Schematic block matrix $\boldsymbol{\theta}_\text{SCBM}$ and, $\boldsymbol{\theta}_1$ when fixing parameter $\mu$ and increasing $\lambda$ in \protect\subref{fig:schem_sweep_fixed_mu} and \protect\subref{fig:schem_sweep_varying_lambda} respectively, or $\boldsymbol{\theta}_2$ when, similarly, fixing $\lambda$ and varying $\mu$ in \protect\subref{fig:schem_sweep_fixed_lambda} and \protect\subref{fig:schem_sweep_varying_mu}}.
\label{fig:schem_sweep}
\end{figure}

Low values of $\mu$ generate assortative community structure (in the sense that most edges are placed within blocks and few between blocks), while $\mu$ close to $0.5$ produce a network close to a random graph. Values of $\lambda$ close to zero generate `clear' core-periphery structure, with most edges being placed within the core and few among peripheral nodes, while $\lambda=0.5$ produces a random graph. In \autoref{fig:schem_sweep} we show the behaviour of $\boldsymbol{\theta}_{\text{SCBM}}$, for fixed $\mu$ and varying $\lambda$ and vice versa. The block matrices in \autoref{fig:schem_sweep_varying_lambda} and \autoref{fig:schem_sweep_varying_mu} illustrate that the cross-partitions for low values of both parameters resemble a nested structure of two communities with internal core-periphery structures, similar to existing work on core-periphery pairs in networks \cite{rombach_core-periphery_2017, yan_multicores-periphery_2019, tunc_unifying_2015, kojaku_finding_2017}. For increasing $\lambda$ while fixing $\mu$, the block densities of the bi-community partition remain constant and the core-periphery structure becomes weaker until, at $\lambda=0.5$, we are left with a bi-community partition as $\boldsymbol{\theta}_2$ now defines a random graph. Similarly, for fixing $\lambda$ and increasing $\mu$ to $\mu=0.5$, we finally reach a core-periphery structure (by reordering the rows and columns of $\boldsymbol{\theta}_{\text{SCBM}}$ in \autoref{fig:schem_sweep_varying_mu}). 

Note that what we are explicitly generating with the SCBM is a simple SBM with the cross-partition induced by the cross-block matrix $\boldsymbol{\theta}_{\text{SCBM}}$ - the connectivity patterns of $\boldsymbol{\theta}_1$ and $\boldsymbol{\theta}_2$ are explicitly satisfied. When we infer the most likely partitions from the generated matrices, we expect that both the explicitly planted cross-partition as well as the two implicitly planted partitions $1$ and $2$ are recovered to some extent, potentially varying across the $(\lambda, \mu)$ space.

\subsection{\label{sec:generatedgraphs} Generated graphs}
To appraise differences in degree distributions, we use the canonical model in one set of simulations and the microcanonical model in another. We thus generate two sets of networks for each expected degree $c$. In the canonical model, node degrees follow a Poisson degree distribution and edge counts within and between blocks are satisfied on average. In the microcanonical case, as described in \autoref{sec:modelvariants}, edge counts do not fluctuate across different runs of the model and we impose further constraints on the node degrees, which we sample from a power law distribution with exponent $\gamma=3$. We use a soft constraint, in the sense that the final network does not have to match the given degree sequence exactly, but only on average. See \aref{app:generated} for a summary of the small deviations of the edge counts in the generated graphs from the planted edge count matrices due to rounding errors.
In both cases, we generate $8$ networks for each $(\lambda, \mu)$-pair, to account for possible fluctuations in the generative process.

\begin{figure}
\centering
\begin{subfigure}{0.5\textwidth}
    \begin{center}
    \includegraphics[width=1\linewidth]{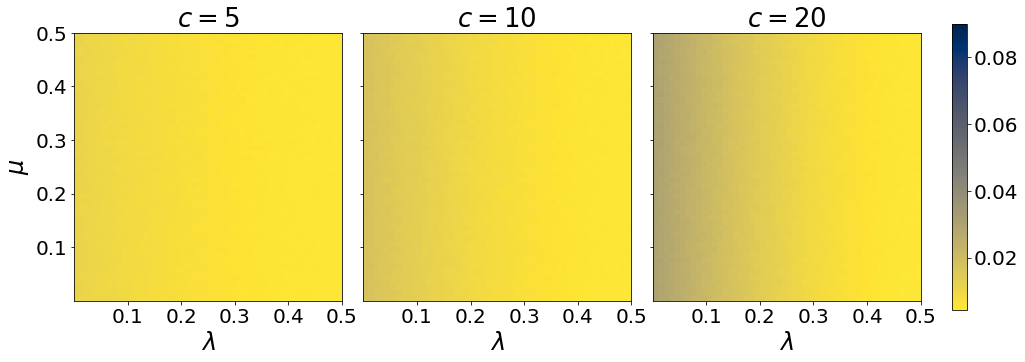}
    \caption{Canonical model.}
    \label{fig:sfig_block_differences_norm_deg_trad}
    \end{center}
\end{subfigure}\\
\begin{subfigure}{0.5\textwidth}
    \centering
    \includegraphics[width=1\linewidth]{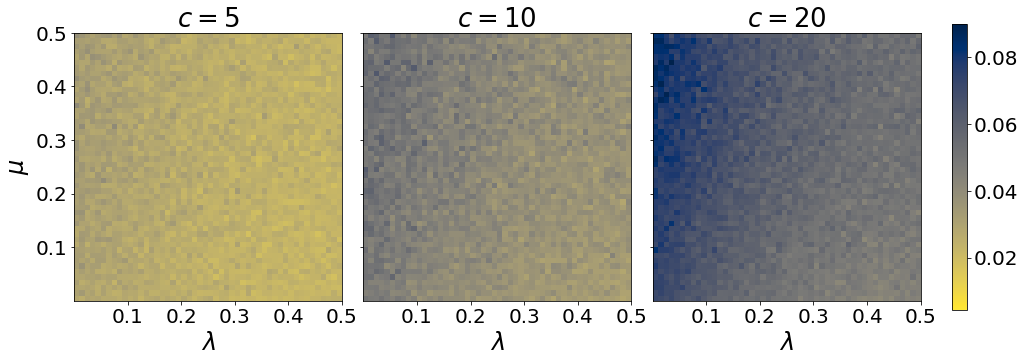}
    \caption{Microcanonical model.}
    \label{fig:sfig_block_differences_norm_deg_mc}
\end{subfigure}
\caption{Mean normalised degree variance for graphs with varying expected degrees $c$.}
\label{fig:sfig_block_differences_norm_deg}
\end{figure}

Before attempting to recover planted partitions in the two sets of graphs, we explore structural characteristics introduced into the networks for different $(\lambda, \mu)$-pairs and through the two different generative processes of the canonical and microcanonical model. 
\autoref{fig:sfig_block_differences_norm_deg} demonstrates that, unsurprisingly, degree variance is highest across the entire $(\lambda, \mu)$ space for graphs generated by the microcanonical model, since we sample a heterogeneous powerlaw degree distribution. It is considerably lower for graphs generated by the canonical version. Notwithstanding, using the canonical model, higher degree heterogeneity is introduced into networks for lower values of $\lambda$, for which we are imposing a strong core-periphery structure (see \autoref{fig:sfig_norm_deg_trad} in the appendix for a rescaled version of the top row of \autoref{fig:sfig_block_differences_norm_deg}).

\begin{figure}
\centering
\begin{subfigure}{0.5\textwidth}
    \centering
    \includegraphics[width=1\linewidth]{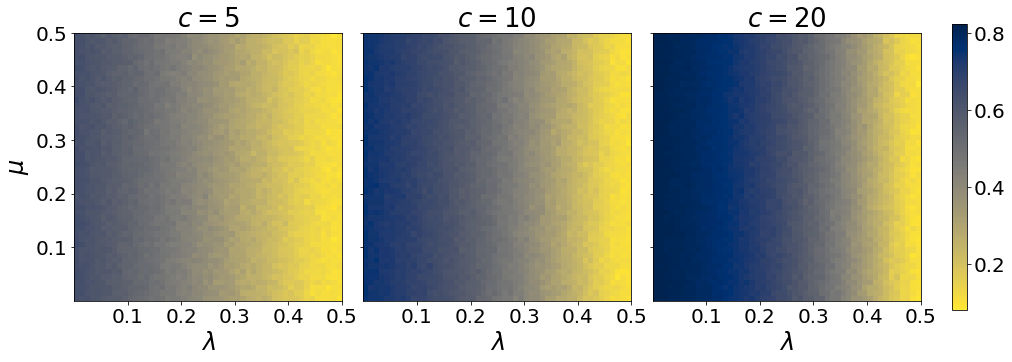}
    \caption{Canonical model.}
    \label{fig:sfig_block_differences_js_dist_trad}
\end{subfigure}\\
\begin{subfigure}{0.5\textwidth}
    \centering
    \includegraphics[width=1\linewidth]{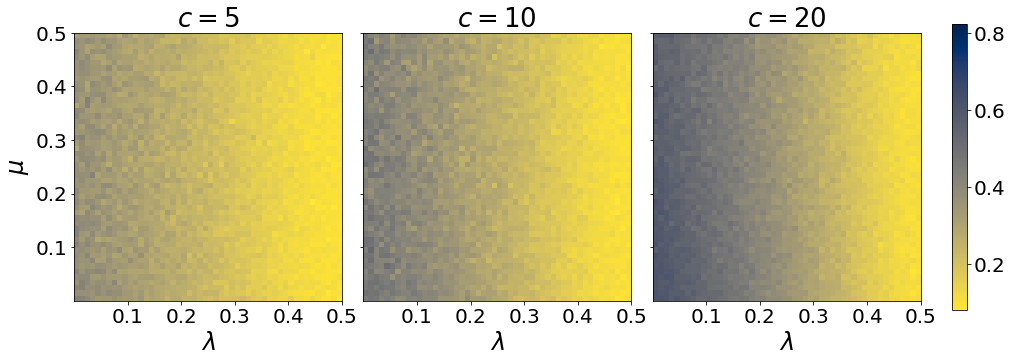}
    \caption{Microcanonical model.}
    \label{fig:sfig_block_differences_js_dist_mc}
\end{subfigure}
\caption{Mean Jensen-Shannon distance between core and periphery node degree distributions for graphs with varying expected degrees $c$.}
\label{fig:sfig_block_differences_js_dist}
\end{figure}

While graphs produced by the microcanonical model exhibit the highest degree heterogeneity overall, \autoref{fig:sfig_block_differences_js_dist} illustrates that their core and periphery blocks are more similar in terms of node degree distributions than in the canonical case, measured by the Jensen-Shannon distance \cite{lin_divergence_1991} between the degree distributions of the core nodes and those of the peripheral nodes. In the canonical case, node degrees are Poisson distributed for the network as a whole, as well as for nodes within each block. However, to accommodate for the core-periphery structure, the mean of the distribution is lower in the periphery than in the core, which in the Poisson case leads to a relatively small overlap between the two distributions \cite{zhang_identification_2015}. In the microcanonical case, we introduce degree heterogeneity through the degree distribution so that planting CP structures does not produce the same differences in the block degree distributions. 

This means that were we to simply assign nodes with above average degree to the core and those with below average degree to the periphery in graphs generated by the canonical model we would retrieve more correctly assigned nodes than in networks produced by the microcanonical model \cite{zhang_identification_2015}, where by `correctly' we refer to the CP block planted by $\mathbf{B}_2$ as defined in \eqref{eq:cp1}. The heterogeneity in degree distributions in the microcanonical case may generate other types of core-periphery patterns than the planted ones. (See \aref{app:generated} for a comparison of the number of correctly classified nodes in this way for the two models and of the degree distributions of the core and the periphery nodes for two example graphs.) 

It is thus worth elaborating on the potential consequences of specifying the degree distribution as an extra parameter in the generative process in particular, additionally to the planted connectivity matrix and block assignments of nodes. While we expect the graphs produced by the canonical model to exhibit structure closely related to what we explicitly plant (\hbox{i.e.} the block connectivity matrices), we may implicitly be introducing additional structure through the constraints imposed on node degrees in the microcanonical case. Heterogeneity in degree distributions, for example, may lead to groups of nodes that display similarities in their connectivity with the rest of the network in terms of their number of connections; it is possible that the dividing lines between these groups do not correspond with those imposed by our planted block structure, which may lead to structures other than those explicitly planted being picked up. These differences in degree heterogeneity introduced through the generative process are thus likely to have an impact on the extent to which SBM variants recover the (coexistence of the) planted partitions in different regions of the $(\lambda, \mu)$ plane, as is confirmed in \autoref{sec:results}.

\subsection{\label{sec:similaritymeasure}Similarity measure}
To quantify the similarity between planted and recovered partitions, we calculate the maximum partition overlap $\omega(p,q)$, namely the proportion of nodes in one partition $p$ assigned to the same, assumed correct, block of the other partition $q$ \citep{peixoto_revealing_2021}. This is calculated by finding the bijection $p' = \zeta(q)$ of the group labels of $q$, so that the number of nodes that have the same block label in $p'$ and $p$ is maximised, so that we have
\begin{equation}
    \omega(p,q) = \frac{1}{N} \max_{\zeta}\ \sum_i \delta_{p_{i},\zeta(q_{i})}.
\end{equation} 
Specifically, this is done by solving the maximum weighted bipartite matching problem for two partitions using a function from the graph-tool python library \citep{peixoto_graph-tool_2014}, which is based on the Kuhn-Munkres \citep{kuhn_hungarian_1955, munkres_algorithms_1957} algorithm. Note that here we use the normalised partition overlap, which is between $0$ and $1$. Therefore, $\omega=1$ when all nodes of two partitions coincide. Note that if both of the compared partitions have two blocks, $\omega=0.5$ is the lower bound and implies that half of the nodes are classified correctly and therefore the two partitions are not correlated. When one partition has more than two blocks, we can have $\omega < 0.5$.

The partition overlap measure is a suitable choice of similarity measure since it is easier to interpret than information theoretic measures, such as those based on mutual information, and since it does not depend on the number and size of blocks in the two partitions being compared, which is an issue for some pair-counting methods such as the rand index \citep{wagner_comparing_2007}. We demonstrate the robustness of our results by calculating the partition similarity for one set of simulations using reduced mutual information \citep{newman_improved_2020} and variation of information, which has also been shown to behave well for unbalanced partitions \citep{meila_comparing_2007} and we show the results in \aref{app:similaritymeasures}. Both similarity measures yield comparable results to those calculated using the partition overlap measure. In particular, detectability thresholds appear to be located identically (or at least extremely similarly) in the $(\lambda, \mu)$ space.

\section{\label{sec:results} Results}
To infer partitions of the generated graphs, we fit two SBM variants (traditional and degree-corrected) using the graph-tool python library \citep{peixoto_graph-tool_2014}. We retrieve a distribution of $50$ partitions for each of the eight graph and therefore a total of $400$ partitions for each combination of $\lambda$ and $\mu$. To ensure that the chosen number of samples is sufficient to explore the posterior we ran the same simulations\footnote{For graphs with $N=400$, $c = 10$ generated by the canonical model and partitions inferred using the non-degree-corrected SBM.} with $4000$ partitions for each $(\lambda, \mu)$-pair (sampling $1000$ partitions for each of four graphs) and found no qualitative difference. In its function as an \emph{inference} method, we from now refer to the traditional SBM as NDC (non-degree-corrected SBM) and to the degree-corrected variant as DC, to avoid confusion with the models (canonical and microcanonical) used to \emph{generate} our networks. 
We finally calculate the partition overlap $\omega$ between inferred partitions and planted partitions for the two planted structures as well as between the inferred partitions and the planted cross-partition.

\subsection{\label{sec:modelfit}Model fit}
\begin{figure}
\centering
\begin{subfigure}{0.5\textwidth}
    \centering
    \includegraphics[width=1\linewidth]{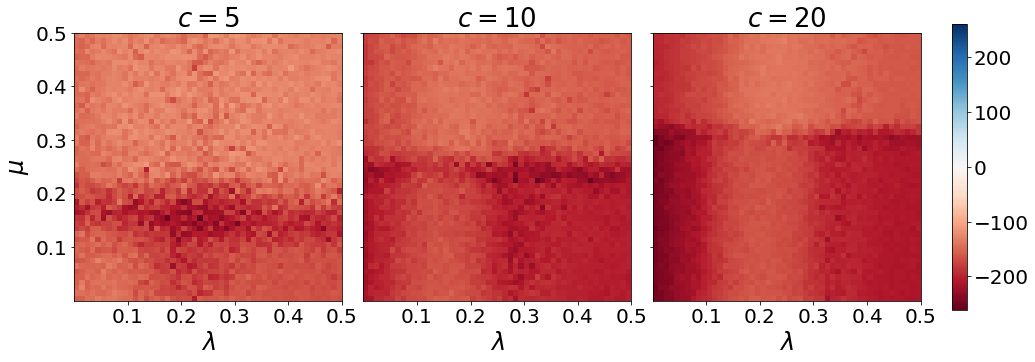}
    \caption{Canonical model.}
    \label{fig:sfig_logev_trad}
\end{subfigure}\\
\begin{subfigure}{0.5\textwidth}
    \centering
    \includegraphics[width=1\linewidth]{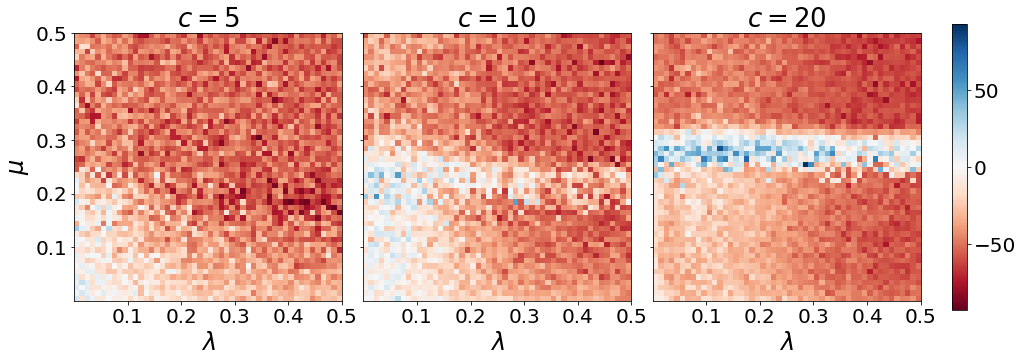}
    \caption{Microcanonical model.}
    \label{fig:sfig_logev_mc}
\end{subfigure}
\caption{Difference between the log-evidence of the DC and NDC model class for graphs with varying expected degrees $c$. Negative values (red) indicate a better fit of the NDC model, positive values (blue) indicate a better fit of the DC model.}
\label{fig:sfig_logev}
\end{figure}

We start by evaluating which of the two SBM variants used for the detection of mesoscale structures provides a better fit to our generated networks. We calculate the (log) model evidence, summed over all partitions for each run, calculated by subtracting the entropy of the posterior distribution from the negative average description length (over all partitions) \cite{peixoto_nonparametric_2017}. \autoref{fig:sfig_logev_trad} demonstrates that for the canonical version NDC is the preferred model across the entire $(\lambda, \mu)$ space; this is unsurprising as edge placement in the generative process is independent of node degree. When we use the microcanonical version (in which we \textit{do} take into account the node degrees in the generative process), we may have expected DC to be a better description of the generated networks across the entire $(\lambda, \mu)$ plane. However, \autoref{fig:sfig_logev_mc} demonstrates that this is not the case: we observe a small region of $\lambda$ and $\mu$ values for which DC has the larger model evidence; for increasing expected degree, this region becomes more pronounced and exists across the entire $\lambda$ range, while restricted to more and more narrow values of $\mu$. Everywhere else, NDC still provides a better model fit. This suggests that the higher complexity of DC is justified only for networks with a high level of heterogeneity in the degree distribution \textit{and} bi-community structure of a certain strength which depends on $c$. It seems that -- in terms of the number of model parameters -- DC provides an overly complex description everywhere else in the $(\lambda, \mu)$ plane, although degree heterogeneity is still high. \footnote{If we allow for multiple edges between node pairs (using the graph-tool python library \citep{peixoto_graph-tool_2014}), we find an additional area in the bottom-left of the $(\lambda, \mu)$ plane -- roughly for values $\lambda < 0.2$ and $\mu < 0.33$ -- in which DC is preferred. A possible explanation could be the larger degree variance introduced in this case, for which the higher complexity of DC is justified. As the partition recovery patterns for these multigraphs are equivalent to the ones we discuss in \autoref{sec:results}, we restrict our analysis to simple graphs.}

\begin{figure*}
\centering
\begin{subfigure}{1\textwidth}
    \centering
    \includegraphics[width=1\linewidth]{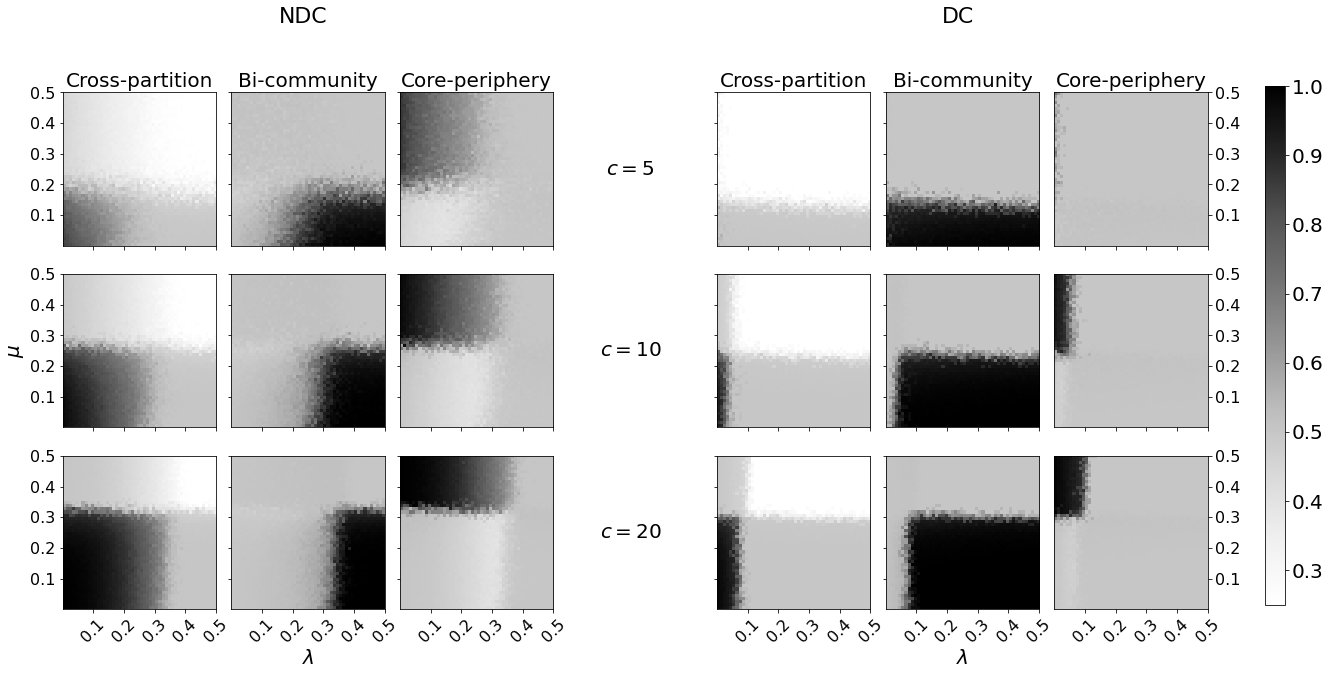}
    \caption{Canonical model.}
    \label{fig:sfig_mean_po_trad}
\end{subfigure}
\begin{subfigure}{1\textwidth}
    \centering
    \includegraphics[width=1\linewidth]{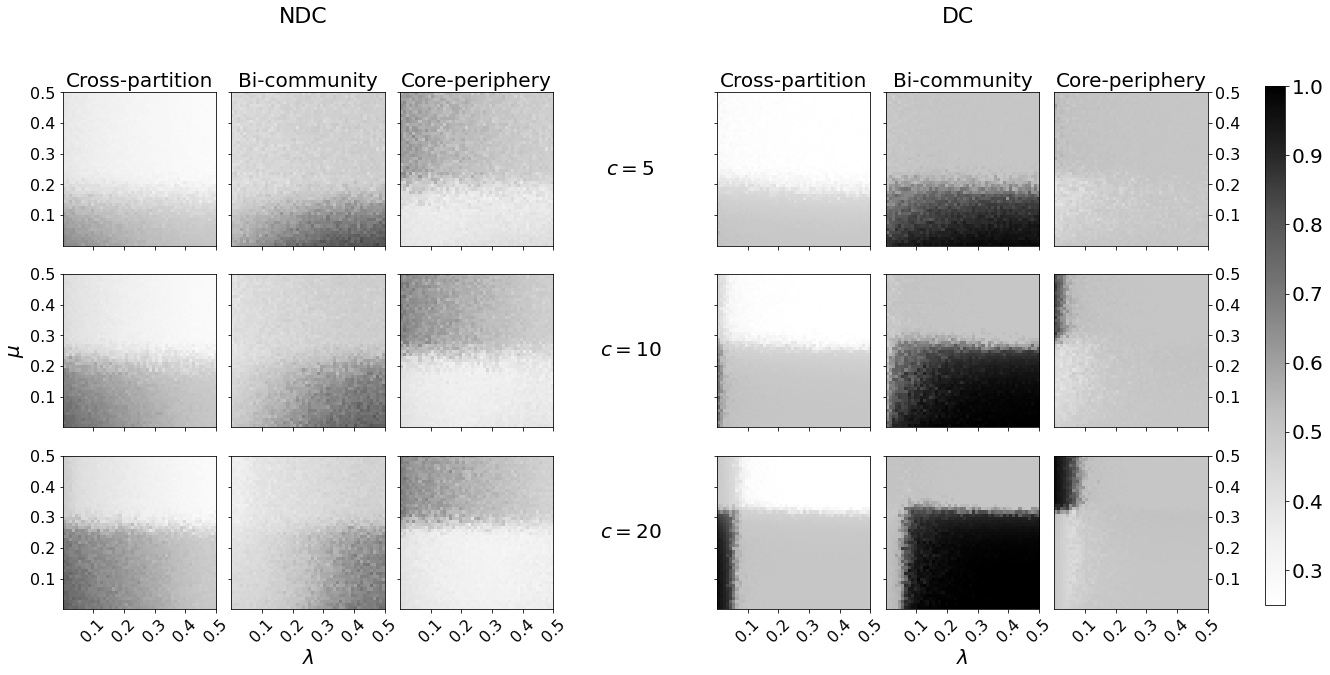}
    \caption{Microcanonical model.}
    \label{fig:sfig_mean_po_mc}
\end{subfigure}
\caption{Mean partition overlap $\langle \omega \rangle$ between the planted partitions and the partitions in the posterior distributions for graphs with varying expected degrees $c$.}
\label{fig:sfig_mean_po_all}
\end{figure*}

\subsection{\label{sec:rec_planted}Recovery of planted structures}
In \autoref{fig:sfig_mean_po_all}, we show the mean partition overlap $\langle \omega \rangle$ for all (explicitly and implicitly) planted partitions in the posterior distribution of inferred partitions for each $(\lambda, \mu)$ pair, for networks generated by two models. The left-most columns in all four quadrants show $\langle \omega \rangle$ between the inferred partitions and the explicitly planted cross-partition $(\boldsymbol{\theta}_\text{SCBM})$; the middle and right columns show the same for the implicitly planted bi-community ($\boldsymbol{\theta}_1$) and core-periphery ($\boldsymbol{\theta}_2$) partitions respectively. Note that here we focus on the detection of the individual partitions and we refer to \autoref{sec:coexistence} for an outline of the detection of partition coexistence -- the extent to which \textit{both} implicitly planted structures appear in the posterior distribution of inferred partitions of a given generated graphs.

As expected, there appears to be some clear thresholds separating areas in the $(\lambda, \mu)$-space in which the cross-partition is recovered from areas in which either of the two implicitly planted structures are detected. The locations of these thresholds vary by expected degree, by generating model and by SBM variant used for the partition inference. The question around the detectability of the two implicitly planted structures thus appears to be related to the detectability of the cross-partition: as lower values of both parameters mean a stronger signal for the cross-partition, it is detected up until some threshold. Only once the signal for the cross-partition becomes weak enough are the implicitly planted partitions favoured.

We first focus on the partitions inferred from the graphs generated by the canonical model, where node degrees follow Poisson distributions. In \autoref{fig:sfig_mean_po_trad}, we show $\langle \omega \rangle$ between planted partitions and those recovered by NDC (on the left-hand side) and DC (on the right-hand side). Both variants detect the bi-community structure frequently up to a certain threshold value of $\mu$, which increases for higher $c$. It turns out that the locations of the thresholds are roughly in line with what is described in existing literature on detectability thresholds for the planted partition model \cite{decelle_inference_2011, decelle_asymptotic_2011}. According to this work, community structure planted by edge count matrix $\mathbf{B}_1$ as defined in \eqref{eq:ac1} is detectable for $c > \frac{1}{(1-2\mu)^2}$; in our case this should place our threshold $\mu_T$ for detectability at $\mu_T \approx 0.276$ ($c = 5$), $\mu_T \approx 0.342$ ($c = 10$), and $\mu_T \approx 0.388$ ($c = 20$). In our simulations, both variants detect bi-communities up to a similar value of $\mu$, which is slightly below $\mu_T$. We observe a second type of threshold for bi-community detection, this time along the $\lambda$ dimension. For low values of $\lambda$, both variants fail to detect the bi-community partition even though $\mu < \mu_T$ and they appear to uncover the cross-partition instead. While the threshold along the $\mu$ dimension is similar for NDC and DC, we find that the $\lambda$ threshold is much higher for NDC than for DC. This means that NDC recovers the cross-partition up to higher values of $\lambda$ than DC, after which bi-communities are detected. The $\lambda$ threshold also increases with growing $c$. 

The detection of the planted core-periphery partition also depends on the expected degree of the networks. In fact, the thresholds of CP detection correspond with those described above for bi-community detection: along the $\mu$ dimension, CP structure is detected once cross-partitions and bi-communities are no longer recovered; along the $\lambda$ axis, CP structure is detected until its structure is too weak, at which point bi-communities are detected. This is somewhat contradictory to the work by \citet{zhang_identification_2015}, who find no evidence for a detectability threshold in the case of CP structures. A likely explanation for the narrow recovery range of the CP structure by DC compared to NDC is that degree correction aims to account for degree heterogeneity in a network in favour of detecting community structure, while NDC has a higher tendency to split networks into blocks of lower and blocks of higher degree \citep{karrer_stochastic_2011}, which here corresponds to the implicitly planted CP structure.

One of the main differences between NDC and DC in the canonical case is therefore the thresholds at which structures do and do not get detected along the $\lambda$ axis. For all values of $c$ and for both SBM variants, the cross-partition is recovered when the bi-community and CP structures are strong. Along each direction, both variants then start picking up the respective bi-block structure once the signal becomes weaker. Since both the bi-community and CP structures are only recovered when the signal for the respective other structure is weak, the coexistence of both structures in the inferred partition distribution is rare; we revisit this in \autoref{sec:coexistence}. As both thresholds (along $\lambda$ and $\mu$) are higher for larger $c$, the cross-partition detection region increases for denser graphs; this phenomenon is more pronounced for NDC, which provides a better model fit than DC across the entire $(\lambda, \mu)$ plane. 

\begin{figure*}
\centering
\begin{subfigure}{0.49\textwidth}
    \centering
    \includegraphics[width=1\textwidth]{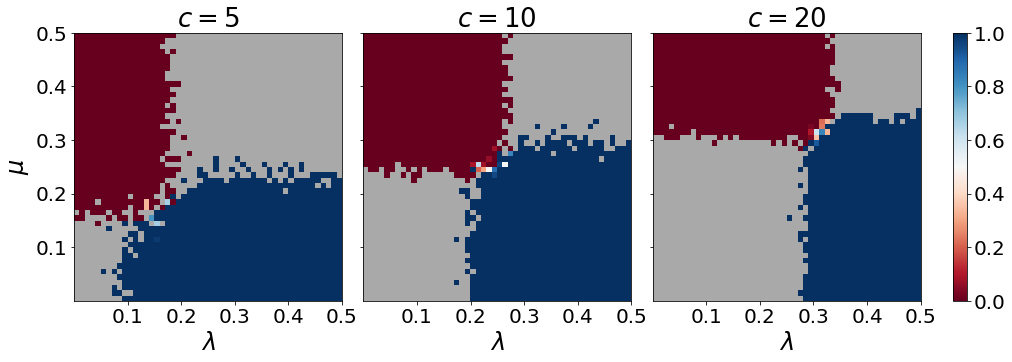}
    \caption{NDC, Canonical model.}
    \label{fig:sfig_both_recovered_ratio_trad_ndc}
\end{subfigure}
\begin{subfigure}{0.49\textwidth}
    \centering
    \includegraphics[width=1\textwidth]{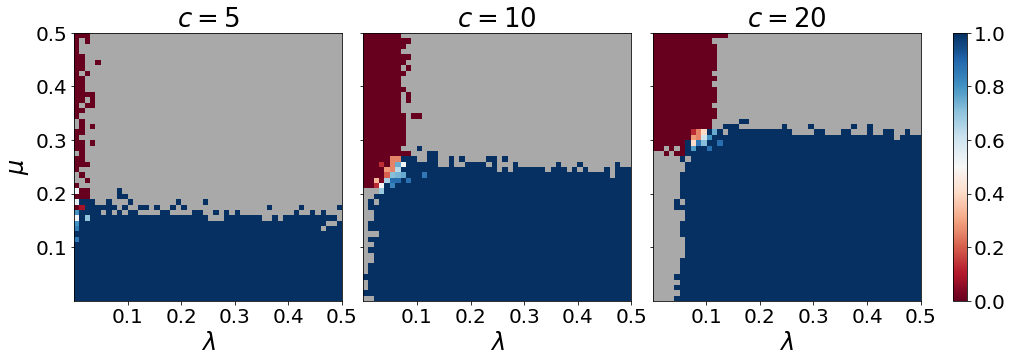}
    \caption{DC, Canonical model.}
    \label{fig:sfig_both_recovered_ratio_trad_dc}
\end{subfigure}
\\
\begin{subfigure}{0.49\textwidth}
    \centering
    \includegraphics[width=1\textwidth]{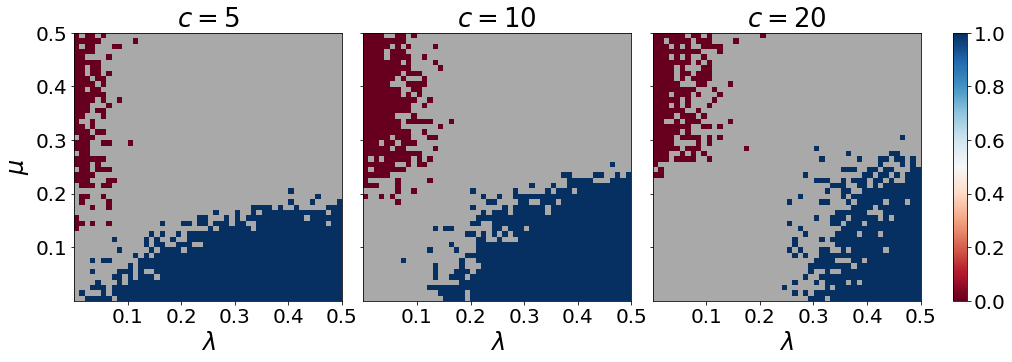}
    \caption{NDC, Microcanonical model.}
    \label{fig:sfig_both_recovered_ratio_mc_ndc}
\end{subfigure}
\begin{subfigure}{0.49\textwidth}
    \centering
    \includegraphics[width=1\textwidth]{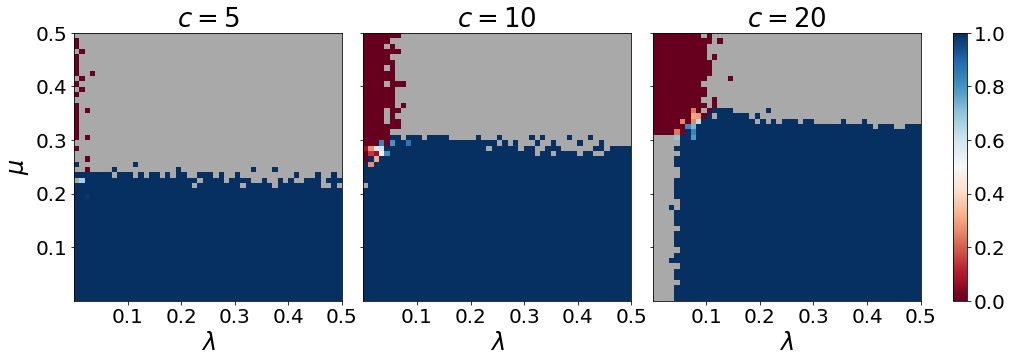}
    \caption{DC, Microcanonical model.}
    \label{fig:sfig_both_recovered_ratio_mc_dc}
\end{subfigure}
\caption{Fraction $\alpha$ of recovered bi-community partitions out of all successfully recovered partitions for varying expected degrees $c$ for $\omega_T=0.75$; at $\alpha=1$ (dark blue) only the bi-community structure is detected, at $\alpha=0$ (dark red), only the CP structure is.}
\label{fig:sfig_both_recovered_ratio}
\end{figure*}

\subsection{\label{sec:rec_mc}Influence of degree distribution} 
To explore the influence of a heterogeneous degree distribution on partition recovery, we fit the two SBM variants to a set of networks generated by the microcanonical model. \autoref{fig:sfig_mean_po_mc} illustrates the mean recovery of partitions in this case by NDC (left) and DC (right). The overall detection patterns resemble those discussed for the canonical case. For DC, all partitions are recovered in similar regions with similar thresholds, albeit slightly more `fuzzy' boundaries on said thresholds. However, we observe a substantial difference in the performance of the NDC variant, for which $\langle \omega \rangle$ is considerably lower for all planted partitions across the entire $(\lambda, \mu)$ plane. In the first instance, this is somewhat surprising, since we have seen in \autoref{sec:modelfit} that even for graphs generated by the microcanonical model NDC provides a better description. A plausible explanation of this phenomenon is the additional structural features that we introduce through the extra constraint on node degrees in our microcanonical model and the thereby imposed degree heterogeneity (see \autoref{sec:generatedgraphs}). It turns out that for relatively strong bi-community structures, the NDC variant recovers layered CP structures nested within each of the two community blocks, both for very strong planted CP structures but also when no explicit CP structure is planted at all (see example graphs in \aref{app:exnetworks}). When CP structures are planted explicitly with strong signal, the layered CP partition recovered by NDC bears some resemblance to the cross-partition; when no explicit CP structure is planted, the layered CP structures within two assortative blocks resemble more the bi-community partition (according to $\omega$ and upon visual inspection of the example networks). As can be seen in \autoref{fig:sfig_mean_po_mc}, the DC variant detects partitions much closer to the planted cross-partition (for lower $\lambda$) and the bi-community (for higher $\lambda$) value than NDC. However, the NDC variant has the better model fit; this implies that by forcing heterogeneous degrees, we may to some extent be `overfeeding' more structure into the network than solely that defined through the block connectivity matrices.

\subsection{\label{sec:coexistence}Structure coexistence}
Other than the individual recovery of the two planted partitions, we are naturally also interested in the appearance of both structures in different regions of the partition landscape detected in a given network, that we denote as coexistence. Specifically, we want to know whether the posterior distribution of partitions inferred by SBM features both planted partitions (rather than only one or the other) for any particular set of $(\lambda, \mu)$ pairs. We measure coexistence recovery by setting the threshold for the partition overlap to $\omega_T=0.75$, for which we consider an inferred partition to be close enough to the planted partition to be considered a `successful recovery'.

To illustrate the coexistence detection, we plot the fraction $\alpha=\frac{q_1}{q_1+q_2}$, where $q_1$ denotes the proportion of partitions in the posterior distribution that resemble the bi-community partition, given $\omega_T$, and $q_2$ denotes the equivalent for CP partitions. The results for each model/variant combination are shown in \autoref{fig:sfig_both_recovered_ratio}. For $\alpha=1$ (dark blue) we only detect the bi-community structure, for $\alpha=0$ (dark red) only the CP structure is present in the posterior distribution, and values close to $\alpha=0.5$ (white) indicate a more balanced posterior distribution, which features both partitions to some extent; such values are found where the detection areas for the two planted partitions appear to be touching or even overlapping. The grey region represents $(\lambda, \mu)$ pairs for which $\alpha$ is undefined since neither of the two structures is recovered successfully.

We observe in \autoref{fig:sfig_both_recovered_ratio_mc_ndc} that -- as expected from the results in \autoref{sec:rec_mc} -- fitting NDC to graphs generated by the microcanonical model does not yield partition distributions anywhere in our $(\lambda, \mu)$-space that feature both bi-community \textit{and} CP partitions. In all other cases, there are small regions for which coexistence is detected, for this relatively generous threshold value of $\omega_T=0.75$. Note that for stricter $\omega_T$ thresholds, we observe reduced recovery regions for each partition in all cases and therefore an even smaller or completely absent region of overlap in which both partitions feature in a given posterior distribution (see \aref{app:extrathresholds}). Clearly, choosing this value of $\omega_T$ to mean `successful recovery' is somewhat arbitrary and one might argue that a maximum of $75\%$ of nodes being assigned to the (implicitly planted) `correct' blocks does not indicate strong partition similarity. In fact, we choose to display the results with this low thresholds here, since it emphasises the finding that even for such a generous threshold coexistence recovery is very small.

\subsection{Discussion}
Overall, we found detectability thresholds for each individual planted structure and we discovered that coexistence of the two structures is only detected in a very small number of cases. We also observed considerable differences in successful partition recovery between the NDC and DC variants, which are more pronounced when they are fitted to graphs generated by the microcanonical model than when fitted to those produced by the canonical version.

We briefly discuss the effect of degree heterogeneity introduced into networks through the generative process of the microcanonical model. We found that, by constraining the degree distributions in this way, we are inadvertently introducing CP divisions \textit{beyond} the explicitly planted CP structure. The additional CP structures are picked up by NDC since it does not correct for node degrees and it thus comparatively `underperforms' at detecting individual planted partitions and coexistence of multiple partitions. These findings are consistent with existing SBM literature, including the original work in which the degree-corrected variant was introduced \cite{karrer_stochastic_2011}. They should serve as a reminder that a network may exhibit multiple conflated structural properties which could in turn complicate the detection of certain types of partitions or even lead to the detection of spurious mesoscale structures. In general, and specifically if previous knowledge exists about structural properties that are likely to be present in a network (\hbox{e.g.,} high degree variance) or about certain types of structures that are of interest, one should consider carefully the SBM variant that is appropriate. Methods which specifically aim to disentangle conflated structural properties \cite{karrer_stochastic_2011, peixoto_disentangling_2022} or to recover certain types of structures, such as assortative communities \cite{zhang_statistical_2020}, could be considered.

Secondly, we focus on the canonical model and the extent to which the two SBM variants recover the bi-community and CP partitions relative to each other, jointly (coexisting in a given posterior distribution) and relative to the cross-partition. Overall, we have found that thresholds for the detection of the individual planted partitions and for the detection of structure coexistence depend on the expected degree of a network as well as the SBM variant used to detect the structure. The NDC variant -- which has higher model evidence -- does better at picking up the cross-partition at the expense of recovery of the CP structure. It recovers coexistence of bi-community and CP partitions to a slightly lesser extent than DC. Since we are \textit{explicitly} planting the cross-partition, and only \textit{implicitly} planting the bi-community and CP structures by making sure the edge probabilities within and between the respective cross-blocks satisfy those within and between the blocks of the two bi-block partitions, it is perhaps not surprising that the variant with the better model fit is the one that favours the cross-partition at the cost of coexistence detection.

Irrespective of whether we use NDC or DC, we find that even in this relatively simple case of planting only two qualitatively different ground-truth partitions, the region in our structural strength landscape for which the coexistence of the bi-community and CP structures is detected is limited to an extremely small area. This is concerning since mesoscale structures in real networks are unlikely to be so simple and a larger number of coexisting network partitions and a multitude of different structures may be present. Clearly, more research is necessary to better understand whether the lack of coexistence recovery is due to a detectability limit after which it is impossible for any algorithm to detect coexistence (similar to the known community detection detectability threshold \cite{decelle_asymptotic_2011, decelle_inference_2011}), or whether some other SBM variant would be able to do a better job at detecting coexistence of multiple planted structures. More work should then also focus on expanding the relatively recent literature on partition diversity \cite{peixoto_revealing_2021, kirkley_representative_2022} by advancing existing methods or developing new tools. The aim should be to enable researchers to reliably explore multiple coexisting ground truth partitions that may have been responsible for the generation of a given network, which we suspect to be the case in real graphs \cite{peel_ground_2017}. In this sense, our framework and findings should be seen as a motivation to test new SBM variants developed for this purpose. The regions of coexistence discovery shown in \autoref{fig:sfig_both_recovered_ratio} may be used as an orientation for the possible locations of detectability thresholds in the case of multiple ground-truth partitions.

More generally, the fact that even for existing methods coexistence \textit{is} detected in certain regions of our structural strength landscape emphasises again the importance of acknowledging the diversity and possible dissensus in partition distributions, and for more researchers in the field of applied network science to consider multiple plausible explanations of the mesoscale of network.

\section{\label{sec:conclusion}Conclusion}
We have proposed the stochastic cross-block model (SCBM), a framework for generative network models that exhibit predefined ambiguity in their mesoscale structure. This framework complements existing generative networks as a \emph{two-ground-truth benchmark} that can be used to measure the extent to which mesoscale structure detection algorithms recover the ambiguity introduced by two simultaneously planted partitions. Our work also generally emphasises the need to explore the question around ambiguity in network structure, in the sense that: the cross-partition that we plant explicitly is what we plant unambiguously, while the ambiguity stems from the two implicitly planted partitions. While we focus on the two-partition case in our simulations, we also outline how our approach can be extended to the multi-partition case, and we encourage future work on more complicated cases of multiple ground truth structures, which are arguably closer to what may occur in real networks. 

We detail a possible way to frame the multiple ground truth problem as a special case of the MMSBM and explain why our method simplifies the MMSBM approach. We found that the coexistence of two qualitatively different partitions (bi-community and core-periphery structure) is only detected in a very small region in our `structural strength' space, which varies in size and shape for different versions of our model and for different SBM variants used for mesoscale structure detection.  Only when both structures are sufficiently strong and neither dominates the other can we recover the existence of both. In the majority of cases, each of the two planted partitions is recovered when the strength of the other structure is weak. We have thus uncovered a type of detectability threshold in the case where multiple types of mesoscale structures influence the network construction. Since the coexistence of more than one plausible explanation for mesoscale structures appears to be a common phenomenon in real networks \cite{peel_ground_2017}, we believe that exposing the presence of such an -- as of yet understudied -- detectability threshold is an important contribution to the SBM literature, especially as most community detection approaches still aim at uncovering a single partition and at validating it against a single ground truth. More work is required to explore the nature of these detectability thresholds analytically, and to appraise detectability limits exhibited by other types of coexisting structures, for example including more than one community partition or bipartite structures; the combination of certain types of structures may be more or less prone to detectability issues than others, especially given the ability (or lack) of certain SBM variants to detect certain types of structures. Other future work may include fitting structure-specific SBM variants to graphs generated by the SCBM benchmark. For example, SBMs designed to detect assortative structure \citep{zhang_statistical_2020} or core-periphery structure \citep{gallagher_clarified_2021}, may be used in conjunction with the minimum description length principle \citep{peixoto_parsimonious_2013} to investigate if certain models provide a better fit for certain parameter choices and to explore the performance of such models in terms of detecting specific implicitly planted structures. Another SBM variant that may be of interest for future research including the SCBM is the hierarchical SBM, \hbox{e.g.} \citep{peixoto_disentangling_2022} which was demonstrated to prevent underfitting of SBMs, to explore whether it would perceive the coexisting partitions as structures nested within each other or in the form of partitions appearing in the same posterior distribution of flat partitions.
Finally, we conclude that future work around the theory of methods for mesoscale structure detection in networks should focus on improving existing methods to be able to identify coexisting structures. More broadly, and in line with recent work \cite{peixoto_revealing_2021, kirkley_representative_2022}, we believe that researchers applying existing methods on real networks should focus on the possibility of discovering multiple dimensions of segmenting the network, rather than accepting unidimensional solutions, that may be even be averaged over `multimodal' partition distributions. In particular, 
possible contexts in which considering multiple plausible network divisions seem particularly important include the field of computational social science which deals with the analysis of interaction dynamics in online public spaces. Appraising the coexistence of multiple types of structures in social media interaction networks, such as community/CP structures or even qualitatively different community partitions (maybe generated through non-aligned political dimensions, as has been recently shown on Twitter affiliations \citep{ramaciotti_morales_multidimensional_2023}), could have considerable benefits for researching online conversation dynamics. In this context, researchers should consider the coexistence of qualitatively different structures and be mindful of the detectability issues addressed here. 
In our simulations, we focus on the special case of equally sized blocks in both the planted partitions as well as the cross-partition and on the community-CP case. However, the benchmark model is flexible to a diverse range of structures of varying block sizes, degree distributions, and planted mesoscale structures.
Future work may use this benchmark to analyse the recovery of ambiguity in networks of different sizes and expected degrees or with other combinations of mesoscale structures. This work may also be extended by testing other types of detection algorithms (beyond SBM) on this benchmark model. Further extensions of the benchmark model itself could allow for more than two blocks in each planted partition, more than two planted partitions or a directed version of the model. 

\begin{acknowledgements}
This work was supported by the “Socsemics” Consolidator grant from the European Research Council (ERC) under the European Union’s Horizon 2020 research and innovation program (grant agreement \hbox{No.} 772743).
\end{acknowledgements}

\newpage

\appendix

\section{\label{app:mmsbmformulation} Detailed MMSBM formulation}
In the original MMSBM formulation \cite{airoldi_mixed_2008}, each node belongs to all latent groups with certain probability, denoted -- rather than by a one-hot binary block membership vector -- by a \textit{mixed membership} vector $\boldsymbol{\pi}_i$ for node $i$, in which element $\pi^k_i$ denotes the probability of the node $i$ belonging to block $k$ and so $\sum_{k} \pi^k_i = 1$. Mixed membership vectors are drawn from a Dirichlet distribution for each node $i$, with some fixed parameter that is equal across all nodes. For each pair of nodes $i$ and $j$, the block memberships $r$ and $s$ are drawn from a multinomial distribution parameterised by the nodes' mixed membership vectors; the existence of an edge between $i$ and $j$ is sampled from a Bernoulli distribution based on the predefined edge probability between blocks $r$ and $s$. The generative process of the network is thus similar to a Bernoulli SBM, except that the block membership of a given node $i$ is drawn repeatedly for each node $j$ it is paired with. This means that nodes may belong to different blocks depending on the pairing that is considered and that they inherit connectivity patterns from multiple blocks. This process leads to the particular feature that the density at the `overlap' of multiple blocks is a weighted average of their individual densities \cite{peixoto_model_2015}.

For ease of explanation, we denote the two blocks in partition $1$ by labels $\{a,b\}$ and the blocks in partition $2$ by labels $\{c,d\}$. We define the block matrices denoting the expected edge densities within and between blocks by  $\boldsymbol{\theta}_1$ and $\boldsymbol{\theta}_2$ in \autoref{eq:theta1} and \autoref{eq:theta2}. Note that block matrices are symmetric since the graphs we are generating are undirected (for example $\theta_{ab} = \theta_{ba}$) and -- according to convention in undirected networks and to simplify calculations -- elements on diagonals denote twice the within-block edge densities.

\begin{subequations}
\begin{empheq}[left=\empheqlbrace]{align}
\label{eq:theta1}
& \boldsymbol{\theta}_1=\begin{pmatrix} \theta_{aa} & \theta_{ab} \\ \theta_{ab} & \theta_{bb}\end{pmatrix} \\
\label{eq:theta2}
& \boldsymbol{\theta}_2=\begin{pmatrix} \theta_{cc} & \theta_{cd} \\ \theta_{cd} & \theta_{dd} \end{pmatrix}
\end{empheq}
\end{subequations}

To fit the MMSBM formulation, we consider the blocks in the two planted partitions as four latent blocks $\{a,b,c,d\}$ but we constrain the generative process in a way that forces certain overlaps to be empty. In particular, the overlaps of block $a$ or $b$ in partition 1 and of block $c$ or $d$ in partition 2 are empty. This constraint is illustrated in \autoref{fig:schem_mmsbm}, where the nodes in the overlaps are coloured according to their block membership colours in the two planted partitions. Our mixed membership vectors thus take on the form $\boldsymbol{\pi}_i = \frac{1}{2}(\mathbf{b}^1_{i},\mathbf{b}^2_{i})$. For example, if $\boldsymbol{\pi}_i = \frac{1}{2}(1,0,1,0)$ then node $i$ belongs to blocks $a$ and $c$ with equal probability $\pi^a_i = \pi^c_i = \frac{1}{2}$. The constraint compared to general mixed membership vectors is therefore that nodes belong to exactly two blocks with equal probability and that $\boldsymbol{\pi}_i = \frac{1}{2}(1,1,0,0)$ and $\pi_i = \frac{1}{2}(0,0,1,1)$ are forbidden, since a node can never belong to block $a$ and $b$ (or to $c$ and $d$). After drawing a mixed membership vector for each node $i$, we draw the block memberships $r$ and $s$ for each node pair $i$ and $j$ independently from their mixed membership vectors. We then place an edge between $i$ and $j$ according to some probability that depends on $r$ and $s$.

\subsection{Normalisation}
A first intuition might be to connect two nodes with probability $\theta_{rs}$ if $r, s \in \{a,b\}$ (or $r, s \in \{c,d\}$) and with zero probability for any other combination of $r$ and $s$. However, since each node receives edges based on the connectivity of two blocks independently, this process will not be consistent with the connectivity of $\boldsymbol{\theta}_1$ and $\boldsymbol{\theta}_2$. If we generate a network according to these probabilities, the expected edge density $\hat{\theta}_{rs}$ between two nodes $i$ and $j$ that belong to blocks $r$ and $s$ respectively is 

\begin{equation}
\label{eq:additiveprobs}
    \hat{\theta}_{rs} = \frac{1}{4} \sum_{r's'} (\pi^{r}_{i}\pi^{s}_{j}\theta_{rs} + \pi^{u}_{i}\pi^{v}_{j}\theta_{r's'}) = \frac{1}{4}\theta_{rs} + \frac{1}{16}\sum_{r's'}\theta_{r's'},
\end{equation}
for $r, s \in \{a,b\}$ and $r',s' \in \{c, d\}$ (or $r, s \in \{c,d\}$ and $r',s' \in \{a,b\}$). In order to accommodate for the connectivity of both planted partitions, we thus need to normalise the edge probabilities appropriately, so that $\hat{\theta}_{rs}=\theta_{rs}$. The additive edge probabilities prevent us from finding a single normalisation constant $x$. Instead we may consider finding $x_{rs}$ for each block pair, so that nodes are connected with probability $x_{rs}\theta_{rs}$. To guarantee $\hat{\theta}_{rs}=\theta_{rs}$ we thus need
\begin{equation}
\label{eq:additiveprobs_norm}
    \theta_{rs} = \frac{1}{4}x_{rs}\theta_{rs} + \frac{1}{16}\sum_{r's'}x_{r's'}\theta_{r's'}.
\end{equation}

Finding suitable $x_{rs}$ requires us to solve an underdetermined system of six equations, that is consistent as long as the sum of the probabilities in $\boldsymbol{\theta}_1$ equals the sum of probabilities in $\boldsymbol{\theta}_2$, and thus has infinitely many solutions. However, it turns out that this system of equations does not have any non-negative solutions for certain combinations of connectivity patterns planted in $\boldsymbol{\theta}_1$ and $\boldsymbol{\theta}_2$, in particular when the differences between block densities in both partitions are large (see \aref{app:farkas} for a proof).

An alternative way to normalise the additive edge probabilities is to determine a normalisation constant for each combination of block pairs that two nodes can occupy in the two partitions. In other words, nodes $i$ and $j$ for which block memberships $r$ and $s$ have been drawn are connected with probability $x_{rsr's'}\theta_{rs}$, where $r'$ and $s'$ are the other two blocks that $i$ and $j$ are also members of. To determine the set of $\{x_{rsr's'}\}$ we rewrite the expected density as
\begin{eqnarray}
    \label{eq:additiveprobs_norm_alt}
    \hat{\theta}_{rs} = \frac{1}{4} \sum_{r's'} x_{rsr's'}(\theta_{rs} + \theta_{r's'})
\end{eqnarray}
and set up a system of six equations so that $\hat{\theta}_{rs}=\theta_{rs}$ is satisfied -- one for each of the upper triangular entries in $\boldsymbol{\theta}_1$ and $\boldsymbol{\theta}_2$ -- and solve it for $\mathbf{x} = \{x_{rsr's'}\}$. This is, again, an underdetermined system of equations with infinitely many solutions and here we can find non-negative solutions for all combinations of $\boldsymbol{\theta}_1$ and $\boldsymbol{\theta}_2$ that we use in our simulations in \autoref{sec:simulations}. We can use a least squares solver to compute a minimum norm solution to the system \cite{lawson_solving_1995}. However, for certain combinations of planted structures, the minimum norm solution returns negative values for some of the $\{x_{rsr's'}\}$. This means that for certain combinations of edge probabilities in the two planted structures, we must either use a non-negative least squares solver -- which may return solutions that include zero values -- or a linear programming solver \citep{dantzig_linear_1963, huangfu_parallelizing_2018} to find strictly positive solutions.

\subsection{\label{app:farkas} Existence of non-negative solutions}
For the simplified case where $\theta_{aa}=\theta_{bb}$ and $\theta_{cc}=\theta_{dd}$, we can use Farkas’ lemma \cite{farkas_theorie_1902} to show that the system does not have any non-negative solutions for a certain combination of densities planted in the two structures --- namely, if and only if $|\theta_{aa}-\theta_{ab}| + |\theta_{cc}-\theta_{cd}| > \frac{E}{n^2}=2\rho$ where $\rho$ is the overall density of the network.

To find the set of normalisation constants, we need to solve equations $\theta_{rs} = \frac{1}{4}x_{rs}\theta_{rs} + \frac{1}{16}\sum_{uv}x_{uv}\theta_{uv}$ for $\mathbf{x}=\{x_{rs}\}$. Without loss of generality, we assume $\theta_{aa}=\theta_{bb}$ and $\theta_{cc}=\theta_{dd}$. We also start by assuming that $\theta_{aa} > \theta_{ab}$ and $\theta_{cc} > \theta_{cd}$. The system becomes $\mathbf{Ax}=\mathbf{y}$ with
\begin{equation}
\mathbf{A} = 
    \begin{pmatrix}
    \frac{1}{4}\theta_{aa}     & 0 & \frac{1}{8}\theta_{cc} & \frac{1}{8}\theta_{cd}      \\
    0     & \frac{1}{4}\theta_{ab} & \frac{1}{8}\theta_{cc} & \frac{1}{8}\theta_{cd}      \\
    \frac{1}{8}\theta_{aa}     & \frac{1}{8}\theta_{ab} & \frac{1}{4}\theta_{cc} & 0      \\
    \frac{1}{8}\theta_{aa}     & \frac{1}{8}\theta_{ab} & 0 & \frac{1}{4}\theta_{cd}      \\
\end{pmatrix} 
\end{equation}
and $\mathbf{y} = (\theta_{aa}, \theta_{ab}, \theta_{cc}, \theta_{cd})$.

Written as a \textit{theorem of alternatives}, Farkas' lemma states that exactly one of the following two statements are true for $\mathbf{A} \in \mathbb{R}^{n \times m}$ and $\mathbf{x} \in \mathbb{R}^n$:
\begin{enumerate}
    \item $\exists \mathbf{x} \in \mathbb{R}^m: \mathbf{Ax}=\mathbf{y}$ and $\mathbf{x} \geq 0$
    \item $\exists \mathbf{v} \in \mathbb{R}^n: \mathbf{v}^T\mathbf{A} \geq 0$ and $\mathbf{v}^T\mathbf{y} \leq 0$
\end{enumerate}

This means that if we can find a vector $\mathbf{v}$ for which the second alternative is always true, this implies that a non-negative solution $\mathbf{x}$ to the system cannot be found. If we choose $\mathbf{v}=(-1, 1, 0, 2)$, we have $\mathbf{v}^T\mathbf{A}=(0, \frac{1}{2}\theta_{ab}, 0, \frac{1}{2}\theta_{cd})$ and $\mathbf{v}^T\mathbf{y}=-\theta_{aa}+\theta_{ab}+2\theta_{cd}$. Therefore, we have $\mathbf{v}^T\mathbf{y} < 0 $ if and only if $\theta_{aa} - \theta_{ab} > 2\theta_{cd}$. Since $\frac{2E}{n^2}=2\theta_{cc}+2\theta_{cd}$, we can write
\begin{subequations}
\begin{empheq}{align}
& \theta_{aa} - \theta_{ab} > 2\theta_{cd} \\
& \theta_{aa} - \theta_{ab} > \frac{E}{n^2} - \theta_{cc} + \theta_{cd} \\
& \theta_{aa} - \theta_{ab} + \theta_{cc} - \theta_{cd} > \frac{E}{n^2} = \frac{4E}{N^2}=2\rho,
\label{eq:densitydiff}
\end{empheq}
\end{subequations}
where $\rho$ is the overall density of the network. Since we have assumed that $\theta_{aa} > \theta_{ab}$ and $\theta_{cc} > \theta_{cd}$, \autoref{eq:densitydiff} states that we do not have any non-negative solutions if the within-block densities are sufficiently larger than the between-block densities in both planted partitions. It is straightforward to find a vector $\mathbf{v}$ for all cases $\theta_{aa} > \theta_{ab}$ and $\theta_{cc} < \theta_{cd}$, $\theta_{aa} < \theta_{ab}$ and $\theta_{cc} < \theta_{cd}$, and $\theta_{aa} < \theta_{ab}$ and $\theta_{cc} > \theta_{cd}$, so that finally we can say that the system does not have any non-negative solutions if and only if $|\theta_{aa}-\theta_{ab}| + |\theta_{cc}-\theta_{cd}| > \frac{E}{n^2}=2\rho$.

In more qualitative terms, this means that when we induce large differences between block densities in both partitions, we cannot find appropriate normalisation constants to solve the system of equations that would enable us to create a network in which the connectivities are consistent with both planted partitions. This limitation would severely restrict our ability to test the detectability of certain combinations of partitions. In fact, it would enable us to explore only half of the parameter space we explore in our simulations in \autoref{sec:simulations}, where we plant a bi-community partition and a core-periphery partition by sweeping two parameters that determine the strength of each structure.

\section{\label{app:connectivity} Constant multiplicative factor}
In the two-partition SCBM with blocks labeled $\{a,b\}$ in partition $1$ and $\{c,d\}$ in partition $2$, each node is assigned to one of the cross-blocks $\boldsymbol{\Gamma} = \{(a, c), (a, d), (b, c), (b, d)\}$. For illustrative purposes, we start by creating the symmetric $4 \times 4$ matrix $\boldsymbol{\theta}'$, whose elements are the products of the elements of $\boldsymbol{\theta}_1$ and $\boldsymbol{\theta}_2$ corresponding to the respective blocks in $1$ and $2$ that make up the cross-blocks in $\boldsymbol{\Gamma}$. Note that we drop the SCBM subscript on our $4 \times 4$ cross-partition matrices for ease of readability. Matrix $\boldsymbol{\theta}'$ is thus the Kronecker product of $\boldsymbol{\theta}_1$ and $\boldsymbol{\theta}_2$, where the order depends on the index set $J$,

\begin{equation}
\boldsymbol{\theta}' = \boldsymbol{\theta}_1 \otimes \boldsymbol{\theta}_2 = \\
    \begin{blockarray}{c @{\hspace{33pt}} cccc}
      & (a,c)     & (a,d)     & (b,c)    & (b,d)          \\
    \begin{block}{c(cccc)}
(a,c) & \theta_{aa} \theta_{cc}   & \theta_{aa} \theta_{cd}   & \theta_{ab} \theta_{cc}  &  \theta_{ab} \theta_{cd}  \\
(a,d) & \theta_{aa} \theta_{cd}   & \theta_{aa} \theta_{dd}   & \theta_{ab} \theta_{cd}  &  \theta_{ab} \theta_{dd}  \\
(b,c) & \theta_{ab} \theta_{cc}   & \theta_{ab} \theta_{cd}   & \theta_{bb} \theta_{cc}  &  \theta_{bb} \theta_{cd} \\
(b,d) & \theta_{ab} \theta_{cd}   & \theta_{ab} \theta_{dd}   & \theta_{bb} \theta_{cd}  &  \theta_{bb} \theta_{dd} \\
    \end{block}
    \end{blockarray}
\label{matrix}
     \end{equation}

The most general form of defining the multiplicative factor is finding $x_{uv}$ for each pair of cross-blocks $u=(r, r')$ and $v=(s,s')$, such that $B_{uv} = x_{uv} \nu_u \nu_v \theta'_{uv}$. If we fix $\nu_u=\nu = \frac{n}{2}$ for all cross-blocks $u$ (and allow for self-loops) we have that the maximum possible number of edges between and within each group is $\nu^2 = \frac{n^2}{4}$, so we have

\begin{equation}
\label{eq:Bsimplecase}
B_{uv} = x_{uv} \nu^2 \theta'_{uv}.
\end{equation}

We want to find the normalisation constants $\{x_{uv}\}$, such that the expected edge counts within and between blocks of both implicitly planted partitions $1$ and $2$ are equal to the sums of the expected edge counts of the overlaps that make up each of the original blocks. Therefore, we require

\begin{equation}
    B_{1_{rs}}=n^2\theta_{1_{rs}} = \sum_{r's'} x_{uv} \nu^2 \theta_{1_{rs}} \theta_{2_{r's'}}
\end{equation}
for blocks $r$ and $s$ in partition $1$ and the same for blocks in partition $2$. We can rewrite this as 
\begin{equation}
    n^2= \frac{1}{4} \sum_{r's'} x_{uv} n^2\theta_{2_{r's'}}
\end{equation}
and therefore also (for the elements of $\mathbf{B}_2$)
\begin{equation}
    n^2= \frac{1}{4} \sum_{rs} x_{uv} n^2\theta_{1_{rs}},
\end{equation}
which is satisfied for a constant $x = x_{uv} = 2n^2/E = 1/\rho$, since
\begin{equation}
    \sum_{rs} n^2\theta_{1_{rs}} = \sum_{r's'} n^2\theta_{2_{r's'}} = 2E
\end{equation}

\section{\label{app:varyingsizes} Varying block sizes}
In the case of equal block sizes in the implicitly planted partitions $1$ and $2$, setting $2n=N$, but allowing for varying cross-block sizes, we can write $\nu_1 = \nu_{ac} = \nu_{bd}$ and $\nu_2 = \nu_{ad} = \nu_{bc}$, where $\nu_{ac}$ is the number of nodes in block $(a,c)$. In this case, we have $\nu_1 + \nu_2 = n$. We no longer have the same maximum possible number of edges within and between each of the block pairs. Therefore, instead of \eqref{eq:Bsimplecase}, we have

\begin{equation}
\label{eq:Bgeneralcase}
B_{uv} = x_{uv} \nu_u \nu_v \theta'_{uv}, 
\end{equation}
where $\nu_u \nu_v$ is one of $\{\nu_1^2, \nu_1 \nu_2, \nu_2^2\}$.

Let us assume that a constant $x=x_{uv}$ does also exist for $n = \frac{N}{2}$, $\nu_1 \neq \nu_2$ and that we are planting non-trivial partitions where the probability of edge placement between blocks is not uniform across block pairs. To satisfy the connectivity in partitions $1$ and $2$, we now have
\begin{equation}
    B_{1_{rs}}=n^2 = x \sum_{r's'} \nu_u \nu_v \theta_{2_{r's'}}.
\end{equation}
Specifically, for the within- and between-block densities of partition $1$ to be satisfied, we need 
\begin{subequations}
\begin{empheq}{align}
\label{eq:p1}
& \frac{n^2}{x} = \theta_{cc} \nu_1^2 + \theta_{dd} \nu_2^2 + 2 \theta_{cd} \nu_1 \nu_2  \\
\label{eq:p2}
& \frac{n^2}{x} = \theta_{cd} \nu_1^2 + \theta_{cd} \nu_2^2 + \theta_{cc} \nu_1 \nu_2 + \theta_{dd} \nu_1 \nu_2 \\
& \frac{n^2}{x} = \theta_{dd} \nu_1^2 + \theta_{cc} \nu_2^2 + 2 \theta_{cd} \nu_1 \nu_2,
\end{empheq}
\end{subequations}
where we have dropped the subscript for partition number $2$ on $\theta$ for easier readability. Setting equal the first and last of these equations, we get
\begin{subequations}
\begin{empheq}{align}
 & (\theta_{cc} - \theta_{dd}) (\nu_1^2 - \nu_2^2) = 0.
\end{empheq}
\end{subequations}
Since $\nu_1 \neq \nu_2$, we have $\theta_{cc} = \theta_{dd}$ and therefore (from equations \eqref{eq:p1} and \eqref{eq:p2}), we then have 
\begin{subequations}
\begin{empheq}{align}
 & \theta_{cc} \nu_1^2 + \theta_{cc} \nu_2^2 + 2 \theta_{cd} \nu_1 \nu_2 = \theta_{cd} \nu_1^2 + \theta_{cd} \nu_2^2 + 2 \theta_{cc} \nu_1 \nu_2 \\
 & (\theta_{cc} - \theta_{cd}) (\nu_1 - \nu_2)^2 = 0,
\end{empheq}
\end{subequations}
and hence $\theta_{cc} = \theta_{cd} = \theta_{dd}$. Clearly, taking the same steps for $\mathbf{B}_2$ gives $\theta_{aa} = \theta_{ab} = \theta_{bb}$. This is a contradiction to our set of assumptions and therefore such a constant $x$ does not exist (apart from the trivial case where we plant a random graph).

\section{\label{app:solving} Solving underdetermined system of equations}
For general cases in which each block in partitions $1$ and $2$ can take on any size, we need to determine a set of normalisation constants $\{x_{uv}\}$ to create the final block matrix $\boldsymbol{\theta}_{\text{SCBM}} = \{\theta_{uv}\} = \{x_{uv}\theta_{1_{rs}}\theta_{2_{r's'}}\}$. In this case, the maximum possible number of edges between and within each block -- both in the two planted partitions as well as the cross-partition -- may differ for each block pair. The elements of the expected edge count matrices are therefore $B_{p_{rs}} = n_r r_s \theta_{p_{rs}}$ for $p \in \{1,2\}$, and $B_{uv} = \nu_u \nu_v \theta_{uv} = \nu_u \nu_v x_{uv} \theta'$. The set of constants $\{x_{uv}\}$ need to be chosen in such a way that the elements of $\mathbf{B}_1$ and $\mathbf{B}_2$ are satisfied (on the diagonal and upper triangular), as is presented in equations \eqref{eq:maa} to \eqref{eq:mdd}. We thus have a total of $2 \times \frac{K_s(K_s+1)}{2}=6$ equations and $\frac{K(K+1)}{2}=10$ unknowns. On the left-hand side, we have $b_{rs} = n_r n_s \theta_{rs} = B_{p_{rs}}$ and we have dropped the subscripts for partitions $1$ and $2$ for legibility. On the right-hand side, we have $\mathbf{B}'_{uv}$, where $B'_{uv} = \{\nu_u \nu_v \theta_{1_{rs}} \theta_{s_{r's'}}\}$ and subscripts are written as the indices of the cross-blocks; element $x_{11}=x_{(a,c)(a,c)}$, for example, is the normalisation constant for within-block edges in block $(a,c)$ (overlap of block $a$ in $s_1$ and block $c$ in $s_2$).
\begin{subequations}
\begin{empheq}[left=\empheqlbrace]{align}
\label{eq:maa}
& b_{aa} = x_{11} \hat{B}_{11} + 2 x_{12} \hat{B}_{12} + x_{22} \hat{B}_{22} \\
& b_{ab} = x_{13} \hat{B}_{13} + x_{14} \hat{B}_{14} + x_{23} \hat{B}_{23} + x_{24} \hat{B}_{24} \\
& b_{bb} = x_{33} \hat{B}_{33} + 2 x_{34} \hat{B}_{34} + x_{44}\hat{B}_{44} \\
& b_{cc} = x_{11} \hat{B}_{11} + 2 x_{13} \hat{B}_{13} + x_{33} \hat{B}_{33} \\
& b_{cd} = x_{12} \hat{B}_{12} + x_{14} \hat{B}_{14} + x_{32} \hat{B}_{32} + x_{34} \hat{B}_{34} \\
\label{eq:mdd}
& b_{dd} = x_{22} \hat{B}_{22} + 2 x_{24} \hat{B}_{24} + x_{44} \hat{B}_{44}.
\end{empheq}
\end{subequations}

Equations \eqref{eq:maa} to \eqref{eq:mdd} are an underdetermined system of six linear equations with ten unknowns. We can write it in matrix form as
\begin{eqnarray}
\label{eq:mat}
\mathbf{A}\mathbf{x}=\mathbf{b}
\end{eqnarray}
with
\begin{eqnarray}
\label{eq:vecs}
& \mathbf{x} = (x_{11}, x_{12}, \dots, x_{32}, x_{33}, x_{44}) \\
& \mathbf{b} = (b_{aa}, b_{ab}, b_{bb}, b_{cc}, b_{cd}, b_{dd})
\end{eqnarray}
so that $\mathbf{x} \in \mathbb{R}^{10}$, $\mathbf{b} \in \mathbb{R}^{6}$ and where $\mathbf{A} = \{a_{ij}\} \in \mathbb{R}^{6 \times 10}$ is the coefficient matrix with elements of $\mathbf{\hat{B}}$ respecting equations \eqref{eq:maa}-\eqref{eq:mdd}. According to the Rouch\'e–Capelli theorem, we know that such an underdetermined system has an infinite number of solutions if and only if the rank of its coefficient matrix is equal to the rank of its augmented matrix $\mathbf{W} = [\mathbf{A}|\mathbf{b}] \in \mathbb{R}^{6 \times 11}$. This is always true for two well-defined connectivity matrices $\mathbf{B}_1$ and $\mathbf{B}_2$; clearly, the total number of edges must be the same in the two planted partitions, and we can show that $\text{rank}(\mathbf{A})=\text{rank}(\mathbf{W})$ if and only if $b_{aa} + b_{ab} + b_{bb} = b_{cc} + b_{cd} + b_{dd}$.

In general, an underdetermined linear system $\mathbf{Ax}=\mathbf{b}$ with $\mathbf{A} \in \mathbb{R}^{m \times n}$ where $m < n$, does not have a unique solution $\mathbf{x}$. Since the system is underconstrained, it has an infinitude of solutions, if it has any solutions at all. A popular method for solving under- (or over-) constrained systems of linear systems of equations is called \textit{least squares} method. The idea behind the least squares method is to find a solution $\mathbf{x}$ which minimises the squared Euclidean norm of the residual $r(\mathbf{x}) = \mathbf{b} - \mathbf{Ax}$. In other words, we want to find $\mathbf{x}$ that minimises $\phi(\mathbf{x}) = ||r(\mathbf{x})||_2^2 = ||\mathbf{b} - \mathbf{Ax}||_2^2$, which can be done by obtaining $\mathbf{x}$ such that $\nabla \phi(\mathbf{x}) = 0$. From this, we obtain the so-called \textit{normal equations} $\mathbf{A}^T \mathbf{Ax} = \mathbf{A}^T \mathbf{b}$ which can be solved analytically if $\mathbf{A}^T \mathbf{A}$ is invertible. In our case, $\mathbf{A}^T \mathbf{A} \in \mathbb{R}^{n \times n}$ has rank at most $m$, where $m < n$, and is therefore singular. This means that -- in this underdetermined case -- the normal equations cannot be solved analytically. Instead, we can find the particular least squares solution that minimises the Euclidean norm $||\mathbf{x}||_2$ (or its square) with the constraint $\mathbf{Ax}=\mathbf{b}$. When there are no other constraints, this \textit{minimum norm solution} $\mathbf{\hat{x}}$ can be found by computing the singular value decomposition (SVD) in order to compute the Moore-Penrose pseudoinverse $\mathbf{A}^+$ of matrix $\mathbf{A}$. The minimum norm solution can then be calculated as $\mathbf{\bar{x}} = \mathbf{A}^+ \mathbf{b}$, always exists and is unique \citep{boyd_convex_2004}.

\begin{figure}
\centering
\begin{subfigure}{0.45\textwidth}
    \centering
    \includegraphics[width=1\textwidth]{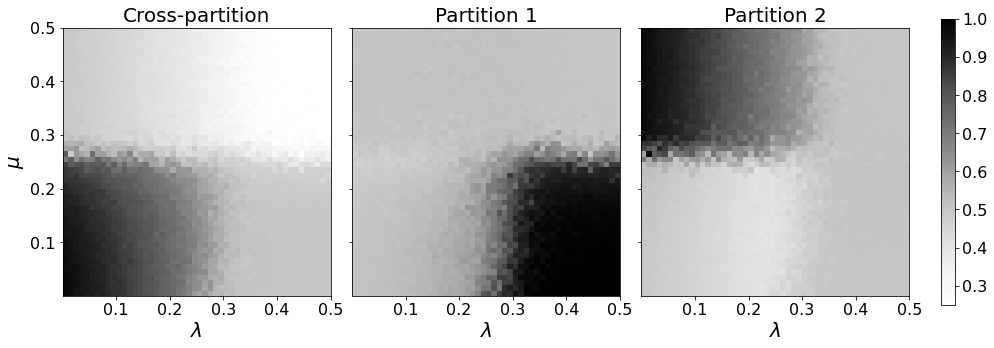}
    \caption{Partition overlap.}
\end{subfigure}
\begin{subfigure}{0.45\textwidth}
    \centering
    \includegraphics[width=1\textwidth]{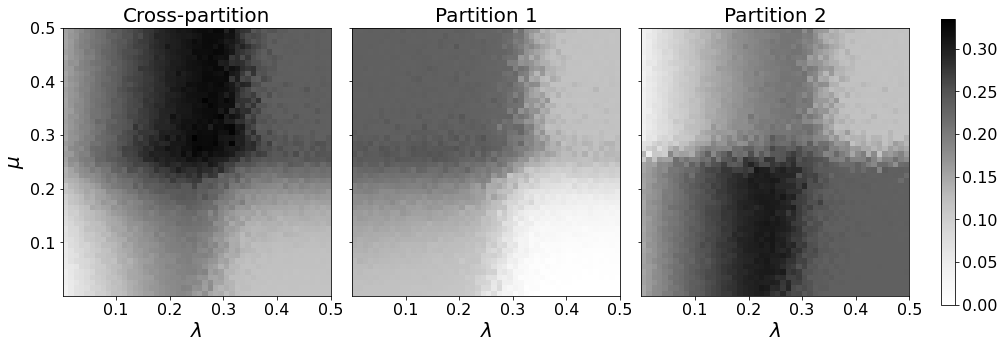}
    \caption{Variation of information.}
\end{subfigure}
\begin{subfigure}{0.45\textwidth}
    \centering
    \includegraphics[width=1\textwidth]{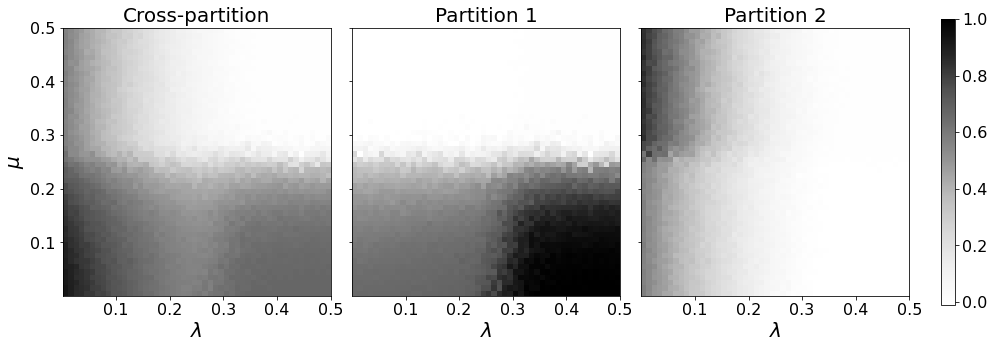}
    \caption{Reduced mutual information.}
\end{subfigure}
\caption{Mean partition overlap, variation of information and mutual information between the planted partitions and the partitions in the posterior distributions for graphs with $c = 10$.}
\label{fig:sfig_compare_similarity_measures}
\end{figure}

\section{\label{app:similaritymeasures} Alternative similarity measures}
To demonstrate the robustness of the partition overlap as our similarity measure, we compare it to two alternative measures, variation of information \citep{meila_comparing_2007} and reduced mutual information \citep{newman_improved_2020}. We use these two measures to calculate the partition similarity (or distance, in the case of variation of information, which is largest for partitions with the largest difference) for one particular set of simulations where $N=400$, $c = 10$ and where graphs were generated by the canonical model and partitions inferred using the non-degree-corrected SBM. We show the results in \autoref{fig:sfig_compare_similarity_measures}. While there are some subtle differences in the mean similarity values, the regions in which the detectability of the different partitions appears to change are located in the same areas of the $(\lambda, \mu)$ space.

\begin{figure}
\centering
\begin{subfigure}{0.4\textwidth}
    \centering
    \includegraphics[width=1.\linewidth]{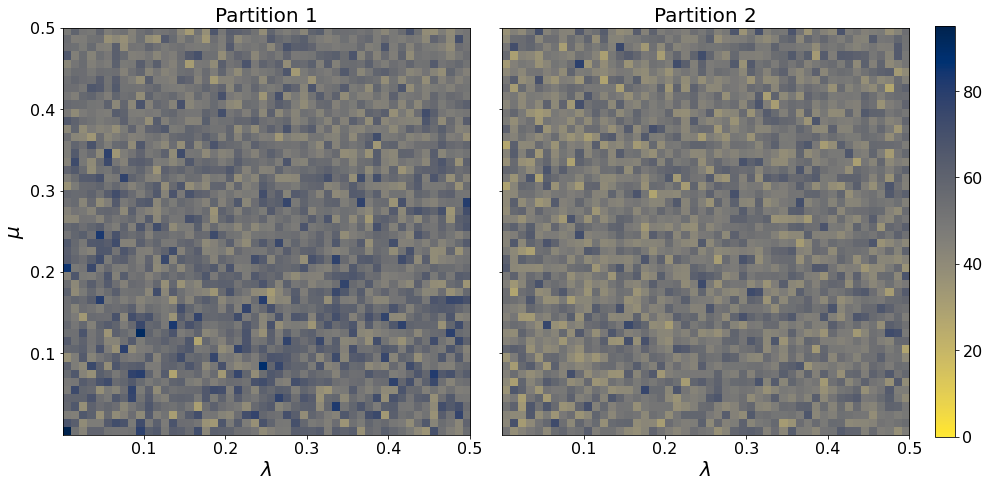}
    \caption{Canonical model.}
    \label{fig:sfig1_gen}
\end{subfigure}
\begin{subfigure}{0.4\textwidth}
    \centering
    \includegraphics[width=1.\linewidth]{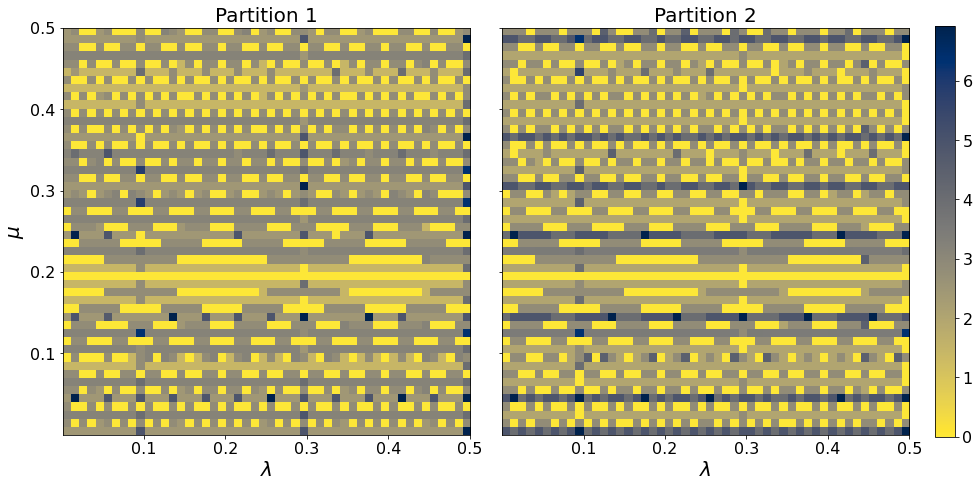}
    \caption{Microcanonical model.}
    \label{fig:sfig3_gen}
\end{subfigure}
\caption{Mean difference (Frobenius norm) between planted and generated edge count matrices in networks, for $c=5$.}
\label{fig:sfig_both_gen}
\end{figure}

\section{\label{app:generated} Characteristics of generated graphs}
To ensure that the variability in the recovered partitions is not in fact due to the graphs we generate, we compare the planted edge count matrices $\mathbf{B}_1$ and $\mathbf{B}_2$ with the actual edge counts in the generated graphs, $\mathbf{M}_1$ and $\mathbf{M}_2$ by calculating $||\mathbf{B}_1 - \mathbf{M}_1||_F$ and $||\mathbf{B}_2 - \mathbf{M}_2||_F$ for all values of $\mu$ and $\lambda$. \autoref{fig:sfig1_gen} shows the mean Frobenius norm for $\mu \in [0.01, 0.5]$ and $\lambda \in [0.01, 0.5]$ for graphs generated by the canonical model. \autoref{fig:sfig3_gen} shows the same plot for the microcanonical SBM. These figures show the distances for graphs with expected degree $c = 5$; we note that the patterns for the higher values of $c$ are similar. We observe that the distances between the planted and generated edge count matrices in the microcanonical case are -- by definition -- much lower than in the traditional case. The non-random patterns we observe in the microcanonical case are due to rounding that is necessary to create the edge count matrix $\mathbf{B}$ which is used to generate networks in this case (while in the canonical case the edge probabilities -- rather than counts -- are used).

\begin{figure}
      \centering
      \includegraphics[width=1.\linewidth]{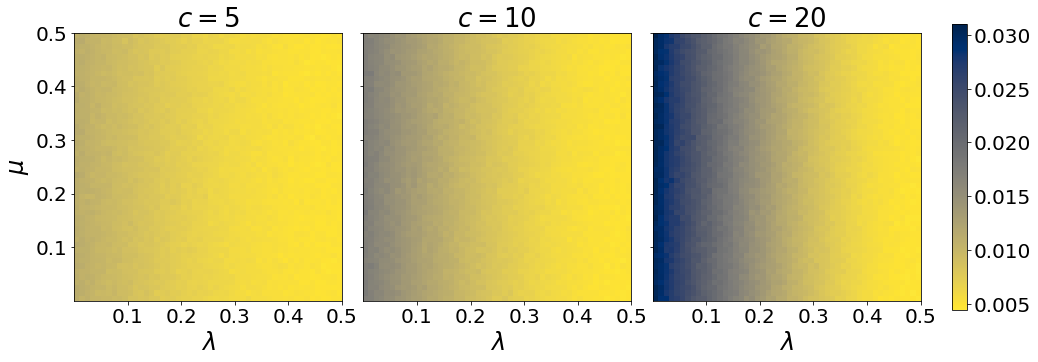}
      \caption{Mean normalised degree variance for graphs with varying expected degrees $c$, only showing the graphs generated by the canonical model to illustrate the degree variance introduced for low values of $\lambda$.}
      \label{fig:sfig_norm_deg_trad}
\end{figure}

\begin{figure}
      \centering
      \includegraphics[width=1.\linewidth]{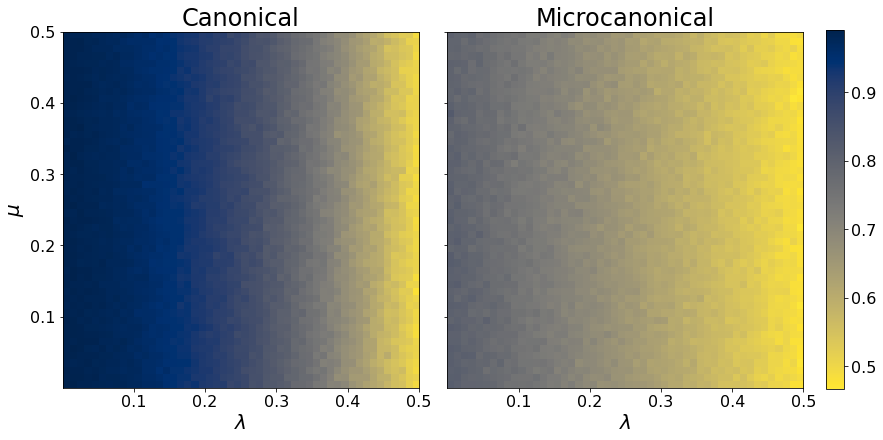}
      \caption{Proportion of nodes correctly classified into core and periphery blocks by assigning nodes with degree higher than the expected degree $c = 20$ to the core and all others to the periphery.}
      \label{fig:sfig_correctly_classified_node_deg}
\end{figure}

\begin{figure}
      \centering
      \includegraphics[width=1.\linewidth]{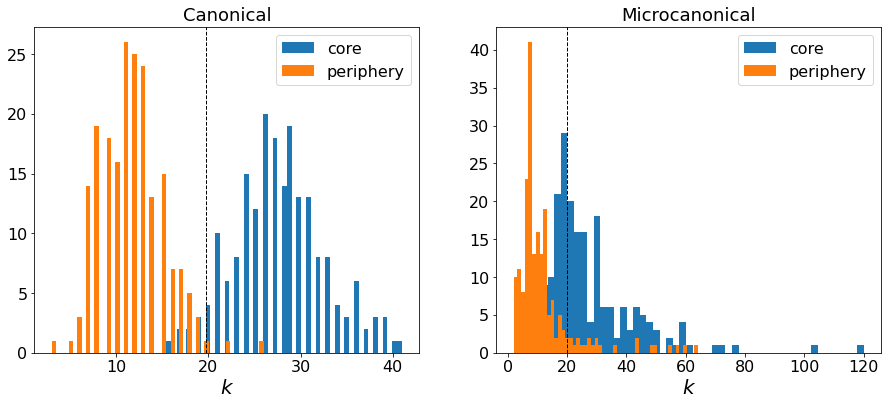}
      \caption{Degree distributions of core nodes vs periphery nodes in the canonical and microcanonical model with overall expected degree (dashed line), for $\mu=0.1$, $\lambda=0.1$ and $c = 20$}
      \label{fig:sfig_block_degree_dist}
\end{figure}

\autoref{fig:sfig_norm_deg_trad} shows the rescaled version of \autoref{fig:sfig_block_differences_norm_deg} in the main text. Here, we plot the mean normalised degree variance for graphs generated by the canonical model, to illustrate the higher variance introduced for lower values of $\lambda$, especially for higher values of $c$.

In \autoref{fig:sfig_correctly_classified_node_deg}, we plot the proportion of nodes correctly classified (according to the planted core and periphery blocks) by assigning nodes with above average degree to the core and those with below average degree to the periphery, for graphs with expected degree $c = 20$. In \autoref{fig:sfig_block_degree_dist}, we show two example degree distributions for $\mu=0.1$, $\lambda=0.1$ and $c = 20$.

\begin{figure}
\centering
\begin{subfigure}{0.45\textwidth}
    \centering
    \includegraphics[width=1\textwidth]{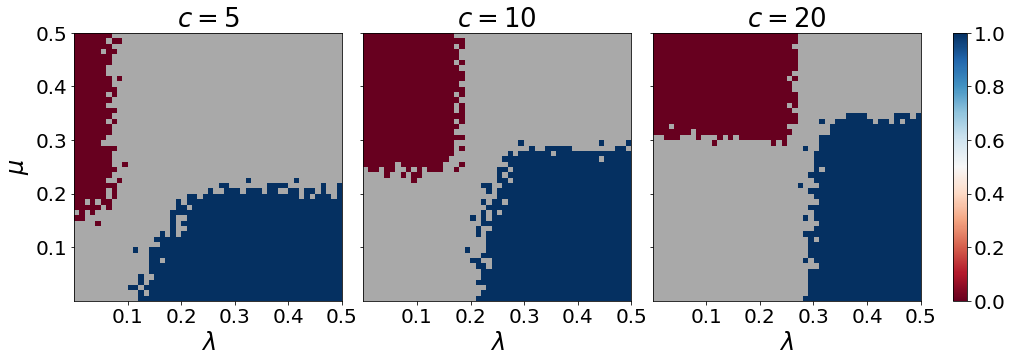}
    \caption{NDC, Canonical model.}
\end{subfigure}
\begin{subfigure}{0.45\textwidth}
    \centering
    \includegraphics[width=1\textwidth]{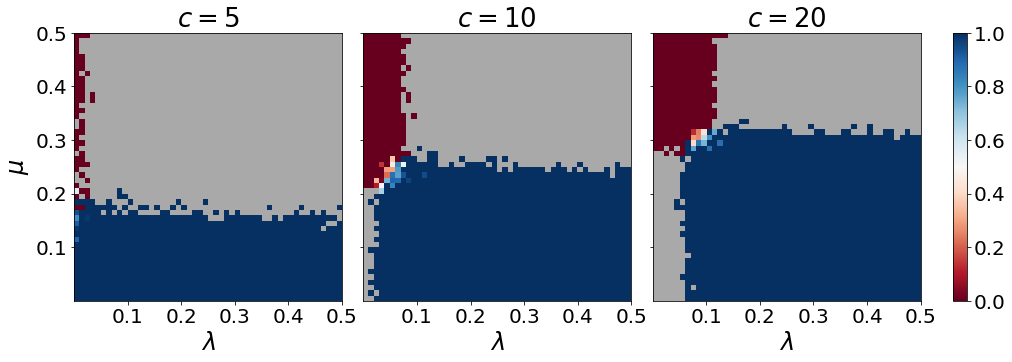}
    \caption{DC, Canonical model.}
\end{subfigure}
\begin{subfigure}{0.45\textwidth}
    \centering
    \includegraphics[width=1\textwidth]{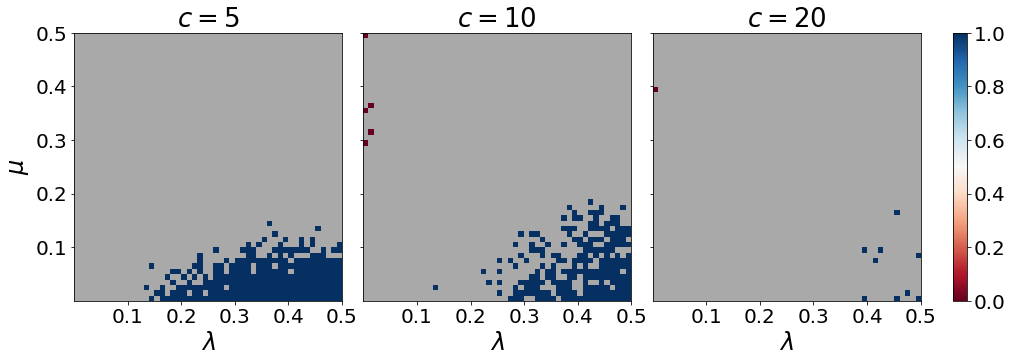}
    \caption{NDC, Microcanonical model.}
\end{subfigure}
\begin{subfigure}{0.45\textwidth}
    \centering
    \includegraphics[width=1\textwidth]{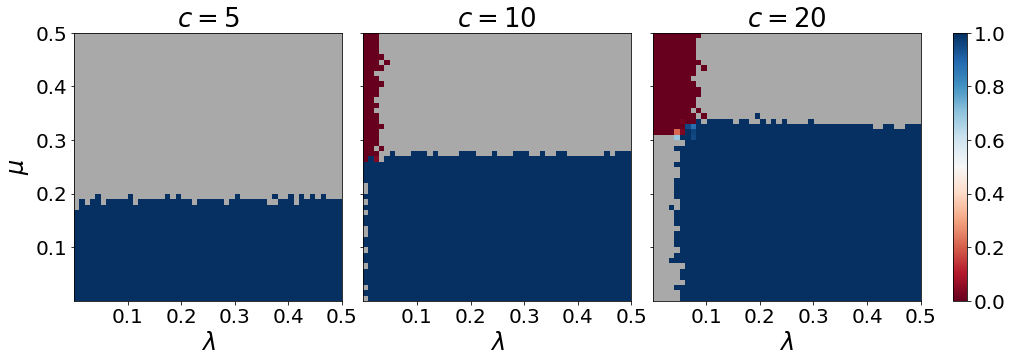}
    \caption{DC, Microcanonical model.}
\end{subfigure}
\caption{Fraction $\alpha$ of recovered bi-community partitions out of all successfully recovered partitions for varying $c$ for $\omega=0.85$; at $\alpha=1$ (resp. $\alpha=0$) only the bi-community (resp. CP) structure is detected.}
\label{fig:sfig_both_recovered_ratio_085}
\end{figure}

\section{\label{app:extrathresholds} Stricter partition-overlap thresholds}
\begin{figure}
\centering
\begin{subfigure}{0.45\textwidth}
    \centering
    \includegraphics[width=1\textwidth]{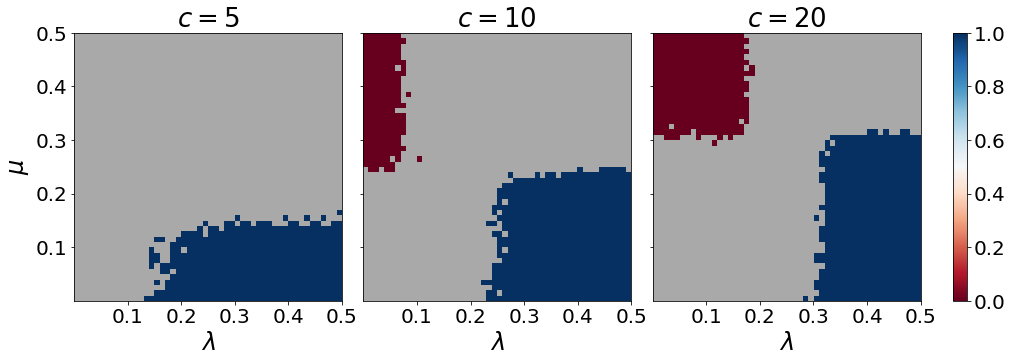}
    \caption{NDC, Canonical model.}
\end{subfigure}
\begin{subfigure}{0.45\textwidth}
    \centering
    \includegraphics[width=1\textwidth]{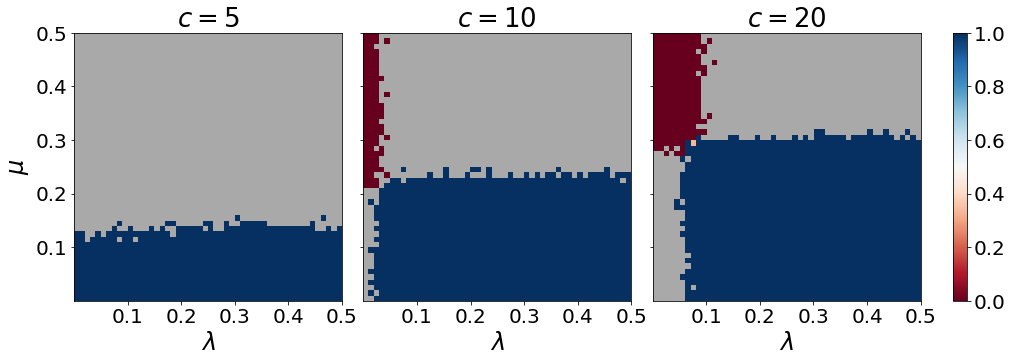}
    \caption{DC, Canonical model.}
\end{subfigure}

\begin{subfigure}{0.45\textwidth}
    \centering
    \includegraphics[width=1\textwidth]{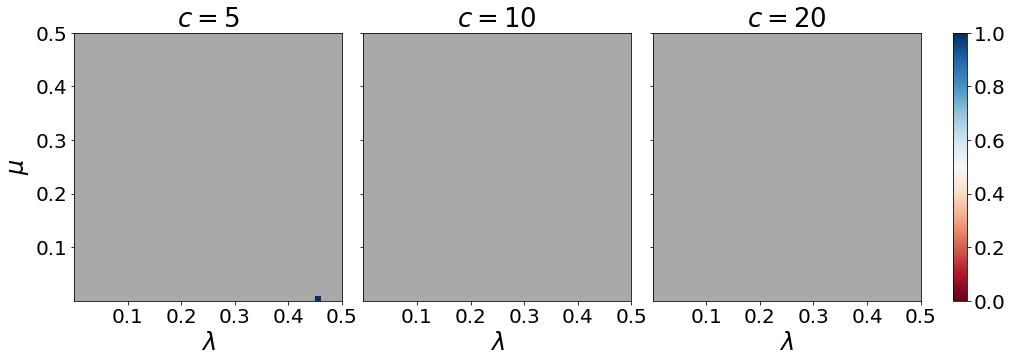}
    \caption{NDC, Microcanonical model.}
\end{subfigure}
\begin{subfigure}{0.45\textwidth}
    \centering
    \includegraphics[width=1\textwidth]{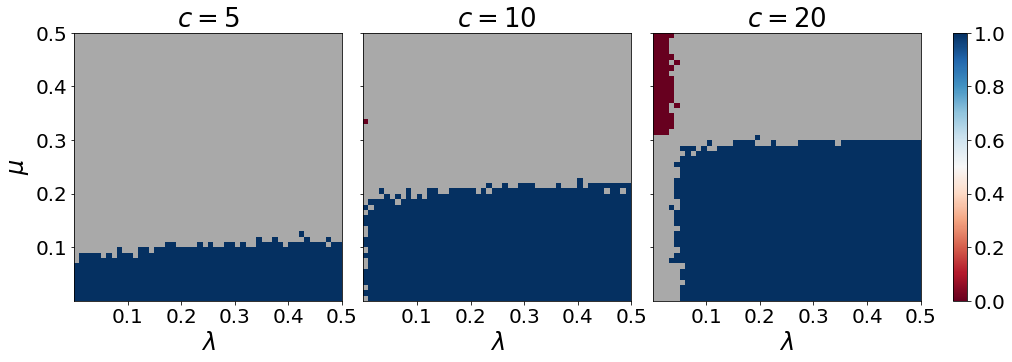}
    \caption{DC, Microcanonical model.}
\end{subfigure}
\caption{Fraction $\alpha$ of recovered bi-community partitions out of all successfully recovered partitions for varying $c$ for $\omega=0.95$; at $\alpha=1$ (resp. $\alpha=0$) only the bi-community (resp. CP) structure is detected.}
\label{fig:sfig_both_recovered_ratio_095}
\end{figure}

\autoref{fig:sfig_both_recovered_ratio_085} and \autoref{fig:sfig_both_recovered_ratio_095} illustrate the detection of the bi-community and CP structures, as well as their coexistence, for the two SBM variants and for partition overlap $\omega=0.85$ and $\omega=0.95$ respectively.

\section{\label{app:exnetworks} Example networks with high degree heterogeneity}
\begin{figure}
\centering
\begin{subfigure}{0.48\columnwidth}
    \centering
    \includegraphics[width=\linewidth]{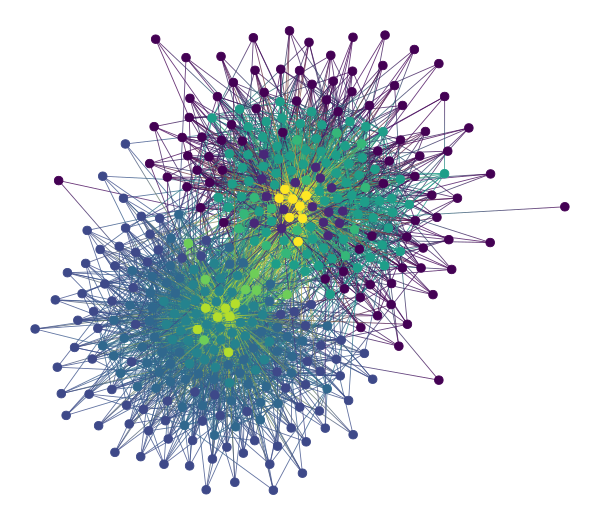}
    \caption{$\mu=0.1$, $\lambda=0.01$, NDC}
    \label{fig:sfig_examplenetworks1_ndc}
\end{subfigure}
\begin{subfigure}{0.48\columnwidth}
    \centering
    \includegraphics[width=\linewidth]{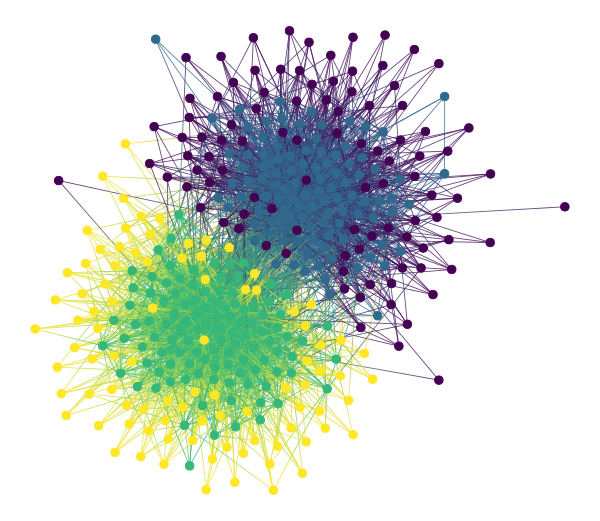}
    \caption{$\mu=0.1$, $\lambda=0.01$, DC}
    \label{fig:sfig_examplenetworks1_dc}
\end{subfigure}

\begin{subfigure}{0.48\columnwidth}
    \centering
    \includegraphics[width=\linewidth]{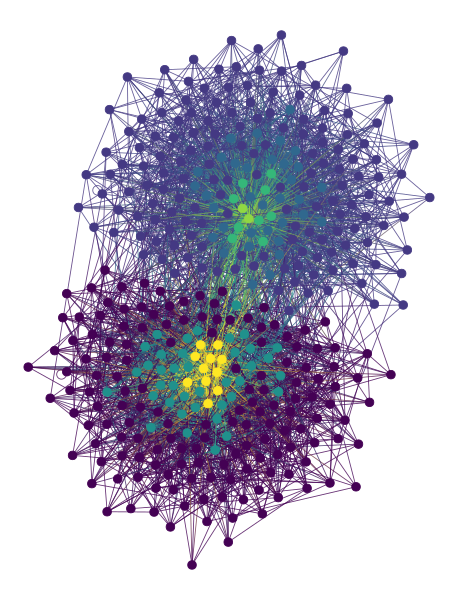}
    \caption{$\mu=0.1$, $\lambda=0.5$, NDC}
    \label{fig:sfig_examplenetworks2_ndc}
\end{subfigure}
\begin{subfigure}{0.48\columnwidth}
    \centering
    \includegraphics[width=\linewidth]{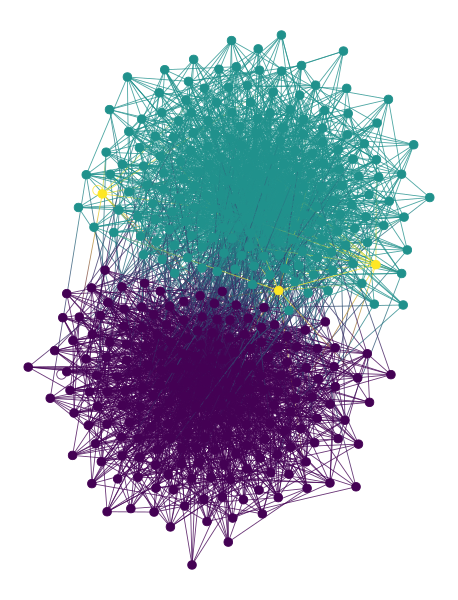}
    \caption{$\mu=0.1$, $\lambda=0.5$, DC}
    \label{fig:sfig_examplenetworks2_dc}
\end{subfigure}
\caption{Example networks for a fixed value of $\mu=0.1$ and varying value of $\lambda=0.01; 0.5$.}
\label{fig:sfig_examplenetworks}
\end{figure}

In \autoref{fig:sfig_examplenetworks}, we plot two example networks generated by the microcanonical model. The network visualisations were generated by the graph-tool python library \citep{peixoto_graph-tool_2014}. We fix $\mu=0.1$ for both networks and we create one graph with a strong planted CP structure ($\lambda=0.01$) and one for which no CP structure at all is planted explicitly through the edge count matrices ($\lambda=0.5$). \autoref{fig:sfig_examplenetworks1_ndc} and \autoref{fig:sfig_examplenetworks2_ndc} show an example of the type of partition frequently recovered by NDC for $\lambda=0.01$ and $\lambda=0.5$ respectively. \autoref{fig:sfig_examplenetworks1_dc} and \autoref{fig:sfig_examplenetworks2_dc} show the same but for the DC variant. We observe that DC recovers the cross-partition for $\lambda=0.01$ and the bi-community partition for $\lambda=0.5$, in line with the equivalent results for graphs generated by the microcanonical model and with what we explicitly planted. NDC, however, (which has higher model evidence) detects a similar structure for $\lambda=0.01$ and $\lambda=0.5$: a two-block partition, where each block contains a core and multiple layered peripheries. The difference between the two detected partitions is the number of layers in the core-periphery structures within each block and the size of the outer periphery. Due to these differences, the partition detected for $\lambda=0.01$ is more similar to the cross-partition, while that detected for $\lambda=0.5$ is more similar to the bi-community partition. 

\clearpage

\bibliography{bibliography}

\begin{thebibliography}{70}%
\makeatletter
\providecommand \@ifxundefined [1]{%
 \@ifx{#1\undefined}
}%
\providecommand \@ifnum [1]{%
 \ifnum #1\expandafter \@firstoftwo
 \else \expandafter \@secondoftwo
 \fi
}%
\providecommand \@ifx [1]{%
 \ifx #1\expandafter \@firstoftwo
 \else \expandafter \@secondoftwo
 \fi
}%
\providecommand \natexlab [1]{#1}%
\providecommand \enquote  [1]{``#1''}%
\providecommand \bibnamefont  [1]{#1}%
\providecommand \bibfnamefont [1]{#1}%
\providecommand \citenamefont [1]{#1}%
\providecommand \href@noop [0]{\@secondoftwo}%
\providecommand \href [0]{\begingroup \@sanitize@url \@href}%
\providecommand \@href[1]{\@@startlink{#1}\@@href}%
\providecommand \@@href[1]{\endgroup#1\@@endlink}%
\providecommand \@sanitize@url [0]{\catcode `\\12\catcode `\$12\catcode
  `\&12\catcode `\#12\catcode `\^12\catcode `\_12\catcode `\%12\relax}%
\providecommand \@@startlink[1]{}%
\providecommand \@@endlink[0]{}%
\providecommand \url  [0]{\begingroup\@sanitize@url \@url }%
\providecommand \@url [1]{\endgroup\@href {#1}{\urlprefix }}%
\providecommand \urlprefix  [0]{URL }%
\providecommand \Eprint [0]{\href }%
\providecommand \doibase [0]{https://doi.org/}%
\providecommand \selectlanguage [0]{\@gobble}%
\providecommand \bibinfo  [0]{\@secondoftwo}%
\providecommand \bibfield  [0]{\@secondoftwo}%
\providecommand \translation [1]{[#1]}%
\providecommand \BibitemOpen [0]{}%
\providecommand \bibitemStop [0]{}%
\providecommand \bibitemNoStop [0]{.\EOS\space}%
\providecommand \EOS [0]{\spacefactor3000\relax}%
\providecommand \BibitemShut  [1]{\csname bibitem#1\endcsname}%
\let\auto@bib@innerbib\@empty
\bibitem [{\citenamefont {Leskovec}\ \emph {et~al.}(2009)\citenamefont
  {Leskovec}, \citenamefont {Lang}, \citenamefont {Dasgupta},\ and\
  \citenamefont {Mahoney}}]{leskovec_community_2009}%
  \BibitemOpen
  \bibfield  {author} {\bibinfo {author} {\bibfnamefont {J.}~\bibnamefont
  {Leskovec}}, \bibinfo {author} {\bibfnamefont {K.~J.}\ \bibnamefont {Lang}},
  \bibinfo {author} {\bibfnamefont {A.}~\bibnamefont {Dasgupta}},\ and\
  \bibinfo {author} {\bibfnamefont {M.~W.}\ \bibnamefont {Mahoney}},\ }\href
  {https://doi.org/10.1080/15427951.2009.10129177} {\bibfield  {journal}
  {\bibinfo  {journal} {Internet Mathematics}\ }\textbf {\bibinfo {volume}
  {6}},\ \bibinfo {pages} {29} (\bibinfo {year} {2009})}\BibitemShut {NoStop}%
\bibitem [{\citenamefont {Rosvall}\ \emph {et~al.}(2019)\citenamefont
  {Rosvall}, \citenamefont {Delvenne}, \citenamefont {Schaub},\ and\
  \citenamefont {Lambiotte}}]{rosvall_different_2019-1}%
  \BibitemOpen
  \bibfield  {author} {\bibinfo {author} {\bibfnamefont {M.}~\bibnamefont
  {Rosvall}}, \bibinfo {author} {\bibfnamefont {J.-C.}\ \bibnamefont
  {Delvenne}}, \bibinfo {author} {\bibfnamefont {M.~T.}\ \bibnamefont
  {Schaub}},\ and\ \bibinfo {author} {\bibfnamefont {R.}~\bibnamefont
  {Lambiotte}},\ }in\ \href {https://doi.org/10.1002/9781119483298.ch4} {\emph
  {\bibinfo {booktitle} {Advances in {{Network Clustering}} and
  {{Blockmodeling}}}}}\ (\bibinfo  {publisher} {{John Wiley \& Sons, Ltd}},\
  \bibinfo {year} {2019})\ Chap.~\bibinfo {chapter} {4}, pp.\ \bibinfo {pages}
  {105--119}\BibitemShut {NoStop}%
\bibitem [{\citenamefont {Rombach}\ \emph {et~al.}(2017)\citenamefont
  {Rombach}, \citenamefont {Porter}, \citenamefont {Fowler},\ and\
  \citenamefont {Mucha}}]{rombach_core-periphery_2017}%
  \BibitemOpen
  \bibfield  {author} {\bibinfo {author} {\bibfnamefont {P.}~\bibnamefont
  {Rombach}}, \bibinfo {author} {\bibfnamefont {M.~A.}\ \bibnamefont {Porter}},
  \bibinfo {author} {\bibfnamefont {J.~H.}\ \bibnamefont {Fowler}},\ and\
  \bibinfo {author} {\bibfnamefont {P.~J.}\ \bibnamefont {Mucha}},\ }\href
  {https://doi.org/10.1137/17M1130046} {\bibfield  {journal} {\bibinfo
  {journal} {SIAM Review}\ }\textbf {\bibinfo {volume} {59}},\ \bibinfo {pages}
  {619} (\bibinfo {year} {2017})}\BibitemShut {NoStop}%
\bibitem [{\citenamefont {Zhang}\ \emph {et~al.}(2015)\citenamefont {Zhang},
  \citenamefont {Martin},\ and\ \citenamefont
  {Newman}}]{zhang_identification_2015}%
  \BibitemOpen
  \bibfield  {author} {\bibinfo {author} {\bibfnamefont {X.}~\bibnamefont
  {Zhang}}, \bibinfo {author} {\bibfnamefont {T.}~\bibnamefont {Martin}},\ and\
  \bibinfo {author} {\bibfnamefont {M.~E.~J.}\ \bibnamefont {Newman}},\ }\href
  {https://doi.org/10.1103/PhysRevE.91.032803} {\bibfield  {journal} {\bibinfo
  {journal} {Physical Review E}\ }\textbf {\bibinfo {volume} {91}},\ \bibinfo
  {pages} {032803} (\bibinfo {year} {2015})}\BibitemShut {NoStop}%
\bibitem [{\citenamefont {Zachary}(1977)}]{zachary_information_1977}%
  \BibitemOpen
  \bibfield  {author} {\bibinfo {author} {\bibfnamefont {W.~W.}\ \bibnamefont
  {Zachary}},\ }\href {https://doi.org/10.1086/jar.33.4.3629752} {\bibfield
  {journal} {\bibinfo  {journal} {Journal of Anthropological Research}\
  }\textbf {\bibinfo {volume} {33}},\ \bibinfo {pages} {452} (\bibinfo {year}
  {1977})}\BibitemShut {NoStop}%
\bibitem [{\citenamefont {Newman}\ and\ \citenamefont
  {Girvan}(2004)}]{newman_finding_2004}%
  \BibitemOpen
  \bibfield  {author} {\bibinfo {author} {\bibfnamefont {M.~E.~J.}\
  \bibnamefont {Newman}}\ and\ \bibinfo {author} {\bibfnamefont
  {M.}~\bibnamefont {Girvan}},\ }\href
  {https://doi.org/10.1103/PhysRevE.69.026113} {\bibfield  {journal} {\bibinfo
  {journal} {Physical Review E}\ }\textbf {\bibinfo {volume} {69}},\ \bibinfo
  {pages} {026113} (\bibinfo {year} {2004})}\BibitemShut {NoStop}%
\bibitem [{\citenamefont {Newman}(2006)}]{newman_modularity_2006}%
  \BibitemOpen
  \bibfield  {author} {\bibinfo {author} {\bibfnamefont {M.~E.~J.}\
  \bibnamefont {Newman}},\ }\href {https://doi.org/10.1073/pnas.0601602103}
  {\bibfield  {journal} {\bibinfo  {journal} {Proceedings of the National
  Academy of Sciences}\ }\textbf {\bibinfo {volume} {103}},\ \bibinfo {pages}
  {8577} (\bibinfo {year} {2006})}\BibitemShut {NoStop}%
\bibitem [{\citenamefont {Duch}\ and\ \citenamefont
  {Arenas}(2005)}]{duch_community_2005}%
  \BibitemOpen
  \bibfield  {author} {\bibinfo {author} {\bibfnamefont {J.}~\bibnamefont
  {Duch}}\ and\ \bibinfo {author} {\bibfnamefont {A.}~\bibnamefont {Arenas}},\
  }\href {https://doi.org/10.1103/PhysRevE.72.027104} {\bibfield  {journal}
  {\bibinfo  {journal} {Physical Review E}\ }\textbf {\bibinfo {volume} {72}},\
  \bibinfo {pages} {027104} (\bibinfo {year} {2005})}\BibitemShut {NoStop}%
\bibitem [{\citenamefont {Blondel}\ \emph {et~al.}(2008)\citenamefont
  {Blondel}, \citenamefont {Guillaume}, \citenamefont {Lambiotte},\ and\
  \citenamefont {Lefebvre}}]{blondel_fast_2008}%
  \BibitemOpen
  \bibfield  {author} {\bibinfo {author} {\bibfnamefont {V.~D.}\ \bibnamefont
  {Blondel}}, \bibinfo {author} {\bibfnamefont {J.-L.}\ \bibnamefont
  {Guillaume}}, \bibinfo {author} {\bibfnamefont {R.}~\bibnamefont
  {Lambiotte}},\ and\ \bibinfo {author} {\bibfnamefont {E.}~\bibnamefont
  {Lefebvre}},\ }\href {https://doi.org/10.1088/1742-5468/2008/10/P10008}
  {\bibfield  {journal} {\bibinfo  {journal} {Journal of Statistical Mechanics:
  Theory and Experiment}\ }\textbf {\bibinfo {volume} {2008}},\ \bibinfo
  {pages} {P10008} (\bibinfo {year} {2008})}\BibitemShut {NoStop}%
\bibitem [{\citenamefont {Evans}(2010)}]{evans_clique_2010}%
  \BibitemOpen
  \bibfield  {author} {\bibinfo {author} {\bibfnamefont {T.~S.}\ \bibnamefont
  {Evans}},\ }\href {https://doi.org/10.1088/1742-5468/2010/12/P12037}
  {\bibfield  {journal} {\bibinfo  {journal} {Journal of Statistical Mechanics:
  Theory and Experiment}\ }\textbf {\bibinfo {volume} {2010}},\ \bibinfo
  {pages} {P12037} (\bibinfo {year} {2010})}\BibitemShut {NoStop}%
\bibitem [{\citenamefont {Peixoto}(2021)}]{peixoto_revealing_2021}%
  \BibitemOpen
  \bibfield  {author} {\bibinfo {author} {\bibfnamefont {T.~P.}\ \bibnamefont
  {Peixoto}},\ }\href {https://doi.org/10.1103/PhysRevX.11.021003} {\bibfield
  {journal} {\bibinfo  {journal} {Physical Review X}\ }\textbf {\bibinfo
  {volume} {11}},\ \bibinfo {pages} {021003} (\bibinfo {year}
  {2021})}\BibitemShut {NoStop}%
\bibitem [{\citenamefont {Peel}\ \emph {et~al.}(2017)\citenamefont {Peel},
  \citenamefont {Larremore},\ and\ \citenamefont {Clauset}}]{peel_ground_2017}%
  \BibitemOpen
  \bibfield  {author} {\bibinfo {author} {\bibfnamefont {L.}~\bibnamefont
  {Peel}}, \bibinfo {author} {\bibfnamefont {D.~B.}\ \bibnamefont
  {Larremore}},\ and\ \bibinfo {author} {\bibfnamefont {A.}~\bibnamefont
  {Clauset}},\ }\href {https://doi.org/10.1126/sciadv.1602548} {\bibfield
  {journal} {\bibinfo  {journal} {Science Advances}\ }\textbf {\bibinfo
  {volume} {3}},\ \bibinfo {pages} {e1602548} (\bibinfo {year}
  {2017})}\BibitemShut {NoStop}%
\bibitem [{\citenamefont {Decelle}\ \emph
  {et~al.}(2011{\natexlab{a}})\citenamefont {Decelle}, \citenamefont
  {Krzakala}, \citenamefont {Moore},\ and\ \citenamefont
  {Zdeborov{\'a}}}]{decelle_asymptotic_2011}%
  \BibitemOpen
  \bibfield  {author} {\bibinfo {author} {\bibfnamefont {A.}~\bibnamefont
  {Decelle}}, \bibinfo {author} {\bibfnamefont {F.}~\bibnamefont {Krzakala}},
  \bibinfo {author} {\bibfnamefont {C.}~\bibnamefont {Moore}},\ and\ \bibinfo
  {author} {\bibfnamefont {L.}~\bibnamefont {Zdeborov{\'a}}},\ }\href
  {https://doi.org/10.1103/PhysRevE.84.066106} {\bibfield  {journal} {\bibinfo
  {journal} {Physical Review E}\ }\textbf {\bibinfo {volume} {84}},\ \bibinfo
  {pages} {066106} (\bibinfo {year} {2011}{\natexlab{a}})}\BibitemShut
  {NoStop}%
\bibitem [{\citenamefont {Decelle}\ \emph
  {et~al.}(2011{\natexlab{b}})\citenamefont {Decelle}, \citenamefont
  {Krzakala}, \citenamefont {Moore},\ and\ \citenamefont
  {Zdeborov{\'a}}}]{decelle_inference_2011}%
  \BibitemOpen
  \bibfield  {author} {\bibinfo {author} {\bibfnamefont {A.}~\bibnamefont
  {Decelle}}, \bibinfo {author} {\bibfnamefont {F.}~\bibnamefont {Krzakala}},
  \bibinfo {author} {\bibfnamefont {C.}~\bibnamefont {Moore}},\ and\ \bibinfo
  {author} {\bibfnamefont {L.}~\bibnamefont {Zdeborov{\'a}}},\ }\href
  {https://doi.org/10.1103/PhysRevLett.107.065701} {\bibfield  {journal}
  {\bibinfo  {journal} {Physical Review Letters}\ }\textbf {\bibinfo {volume}
  {107}},\ \bibinfo {pages} {065701} (\bibinfo {year}
  {2011}{\natexlab{b}})}\BibitemShut {NoStop}%
\bibitem [{\citenamefont {Condon}\ and\ \citenamefont
  {Karp}(2001)}]{condon_algorithms_2001}%
  \BibitemOpen
  \bibfield  {author} {\bibinfo {author} {\bibfnamefont {A.}~\bibnamefont
  {Condon}}\ and\ \bibinfo {author} {\bibfnamefont {R.~M.}\ \bibnamefont
  {Karp}},\ }\href
  {https://doi.org/10.1002/1098-2418(200103)18:2<116::AID-RSA1001>3.0.CO;2-2}
  {\bibfield  {journal} {\bibinfo  {journal} {Random Structures \& Algorithms}\
  }\textbf {\bibinfo {volume} {18}},\ \bibinfo {pages} {116} (\bibinfo {year}
  {2001})}\BibitemShut {NoStop}%
\bibitem [{\citenamefont {Alba}(1973)}]{alba_graph-theoretic_1973}%
  \BibitemOpen
  \bibfield  {author} {\bibinfo {author} {\bibfnamefont {R.~D.}\ \bibnamefont
  {Alba}},\ }\href {https://doi.org/10.1080/0022250X.1973.9989826} {\bibfield
  {journal} {\bibinfo  {journal} {Journal of Mathematical Sociology}\ }\textbf
  {\bibinfo {volume} {3}},\ \bibinfo {pages} {113} (\bibinfo {year}
  {1973})}\BibitemShut {NoStop}%
\bibitem [{\citenamefont {Fortunato}(2010)}]{fortunato_community_2010}%
  \BibitemOpen
  \bibfield  {author} {\bibinfo {author} {\bibfnamefont {S.}~\bibnamefont
  {Fortunato}},\ }\href {https://doi.org/10.1016/j.physrep.2009.11.002}
  {\bibfield  {journal} {\bibinfo  {journal} {Physics Reports}\ }\textbf
  {\bibinfo {volume} {486}},\ \bibinfo {pages} {75} (\bibinfo {year}
  {2010})}\BibitemShut {NoStop}%
\bibitem [{\citenamefont {Newman}(2013)}]{newman_spectral_2013}%
  \BibitemOpen
  \bibfield  {author} {\bibinfo {author} {\bibfnamefont {M.~E.~J.}\
  \bibnamefont {Newman}},\ }\href {https://doi.org/10.1103/PhysRevE.88.042822}
  {\bibfield  {journal} {\bibinfo  {journal} {Physical Review E}\ }\textbf
  {\bibinfo {volume} {88}},\ \bibinfo {pages} {042822} (\bibinfo {year}
  {2013})}\BibitemShut {NoStop}%
\bibitem [{\citenamefont {Karrer}\ and\ \citenamefont
  {Newman}(2011)}]{karrer_stochastic_2011}%
  \BibitemOpen
  \bibfield  {author} {\bibinfo {author} {\bibfnamefont {B.}~\bibnamefont
  {Karrer}}\ and\ \bibinfo {author} {\bibfnamefont {M.~E.~J.}\ \bibnamefont
  {Newman}},\ }\href {https://doi.org/10.1103/PhysRevE.83.016107} {\bibfield
  {journal} {\bibinfo  {journal} {Physical Review E}\ }\textbf {\bibinfo
  {volume} {83}},\ \bibinfo {pages} {016107} (\bibinfo {year}
  {2011})}\BibitemShut {NoStop}%
\bibitem [{\citenamefont {Rosvall}\ and\ \citenamefont
  {Bergstrom}(2008)}]{rosvall_maps_2008}%
  \BibitemOpen
  \bibfield  {author} {\bibinfo {author} {\bibfnamefont {M.}~\bibnamefont
  {Rosvall}}\ and\ \bibinfo {author} {\bibfnamefont {C.~T.}\ \bibnamefont
  {Bergstrom}},\ }\href {https://doi.org/10.1073/pnas.0706851105} {\bibfield
  {journal} {\bibinfo  {journal} {Proceedings of the National Academy of
  Sciences}\ }\textbf {\bibinfo {volume} {105}},\ \bibinfo {pages} {1118}
  (\bibinfo {year} {2008})}\BibitemShut {NoStop}%
\bibitem [{\citenamefont {Borgatti}\ and\ \citenamefont
  {Everett}(2000)}]{borgatti_models_2000}%
  \BibitemOpen
  \bibfield  {author} {\bibinfo {author} {\bibfnamefont {S.~P.}\ \bibnamefont
  {Borgatti}}\ and\ \bibinfo {author} {\bibfnamefont {M.~G.}\ \bibnamefont
  {Everett}},\ }\href@noop {} {\bibfield  {journal} {\bibinfo  {journal}
  {Social networks}\ }\textbf {\bibinfo {volume} {21}},\ \bibinfo {pages} {375}
  (\bibinfo {year} {2000})}\BibitemShut {NoStop}%
\bibitem [{\citenamefont {Holme}(2005)}]{holme_core-periphery_2005}%
  \BibitemOpen
  \bibfield  {author} {\bibinfo {author} {\bibfnamefont {P.}~\bibnamefont
  {Holme}},\ }\href@noop {} {\bibfield  {journal} {\bibinfo  {journal}
  {Physical Review E}\ }\textbf {\bibinfo {volume} {72}},\ \bibinfo {pages}
  {046111} (\bibinfo {year} {2005})}\BibitemShut {NoStop}%
\bibitem [{\citenamefont {Lee}\ \emph {et~al.}(2014)\citenamefont {Lee},
  \citenamefont {Cucuringu},\ and\ \citenamefont
  {Porter}}]{lee_density-based_2014}%
  \BibitemOpen
  \bibfield  {author} {\bibinfo {author} {\bibfnamefont {S.~H.}\ \bibnamefont
  {Lee}}, \bibinfo {author} {\bibfnamefont {M.}~\bibnamefont {Cucuringu}},\
  and\ \bibinfo {author} {\bibfnamefont {M.~A.}\ \bibnamefont {Porter}},\
  }\href@noop {} {\bibfield  {journal} {\bibinfo  {journal} {Physical Review
  E}\ }\textbf {\bibinfo {volume} {89}},\ \bibinfo {pages} {032810} (\bibinfo
  {year} {2014})}\BibitemShut {NoStop}%
\bibitem [{\citenamefont {Cucuringu}\ \emph {et~al.}(2016)\citenamefont
  {Cucuringu}, \citenamefont {Rombach}, \citenamefont {Lee},\ and\
  \citenamefont {Porter}}]{cucuringu_detection_2016}%
  \BibitemOpen
  \bibfield  {author} {\bibinfo {author} {\bibfnamefont {M.}~\bibnamefont
  {Cucuringu}}, \bibinfo {author} {\bibfnamefont {P.}~\bibnamefont {Rombach}},
  \bibinfo {author} {\bibfnamefont {S.~H.}\ \bibnamefont {Lee}},\ and\ \bibinfo
  {author} {\bibfnamefont {M.~A.}\ \bibnamefont {Porter}},\ }\href@noop {}
  {\bibfield  {journal} {\bibinfo  {journal} {European Journal of Applied
  Mathematics}\ }\textbf {\bibinfo {volume} {27}},\ \bibinfo {pages} {846}
  (\bibinfo {year} {2016})}\BibitemShut {NoStop}%
\bibitem [{\citenamefont {Gallagher}\ \emph {et~al.}(2021)\citenamefont
  {Gallagher}, \citenamefont {Young},\ and\ \citenamefont
  {Welles}}]{gallagher_clarified_2021}%
  \BibitemOpen
  \bibfield  {author} {\bibinfo {author} {\bibfnamefont {R.~J.}\ \bibnamefont
  {Gallagher}}, \bibinfo {author} {\bibfnamefont {J.-G.}\ \bibnamefont
  {Young}},\ and\ \bibinfo {author} {\bibfnamefont {B.~F.}\ \bibnamefont
  {Welles}},\ }\href {https://doi.org/10.1126/sciadv.abc9800} {\bibfield
  {journal} {\bibinfo  {journal} {Science Advances}\ }\textbf {\bibinfo
  {volume} {7}},\ \bibinfo {pages} {eabc9800} (\bibinfo {year}
  {2021})}\BibitemShut {NoStop}%
\bibitem [{\citenamefont {Yan}\ and\ \citenamefont
  {Luo}(2019)}]{yan_multicores-periphery_2019}%
  \BibitemOpen
  \bibfield  {author} {\bibinfo {author} {\bibfnamefont {B.}~\bibnamefont
  {Yan}}\ and\ \bibinfo {author} {\bibfnamefont {J.}~\bibnamefont {Luo}},\
  }\href@noop {} {\bibfield  {journal} {\bibinfo  {journal} {Network Science}\
  }\textbf {\bibinfo {volume} {7}},\ \bibinfo {pages} {70} (\bibinfo {year}
  {2019})}\BibitemShut {NoStop}%
\bibitem [{\citenamefont {Tun{\c{c}}}\ and\ \citenamefont
  {Verma}(2015)}]{tunc_unifying_2015}%
  \BibitemOpen
  \bibfield  {author} {\bibinfo {author} {\bibfnamefont {B.}~\bibnamefont
  {Tun{\c{c}}}}\ and\ \bibinfo {author} {\bibfnamefont {R.}~\bibnamefont
  {Verma}},\ }\href@noop {} {\bibfield  {journal} {\bibinfo  {journal} {PloS
  one}\ }\textbf {\bibinfo {volume} {10}},\ \bibinfo {pages} {e0143133}
  (\bibinfo {year} {2015})}\BibitemShut {NoStop}%
\bibitem [{\citenamefont {Kojaku}\ and\ \citenamefont
  {Masuda}(2017)}]{kojaku_finding_2017}%
  \BibitemOpen
  \bibfield  {author} {\bibinfo {author} {\bibfnamefont {S.}~\bibnamefont
  {Kojaku}}\ and\ \bibinfo {author} {\bibfnamefont {N.}~\bibnamefont
  {Masuda}},\ }\href@noop {} {\bibfield  {journal} {\bibinfo  {journal}
  {Physical Review E}\ }\textbf {\bibinfo {volume} {96}},\ \bibinfo {pages}
  {052313} (\bibinfo {year} {2017})}\BibitemShut {NoStop}%
\bibitem [{\citenamefont {Luce}\ and\ \citenamefont
  {Perry}(1949)}]{luce_method_1949}%
  \BibitemOpen
  \bibfield  {author} {\bibinfo {author} {\bibfnamefont {R.~D.}\ \bibnamefont
  {Luce}}\ and\ \bibinfo {author} {\bibfnamefont {A.~D.}\ \bibnamefont
  {Perry}},\ }\href {https://doi.org/10.1007/BF02289146} {\bibfield  {journal}
  {\bibinfo  {journal} {Psychometrika}\ }\textbf {\bibinfo {volume} {14}},\
  \bibinfo {pages} {95} (\bibinfo {year} {1949})}\BibitemShut {NoStop}%
\bibitem [{\citenamefont {Seidman}(1983)}]{seidman_network_1983}%
  \BibitemOpen
  \bibfield  {author} {\bibinfo {author} {\bibfnamefont {S.~B.}\ \bibnamefont
  {Seidman}},\ }\href {https://doi.org/10.1016/0378-8733(83)90028-X} {\bibfield
   {journal} {\bibinfo  {journal} {Social networks}\ }\textbf {\bibinfo
  {volume} {5}},\ \bibinfo {pages} {269} (\bibinfo {year} {1983})}\BibitemShut
  {NoStop}%
\bibitem [{\citenamefont {Breiger}\ \emph {et~al.}(1975)\citenamefont
  {Breiger}, \citenamefont {Boorman},\ and\ \citenamefont
  {Arabie}}]{breiger_algorithm_1975}%
  \BibitemOpen
  \bibfield  {author} {\bibinfo {author} {\bibfnamefont {R.~L.}\ \bibnamefont
  {Breiger}}, \bibinfo {author} {\bibfnamefont {S.~A.}\ \bibnamefont
  {Boorman}},\ and\ \bibinfo {author} {\bibfnamefont {P.}~\bibnamefont
  {Arabie}},\ }\href {https://doi.org/10.1016/0022-2496(75)90028-0} {\bibfield
  {journal} {\bibinfo  {journal} {Journal of mathematical psychology}\ }\textbf
  {\bibinfo {volume} {12}},\ \bibinfo {pages} {328} (\bibinfo {year}
  {1975})}\BibitemShut {NoStop}%
\bibitem [{\citenamefont {Girvan}\ and\ \citenamefont
  {Newman}(2002)}]{girvan_community_2002}%
  \BibitemOpen
  \bibfield  {author} {\bibinfo {author} {\bibfnamefont {M.}~\bibnamefont
  {Girvan}}\ and\ \bibinfo {author} {\bibfnamefont {M.~E.}\ \bibnamefont
  {Newman}},\ }\href {https://doi.org/10.1073/pnas.122653799} {\bibfield
  {journal} {\bibinfo  {journal} {Proceedings of the national academy of
  sciences}\ }\textbf {\bibinfo {volume} {99}},\ \bibinfo {pages} {7821}
  (\bibinfo {year} {2002})}\BibitemShut {NoStop}%
\bibitem [{\citenamefont {Holland}\ \emph {et~al.}(1983)\citenamefont
  {Holland}, \citenamefont {Laskey},\ and\ \citenamefont
  {Leinhardt}}]{holland_stochastic_1983}%
  \BibitemOpen
  \bibfield  {author} {\bibinfo {author} {\bibfnamefont {P.~W.}\ \bibnamefont
  {Holland}}, \bibinfo {author} {\bibfnamefont {K.~B.}\ \bibnamefont
  {Laskey}},\ and\ \bibinfo {author} {\bibfnamefont {S.}~\bibnamefont
  {Leinhardt}},\ }\href {https://doi.org/10.1016/0378-8733(83)90021-7}
  {\bibfield  {journal} {\bibinfo  {journal} {Social Networks}\ }\textbf
  {\bibinfo {volume} {5}},\ \bibinfo {pages} {109} (\bibinfo {year}
  {1983})}\BibitemShut {NoStop}%
\bibitem [{\citenamefont {Peixoto}(2017)}]{peixoto_nonparametric_2017}%
  \BibitemOpen
  \bibfield  {author} {\bibinfo {author} {\bibfnamefont {T.~P.}\ \bibnamefont
  {Peixoto}},\ }\href {https://doi.org/10.1103/PhysRevE.95.012317} {\bibfield
  {journal} {\bibinfo  {journal} {Physical Review E}\ }\textbf {\bibinfo
  {volume} {95}},\ \bibinfo {pages} {012317} (\bibinfo {year}
  {2017})}\BibitemShut {NoStop}%
\bibitem [{\citenamefont {Lancichinetti}\ and\ \citenamefont
  {Fortunato}(2012)}]{lancichinetti_consensus_2012}%
  \BibitemOpen
  \bibfield  {author} {\bibinfo {author} {\bibfnamefont {A.}~\bibnamefont
  {Lancichinetti}}\ and\ \bibinfo {author} {\bibfnamefont {S.}~\bibnamefont
  {Fortunato}},\ }\href {https://doi.org/10.1038/srep00336} {\bibfield
  {journal} {\bibinfo  {journal} {Scientific reports}\ }\textbf {\bibinfo
  {volume} {2}},\ \bibinfo {pages} {1} (\bibinfo {year} {2012})}\BibitemShut
  {NoStop}%
\bibitem [{\citenamefont {Tandon}\ \emph {et~al.}(2019)\citenamefont {Tandon},
  \citenamefont {Albeshri}, \citenamefont {Thayananthan}, \citenamefont
  {Alhalabi},\ and\ \citenamefont {Fortunato}}]{tandon_fast_2019}%
  \BibitemOpen
  \bibfield  {author} {\bibinfo {author} {\bibfnamefont {A.}~\bibnamefont
  {Tandon}}, \bibinfo {author} {\bibfnamefont {A.}~\bibnamefont {Albeshri}},
  \bibinfo {author} {\bibfnamefont {V.}~\bibnamefont {Thayananthan}}, \bibinfo
  {author} {\bibfnamefont {W.}~\bibnamefont {Alhalabi}},\ and\ \bibinfo
  {author} {\bibfnamefont {S.}~\bibnamefont {Fortunato}},\ }\href
  {https://doi.org/10.1103/PhysRevE.99.042301} {\bibfield  {journal} {\bibinfo
  {journal} {Physical Review E}\ }\textbf {\bibinfo {volume} {99}},\ \bibinfo
  {pages} {042301} (\bibinfo {year} {2019})}\BibitemShut {NoStop}%
\bibitem [{\citenamefont {Kirkley}\ and\ \citenamefont
  {Newman}(2022)}]{kirkley_representative_2022}%
  \BibitemOpen
  \bibfield  {author} {\bibinfo {author} {\bibfnamefont {A.}~\bibnamefont
  {Kirkley}}\ and\ \bibinfo {author} {\bibfnamefont {M.~E.~J.}\ \bibnamefont
  {Newman}},\ }\href {https://doi.org/10.1038/s42005-022-00816-3} {\bibfield
  {journal} {\bibinfo  {journal} {Communications Physics}\ }\textbf {\bibinfo
  {volume} {5}},\ \bibinfo {pages} {1} (\bibinfo {year} {2022})}\BibitemShut
  {NoStop}%
\bibitem [{\citenamefont {Moore}(2017)}]{moore_computer_2017}%
  \BibitemOpen
  \bibfield  {author} {\bibinfo {author} {\bibfnamefont {C.}~\bibnamefont
  {Moore}},\ }\href@noop {} {\bibinfo {title} {The {{Computer Science}} and
  {{Physics}} of {{Community Detection}}: {{Landscapes}}, {{Phase
  Transitions}}, and {{Hardness}}}} (\bibinfo {year} {2017}),\ \Eprint
  {https://arxiv.org/abs/1702.00467} {arXiv:1702.00467} \BibitemShut {NoStop}%
\bibitem [{\citenamefont {Abbe}(2017)}]{abbe_community_2017}%
  \BibitemOpen
  \bibfield  {author} {\bibinfo {author} {\bibfnamefont {E.}~\bibnamefont
  {Abbe}},\ }\href@noop {} {\bibfield  {journal} {\bibinfo  {journal} {The
  Journal of Machine Learning Research}\ }\textbf {\bibinfo {volume} {18}},\
  \bibinfo {pages} {6446} (\bibinfo {year} {2017})}\BibitemShut {NoStop}%
\bibitem [{\citenamefont {Reichardt}\ and\ \citenamefont
  {Leone}(2008)}]{reichardt__2008}%
  \BibitemOpen
  \bibfield  {author} {\bibinfo {author} {\bibfnamefont {J.}~\bibnamefont
  {Reichardt}}\ and\ \bibinfo {author} {\bibfnamefont {M.}~\bibnamefont
  {Leone}},\ }\href {https://doi.org/10.1103/PhysRevLett.101.078701} {\bibfield
   {journal} {\bibinfo  {journal} {Physical review letters}\ }\textbf {\bibinfo
  {volume} {101}},\ \bibinfo {pages} {078701} (\bibinfo {year}
  {2008})}\BibitemShut {NoStop}%
\bibitem [{\citenamefont {Nadakuditi}\ and\ \citenamefont
  {Newman}(2012)}]{nadakuditi_graph_2012}%
  \BibitemOpen
  \bibfield  {author} {\bibinfo {author} {\bibfnamefont {R.~R.}\ \bibnamefont
  {Nadakuditi}}\ and\ \bibinfo {author} {\bibfnamefont {M.~E.~J.}\ \bibnamefont
  {Newman}},\ }\href {https://doi.org/10.1103/PhysRevLett.108.188701}
  {\bibfield  {journal} {\bibinfo  {journal} {Physical Review Letters}\
  }\textbf {\bibinfo {volume} {108}},\ \bibinfo {pages} {188701} (\bibinfo
  {year} {2012})}\BibitemShut {NoStop}%
\bibitem [{\citenamefont {Zhang}\ \emph {et~al.}(2014)\citenamefont {Zhang},
  \citenamefont {Nadakuditi},\ and\ \citenamefont
  {Newman}}]{zhang_spectra_2014}%
  \BibitemOpen
  \bibfield  {author} {\bibinfo {author} {\bibfnamefont {X.}~\bibnamefont
  {Zhang}}, \bibinfo {author} {\bibfnamefont {R.~R.}\ \bibnamefont
  {Nadakuditi}},\ and\ \bibinfo {author} {\bibfnamefont {M.~E.~J.}\
  \bibnamefont {Newman}},\ }\href {https://doi.org/10.1103/PhysRevE.89.042816}
  {\bibfield  {journal} {\bibinfo  {journal} {Physical Review E}\ }\textbf
  {\bibinfo {volume} {89}},\ \bibinfo {pages} {042816} (\bibinfo {year}
  {2014})}\BibitemShut {NoStop}%
\bibitem [{\citenamefont {Radicchi}(2013)}]{radicchi_detectability_2013}%
  \BibitemOpen
  \bibfield  {author} {\bibinfo {author} {\bibfnamefont {F.}~\bibnamefont
  {Radicchi}},\ }\href {https://doi.org/10.1103/PhysRevE.88.010801} {\bibfield
  {journal} {\bibinfo  {journal} {Physical Review E}\ }\textbf {\bibinfo
  {volume} {88}},\ \bibinfo {pages} {010801} (\bibinfo {year}
  {2013})}\BibitemShut {NoStop}%
\bibitem [{\citenamefont {Lorrain}\ and\ \citenamefont
  {White}(1971)}]{lorrain_structural_1971}%
  \BibitemOpen
  \bibfield  {author} {\bibinfo {author} {\bibfnamefont {F.}~\bibnamefont
  {Lorrain}}\ and\ \bibinfo {author} {\bibfnamefont {H.~C.}\ \bibnamefont
  {White}},\ }\href@noop {} {\bibfield  {journal} {\bibinfo  {journal} {The
  Journal of mathematical sociology}\ }\textbf {\bibinfo {volume} {1}},\
  \bibinfo {pages} {49} (\bibinfo {year} {1971})}\BibitemShut {NoStop}%
\bibitem [{\citenamefont {White}\ \emph {et~al.}(1976)\citenamefont {White},
  \citenamefont {Boorman},\ and\ \citenamefont {Breiger}}]{white_social_1976}%
  \BibitemOpen
  \bibfield  {author} {\bibinfo {author} {\bibfnamefont {H.~C.}\ \bibnamefont
  {White}}, \bibinfo {author} {\bibfnamefont {S.~A.}\ \bibnamefont {Boorman}},\
  and\ \bibinfo {author} {\bibfnamefont {R.~L.}\ \bibnamefont {Breiger}},\
  }\href@noop {} {\bibfield  {journal} {\bibinfo  {journal} {American journal
  of sociology}\ }\textbf {\bibinfo {volume} {81}},\ \bibinfo {pages} {730}
  (\bibinfo {year} {1976})}\BibitemShut {NoStop}%
\bibitem [{\citenamefont {Everett}\ and\ \citenamefont
  {Borgatti}(1994)}]{everett_regular_1994}%
  \BibitemOpen
  \bibfield  {author} {\bibinfo {author} {\bibfnamefont {M.~G.}\ \bibnamefont
  {Everett}}\ and\ \bibinfo {author} {\bibfnamefont {S.~P.}\ \bibnamefont
  {Borgatti}},\ }\href@noop {} {\bibfield  {journal} {\bibinfo  {journal}
  {Journal of mathematical sociology}\ }\textbf {\bibinfo {volume} {19}},\
  \bibinfo {pages} {29} (\bibinfo {year} {1994})}\BibitemShut {NoStop}%
\bibitem [{\citenamefont {Hastings}(2006)}]{hastings_community_2006}%
  \BibitemOpen
  \bibfield  {author} {\bibinfo {author} {\bibfnamefont {M.~B.}\ \bibnamefont
  {Hastings}},\ }\href {https://doi.org/10.1103/PhysRevE.74.035102} {\bibfield
  {journal} {\bibinfo  {journal} {Physical Review E}\ }\textbf {\bibinfo
  {volume} {74}},\ \bibinfo {pages} {035102} (\bibinfo {year}
  {2006})}\BibitemShut {NoStop}%
\bibitem [{\citenamefont
  {Peixoto}(2014{\natexlab{a}})}]{peixoto_hierarchical_2014}%
  \BibitemOpen
  \bibfield  {author} {\bibinfo {author} {\bibfnamefont {T.~P.}\ \bibnamefont
  {Peixoto}},\ }\href {https://doi.org/10.1103/PhysRevX.4.011047} {\bibfield
  {journal} {\bibinfo  {journal} {Physical Review X}\ }\textbf {\bibinfo
  {volume} {4}},\ \bibinfo {pages} {011047} (\bibinfo {year}
  {2014}{\natexlab{a}})}\BibitemShut {NoStop}%
\bibitem [{\citenamefont {Airoldi}\ \emph {et~al.}(2008)\citenamefont
  {Airoldi}, \citenamefont {Blei}, \citenamefont {Fienberg},\ and\
  \citenamefont {Xing}}]{airoldi_mixed_2008}%
  \BibitemOpen
  \bibfield  {author} {\bibinfo {author} {\bibfnamefont {E.~M.}\ \bibnamefont
  {Airoldi}}, \bibinfo {author} {\bibfnamefont {D.~M.}\ \bibnamefont {Blei}},
  \bibinfo {author} {\bibfnamefont {S.~E.}\ \bibnamefont {Fienberg}},\ and\
  \bibinfo {author} {\bibfnamefont {E.~P.}\ \bibnamefont {Xing}},\ }\href
  {https://proceedings.neurips.cc/paper/2008/file/8613985ec49eb8f757ae6439e879bb2a-Paper.pdf}
  {\bibfield  {journal} {\bibinfo  {journal} {Advances in neural information
  processing systems}\ }\textbf {\bibinfo {volume} {21}} (\bibinfo {year}
  {2008})}\BibitemShut {NoStop}%
\bibitem [{\citenamefont {Peixoto}(2015{\natexlab{a}})}]{peixoto_model_2015}%
  \BibitemOpen
  \bibfield  {author} {\bibinfo {author} {\bibfnamefont {T.~P.}\ \bibnamefont
  {Peixoto}},\ }\href {https://doi.org/10.1103/PhysRevX.5.011033} {\bibfield
  {journal} {\bibinfo  {journal} {Physical Review X}\ }\textbf {\bibinfo
  {volume} {5}},\ \bibinfo {pages} {011033} (\bibinfo {year}
  {2015}{\natexlab{a}})}\BibitemShut {NoStop}%
\bibitem [{\citenamefont
  {Peixoto}(2015{\natexlab{b}})}]{peixoto_inferring_2015}%
  \BibitemOpen
  \bibfield  {author} {\bibinfo {author} {\bibfnamefont {T.~P.}\ \bibnamefont
  {Peixoto}},\ }\href {https://doi.org/10.1103/PhysRevE.92.042807} {\bibfield
  {journal} {\bibinfo  {journal} {Physical Review E}\ }\textbf {\bibinfo
  {volume} {92}},\ \bibinfo {pages} {042807} (\bibinfo {year}
  {2015}{\natexlab{b}})}\BibitemShut {NoStop}%
\bibitem [{\citenamefont {Lancichinetti}\ and\ \citenamefont
  {Fortunato}(2009)}]{lancichinetti_community_2009}%
  \BibitemOpen
  \bibfield  {author} {\bibinfo {author} {\bibfnamefont {A.}~\bibnamefont
  {Lancichinetti}}\ and\ \bibinfo {author} {\bibfnamefont {S.}~\bibnamefont
  {Fortunato}},\ }\href {https://doi.org/10.1103/PhysRevE.80.056117} {\bibfield
   {journal} {\bibinfo  {journal} {Physical Review E}\ }\textbf {\bibinfo
  {volume} {80}},\ \bibinfo {pages} {056117} (\bibinfo {year}
  {2009})}\BibitemShut {NoStop}%
\bibitem [{\citenamefont {Bazzi}\ \emph {et~al.}(2020)\citenamefont {Bazzi},
  \citenamefont {Jeub}, \citenamefont {Arenas}, \citenamefont {Howison},\ and\
  \citenamefont {Porter}}]{bazzi_framework_2020}%
  \BibitemOpen
  \bibfield  {author} {\bibinfo {author} {\bibfnamefont {M.}~\bibnamefont
  {Bazzi}}, \bibinfo {author} {\bibfnamefont {L.~G.~S.}\ \bibnamefont {Jeub}},
  \bibinfo {author} {\bibfnamefont {A.}~\bibnamefont {Arenas}}, \bibinfo
  {author} {\bibfnamefont {S.~D.}\ \bibnamefont {Howison}},\ and\ \bibinfo
  {author} {\bibfnamefont {M.~A.}\ \bibnamefont {Porter}},\ }\href
  {https://doi.org/10.1103/PhysRevResearch.2.023100} {\bibfield  {journal}
  {\bibinfo  {journal} {Physical Review Research}\ }\textbf {\bibinfo {volume}
  {2}},\ \bibinfo {pages} {023100} (\bibinfo {year} {2020})}\BibitemShut
  {NoStop}%
\bibitem [{\citenamefont {Fosdick}\ \emph {et~al.}(2018)\citenamefont
  {Fosdick}, \citenamefont {Larremore}, \citenamefont {Nishimura},\ and\
  \citenamefont {Ugander}}]{fosdick_configuring_2018}%
  \BibitemOpen
  \bibfield  {author} {\bibinfo {author} {\bibfnamefont {B.~K.}\ \bibnamefont
  {Fosdick}}, \bibinfo {author} {\bibfnamefont {D.~B.}\ \bibnamefont
  {Larremore}}, \bibinfo {author} {\bibfnamefont {J.}~\bibnamefont
  {Nishimura}},\ and\ \bibinfo {author} {\bibfnamefont {J.}~\bibnamefont
  {Ugander}},\ }\href {https://doi.org/10.1137/16M1087175} {\bibfield
  {journal} {\bibinfo  {journal} {SIAM Review}\ }\textbf {\bibinfo {volume}
  {60}},\ \bibinfo {pages} {315} (\bibinfo {year} {2018})}\BibitemShut
  {NoStop}%
\bibitem [{\citenamefont {Lin}(1991)}]{lin_divergence_1991}%
  \BibitemOpen
  \bibfield  {author} {\bibinfo {author} {\bibfnamefont {J.}~\bibnamefont
  {Lin}},\ }\href {https://doi.org/10.1109/18.61115} {\bibfield  {journal}
  {\bibinfo  {journal} {IEEE Transactions on Information theory}\ }\textbf
  {\bibinfo {volume} {37}},\ \bibinfo {pages} {145} (\bibinfo {year}
  {1991})}\BibitemShut {NoStop}%
\bibitem [{\citenamefont
  {Peixoto}(2014{\natexlab{b}})}]{peixoto_graph-tool_2014}%
  \BibitemOpen
  \bibfield  {author} {\bibinfo {author} {\bibfnamefont {T.~P.}\ \bibnamefont
  {Peixoto}},\ }\bibfield  {journal} {\bibinfo  {journal} {figshare}\ }\href
  {https://doi.org/10.6084/m9.figshare.1164194} {10.6084/m9.figshare.1164194}
  (\bibinfo {year} {2014}{\natexlab{b}})\BibitemShut {NoStop}%
\bibitem [{\citenamefont {Kuhn}(1955)}]{kuhn_hungarian_1955}%
  \BibitemOpen
  \bibfield  {author} {\bibinfo {author} {\bibfnamefont {H.~W.}\ \bibnamefont
  {Kuhn}},\ }\href {https://doi.org/10.1002/nav.3800020109} {\bibfield
  {journal} {\bibinfo  {journal} {Naval research logistics quarterly}\ }\textbf
  {\bibinfo {volume} {2}},\ \bibinfo {pages} {83} (\bibinfo {year}
  {1955})}\BibitemShut {NoStop}%
\bibitem [{\citenamefont {Munkres}(1957)}]{munkres_algorithms_1957}%
  \BibitemOpen
  \bibfield  {author} {\bibinfo {author} {\bibfnamefont {J.}~\bibnamefont
  {Munkres}},\ }\href {https://doi.org/10.1137/0105003} {\bibfield  {journal}
  {\bibinfo  {journal} {Journal of the society for industrial and applied
  mathematics}\ }\textbf {\bibinfo {volume} {5}},\ \bibinfo {pages} {32}
  (\bibinfo {year} {1957})}\BibitemShut {NoStop}%
\bibitem [{\citenamefont {Wagner}\ and\ \citenamefont
  {Wagner}(2007)}]{wagner_comparing_2007}%
  \BibitemOpen
  \bibfield  {author} {\bibinfo {author} {\bibfnamefont {S.}~\bibnamefont
  {Wagner}}\ and\ \bibinfo {author} {\bibfnamefont {D.}~\bibnamefont
  {Wagner}},\ }\href {https://doi.org/10.5445/IR/1000011477} {\bibinfo {title}
  {{Comparing Clusterings - An Overview}}},\ \bibinfo {howpublished}
  {https://publikationen.bibliothek.kit.edu/1000011477} (\bibinfo {year}
  {2007})\BibitemShut {NoStop}%
\bibitem [{\citenamefont {Newman}\ \emph {et~al.}(2020)\citenamefont {Newman},
  \citenamefont {Cantwell},\ and\ \citenamefont
  {Young}}]{newman_improved_2020}%
  \BibitemOpen
  \bibfield  {author} {\bibinfo {author} {\bibfnamefont {M.~E.}\ \bibnamefont
  {Newman}}, \bibinfo {author} {\bibfnamefont {G.~T.}\ \bibnamefont
  {Cantwell}},\ and\ \bibinfo {author} {\bibfnamefont {J.-G.}\ \bibnamefont
  {Young}},\ }\href@noop {} {\bibfield  {journal} {\bibinfo  {journal}
  {Physical Review E}\ }\textbf {\bibinfo {volume} {101}},\ \bibinfo {pages}
  {042304} (\bibinfo {year} {2020})}\BibitemShut {NoStop}%
\bibitem [{\citenamefont {Meil{\u{a}}}(2007)}]{meila_comparing_2007}%
  \BibitemOpen
  \bibfield  {author} {\bibinfo {author} {\bibfnamefont {M.}~\bibnamefont
  {Meil{\u{a}}}},\ }\href@noop {} {\bibfield  {journal} {\bibinfo  {journal}
  {Journal of multivariate analysis}\ }\textbf {\bibinfo {volume} {98}},\
  \bibinfo {pages} {873} (\bibinfo {year} {2007})}\BibitemShut {NoStop}%
\bibitem [{\citenamefont {Peixoto}(2022)}]{peixoto_disentangling_2022}%
  \BibitemOpen
  \bibfield  {author} {\bibinfo {author} {\bibfnamefont {T.~P.}\ \bibnamefont
  {Peixoto}},\ }\href {https://doi.org/10.1103/PhysRevX.12.011004} {\bibfield
  {journal} {\bibinfo  {journal} {Physical Review X}\ }\textbf {\bibinfo
  {volume} {12}},\ \bibinfo {pages} {011004} (\bibinfo {year}
  {2022})}\BibitemShut {NoStop}%
\bibitem [{\citenamefont {Zhang}\ and\ \citenamefont
  {Peixoto}(2020)}]{zhang_statistical_2020}%
  \BibitemOpen
  \bibfield  {author} {\bibinfo {author} {\bibfnamefont {L.}~\bibnamefont
  {Zhang}}\ and\ \bibinfo {author} {\bibfnamefont {T.~P.}\ \bibnamefont
  {Peixoto}},\ }\href {https://doi.org/10.1103/PhysRevResearch.2.043271}
  {\bibfield  {journal} {\bibinfo  {journal} {Physical Review Research}\
  }\textbf {\bibinfo {volume} {2}},\ \bibinfo {pages} {043271} (\bibinfo {year}
  {2020})}\BibitemShut {NoStop}%
\bibitem [{\citenamefont {Peixoto}(2013)}]{peixoto_parsimonious_2013}%
  \BibitemOpen
  \bibfield  {author} {\bibinfo {author} {\bibfnamefont {T.~P.}\ \bibnamefont
  {Peixoto}},\ }\href@noop {} {\bibfield  {journal} {\bibinfo  {journal}
  {Physical review letters}\ }\textbf {\bibinfo {volume} {110}},\ \bibinfo
  {pages} {148701} (\bibinfo {year} {2013})}\BibitemShut {NoStop}%
\bibitem [{\citenamefont
  {Ramaciotti~Morales}(2023)}]{ramaciotti_morales_multidimensional_2023}%
  \BibitemOpen
  \bibfield  {author} {\bibinfo {author} {\bibfnamefont {P.}~\bibnamefont
  {Ramaciotti~Morales}},\ }in\ \href
  {https://doi.org/10.1007/978-3-031-21127-0_15} {\emph {\bibinfo {booktitle}
  {Complex {{Networks}} and {{Their Applications XI}}}}}\ (\bibinfo
  {publisher} {{Springer International Publishing}},\ \bibinfo {address}
  {{Cham}},\ \bibinfo {year} {2023})\ pp.\ \bibinfo {pages}
  {176--189}\BibitemShut {NoStop}%
\bibitem [{\citenamefont {Lawson}\ and\ \citenamefont
  {Hanson}(1995)}]{lawson_solving_1995}%
  \BibitemOpen
  \bibfield  {author} {\bibinfo {author} {\bibfnamefont {C.~L.}\ \bibnamefont
  {Lawson}}\ and\ \bibinfo {author} {\bibfnamefont {R.~J.}\ \bibnamefont
  {Hanson}},\ }\href@noop {} {\emph {\bibinfo {title} {Solving least squares
  problems}}}\ (\bibinfo  {publisher} {SIAM},\ \bibinfo {year}
  {1995})\BibitemShut {NoStop}%
\bibitem [{\citenamefont {Dantzig}(1963)}]{dantzig_linear_1963}%
  \BibitemOpen
  \bibfield  {author} {\bibinfo {author} {\bibfnamefont {G.}~\bibnamefont
  {Dantzig}},\ }\href@noop {} {\emph {\bibinfo {title} {Linear programming and
  extensions}}}\ (\bibinfo  {publisher} {Princeton university press},\ \bibinfo
  {year} {1963})\BibitemShut {NoStop}%
\bibitem [{\citenamefont {Huangfu}\ and\ \citenamefont
  {Hall}(2018)}]{huangfu_parallelizing_2018}%
  \BibitemOpen
  \bibfield  {author} {\bibinfo {author} {\bibfnamefont {Q.}~\bibnamefont
  {Huangfu}}\ and\ \bibinfo {author} {\bibfnamefont {J.~J.}\ \bibnamefont
  {Hall}},\ }\href@noop {} {\bibfield  {journal} {\bibinfo  {journal}
  {Mathematical Programming Computation}\ }\textbf {\bibinfo {volume} {10}},\
  \bibinfo {pages} {119} (\bibinfo {year} {2018})}\BibitemShut {NoStop}%
\bibitem [{\citenamefont {Farkas}(1902)}]{farkas_theorie_1902}%
  \BibitemOpen
  \bibfield  {author} {\bibinfo {author} {\bibfnamefont {J.}~\bibnamefont
  {Farkas}},\ }\href@noop {} {\bibfield  {journal} {\bibinfo  {journal}
  {Journal f{\"u}r die reine und angewandte Mathematik (Crelles Journal)}\
  }\textbf {\bibinfo {volume} {1902}},\ \bibinfo {pages} {1} (\bibinfo {year}
  {1902})}\BibitemShut {NoStop}%
\bibitem [{\citenamefont {Boyd}\ \emph {et~al.}(2004)\citenamefont {Boyd},
  \citenamefont {Boyd},\ and\ \citenamefont {Vandenberghe}}]{boyd_convex_2004}%
  \BibitemOpen
  \bibfield  {author} {\bibinfo {author} {\bibfnamefont {S.}~\bibnamefont
  {Boyd}}, \bibinfo {author} {\bibfnamefont {S.~P.}\ \bibnamefont {Boyd}},\
  and\ \bibinfo {author} {\bibfnamefont {L.}~\bibnamefont {Vandenberghe}},\
  }\href@noop {} {\emph {\bibinfo {title} {Convex Optimization}}}\ (\bibinfo
  {publisher} {{Cambridge university press}},\ \bibinfo {year}
  {2004})\BibitemShut {NoStop}%
\end{thebibliography}%

\end{document}